\documentclass[twocolappendix,revtex4]{emulateapj}
\usepackage{apjfonts,color,amsmath,textcomp,url,graphicx,subfigure}
\usepackage{pdflscape}

\newcommand{\kms}{\hbox{km\,s$^{-1}$}}
\newcommand{\MJup}{$M_{\mathrm{Jup}}$}
\newcommand{\RJup}{$R_{\mathrm{Jup}}$}
\newcommand{\Msol}{$M_{\odot}$}
\newcommand{\masyr}{$\mathrm{mas}\,\mathrm{yr}^{-1}$}
\newcommand{\Teff}{\ensuremath{T_{\mathrm{eff}}}}
\newcommand{\besancon}{Besan\c con}

\slugcomment{To appear in ApJ.}

\shorttitle{BANYAN. VII. Young Substellar Moving Group Candidate Members}
\shortauthors{Gagn\'e et al.}

\begin{document}

\title{BANYAN. VII. A NEW POPULATION OF YOUNG SUBSTELLAR CANDIDATE MEMBERS OF NEARBY MOVING GROUPS FROM THE BASS SURVEY}

\author{Jonathan Gagn\'e\altaffilmark{1},\, Jacqueline K. Faherty\altaffilmark{2,3,4},\, Kelle L. Cruz\altaffilmark{3,5,6},\, David Lafreni\`ere\altaffilmark{1},\, Ren\'e Doyon\altaffilmark{1},\, Lison Malo\altaffilmark{7,1},\, Adam J. Burgasser\altaffilmark{8},\, Marie-Eve Naud\altaffilmark{1},\, \'Etienne Artigau\altaffilmark{1},\, Sandie Bouchard\altaffilmark{1},\, John E. Gizis\altaffilmark{9},\, and Lo\"ic Albert\altaffilmark{1}}
\affil{\altaffilmark{1} Institut de Recherche sur les Exoplan\`etes (iREx), Universit\'e de Montr\'eal, D\'epartement de Physique, C.P.~6128 Succ. Centre-ville, Montr\'eal, QC H3C~3J7, Canada. jonathan.gagne@astro.umontreal.ca}
\affil{\altaffilmark{2} Department of Terrestrial Magnetism, Carnegie Institution of Washington, Washington, DC~20015, USA}
\affil{\altaffilmark{3} Department of Astrophysics, American Museum of Natural History, Central Park West at 79th Street, New York, NY~10024.}
\affil{\altaffilmark{4} Hubble Fellow}
\affil{\altaffilmark{5} Department of Physics \& Astronomy, Hunter College, City University of New York, 695 Park Avenue, NY~10065, USA.}
\affil{\altaffilmark{6} Department of Physics, Graduate Center, City University of New York, 365 Fifth Avenue, New York, NY 10016, USA}
\affil{\altaffilmark{7} Canada-France-Hawaii Telescope, 65-1238 Mamalahoa Hwy, Kamuela, HI~96743, USA}
\affil{\altaffilmark{8} Center for Astrophysics and Space Sciences, University of California, San Diego, 9500 Gilman Dr., Mail Code 0424, La Jolla, CA~92093, USA.}
\affil{\altaffilmark{9} Department of Physics and Astronomy, University of Delaware, 104 The Green, Newark, DE~19716, USA.}

\begin{abstract}

We present the results of a near-infrared (NIR) spectroscopic follow-up survey of 182 M4--L7 low-mass stars and brown dwarfs (BDs) from the \emph{BANYAN All-Sky Survey} (\emph{BASS}) for candidate members of nearby, young moving groups (YMGs). We confirm signs of low-gravity for 42 new BD discoveries with estimated masses between 8--75\,\MJup\ and identify previously unrecognized signs of low gravity for 24 known BDs. This allows us to refine the fraction of low-gravity dwarfs in the high-probability \emph{BASS} sample to $\sim$\,82\%. We use this unique sample of 66 young BDs, supplemented with 22 young BDs from the literature, to construct new empirical NIR absolute magnitude and color sequences for low-gravity BDs. We show that low-resolution NIR spectroscopy alone cannot differentiate between the ages of YMGs younger than $\sim$\,120\,Myr, and that the BT-Settl atmosphere models do not reproduce well the dust clouds in field or low-gravity L-type dwarfs. We obtain a spectroscopic confirmation of low-gravity for 2MASS~J14252798--3650229, which is a new $\sim$\,27\,\MJup, L4\,$\gamma$ bona fide member of AB~Doradus. We identify in this work a total of 19 new low-gravity candidate members of YMGs with estimated masses below 13\,\MJup, seven of which have kinematically estimated distances within 40\,pc. These objects will be valuable benchmarks for a detailed atmospheric characterization of planetary-mass objects with the next generation of instruments such as the James Webb Space Telescope. We find 16 strong candidate members of the Tucana-Horologium association with estimated masses between 12.5--14\,\MJup, a regime where our study was particularly sensitive. This would indicate that for this association there is at least one isolated object in this mass range for every $17.5_{-5.0}^{+6.6}$ main-sequence stellar member, a number significantly higher than expected based on standard log-normal initial mass function, however in the absence of radial velocity and parallax measurements for all of them, it is likely that this over-density is caused by a number of young interlopers from other moving groups. Finally, as a byproduct of this project, we identify 12 new L0--L5 field BDs, seven of which display peculiar spectroscopic properties.

\end{abstract}


\section{INTRODUCTION}

Young moving groups (YMGs) consist of stars that formed recently ($\lesssim$\,120\,Myr) from a molecular cloud and that are too young to have experienced significant gravitational perturbations from their environment. The members of YMGs share similar galactic velocities within a few \kms. The closest and youngest moving groups include the TW~Hydrae association (TWA; 5--15\,Myr; \citealp{1989ApJ...343L..61D,1997Sci...277...67K,2004ARA&A..42..685Z,2013ApJ...762..118W}), $\beta$~Pictoris ($\beta$PMG; 20--26\,Myr; \citealp{2001ApJ...562L..87Z,2014ApJ...792...37M,2014MNRAS.438L..11B}), Tucana-Horologium (THA; 20--40\,Myr; \citealp{2000AJ....120.1410T,2000ApJ...535..959Z,2001ApJ...559..388Z,2014AJ....147..146K}), Carina (CAR; 20--40\,Myr; \citealp{2008hsf2.book..757T}), Columba (COL; 20--40\,Myr; \citealp{2008hsf2.book..757T}),  Argus (ARG; 30--50\,Myr; \citealp{2000MNRAS.317..289M}) and AB~Doradus (ABDMG; 110--130\,Myr; \citealp{2004ApJ...613L..65Z,2005ApJ...628L..69L,2013ApJ...766....6B}). The YMGs are ideal laboratories to measure fundamental properties of star formation such as the initial mass function (IMF) because their members are coeval. This is of particular interest in the case of very low-mass stars and substellar-mass objects (spectral types $\geq$\,M5) since these populations are still poorly characterized. The massive, bright population of YMGs has already been explored, thanks to the \emph{Hipparcos} survey \citep{1997A&A...323L..49P}. However, fainter members are hard to identify mainly because of the lack of radial velocity (RV) and trigonometric distance measurements that are necessary to obtain their spacial velocities and galactic positions. Several efforts have been made to identify the very low-mass members of YMGs \citep{2011MNRAS.411..117K,2012AJ....144..109S,2012ApJ...758...56S,2012ApJ...752...56F,2013AJ....145....2F,2013ApJ...774..101R,2013ApJ...777L..20L,2014ApJ...788...81M,2014AJ....147..146K,2014AJ....147...85R,2015MNRAS.447.1267M}; however, as of today it is likely that most of them still remain to be identified.

The Bayesian Analysis for Nearby Young AssociatioNs tool\footnote{Publicly available at \url{http://www.astro.umontreal.ca/\textasciitilde malo/banyan.php}} (BANYAN; \citealp{2013ApJ...762...88M}), which is based on naive Bayesian inference, identified promising candidate members of YMGs among a sample of low-mass stars that do not have prior RV or parallax measurements. The BANYAN~II tool\footnote{Publicly available at \url{http://www.astro.umontreal.ca/\textasciitilde gagne/banyanII.php}} (\citealt{2014ApJ...783..121G}; Paper~II hereafter) was subsequently developed to identify substellar candidate members with a similar but improved algorithm. BANYAN~II is an expansion on BANYAN~I that is focused on very-low mass stars and brown dwarfs (BDs) with spectral types $\geq$\,M5. The \emph{BANYAN All-Sky Survey} (\emph{BASS}; \citealt{2015ApJ...798...73G}; Paper~V hereafter) was initiated by our team to search for the elusive late-type ($\geq$\,M5) members of YMGs, using the BANYAN~II tool on an all-sky cross-match of the \emph{Two Micron All-Sky Survey} (\emph{2MASS};  \citealp{2006AJ....131.1163S}) with the \emph{AllWISE} survey \citep{2014ApJ...783..122K}. The \emph{AllWISE} survey is based on a combination of the cryogenic phase of the \emph{Wide-Field Survey Explorer mission} (\emph{WISE}; \citealp{2010AJ....140.1868W}) and the \emph{Near-Earth Object WISE} (\emph{NEOWISE}; \citealp{2011ApJ...743..156M}) post-cryogenic phase.

We present here the results of a near-infrared (NIR) spectroscopic follow-up survey of substellar candidate members of YMGs identified in \emph{BASS}. In Section~\ref{sec:cand}, we summarize \emph{BASS} and the method that we used to build the sample of candidate members from a cross-match of \emph{2MASS} and \emph{AllWISE}. We detail our NIR spectroscopic follow-up and its motivation in Section~\ref{sec:obs}. In Section~\ref{sec:sptclass}, we present our method to assign a spectral and gravity classification. We present the resulting spectral types and updated YMG membership probability for our sample in Section~\ref{sec:results}. In Section~\ref{sec:discussion}, we use new discoveries presented here and other known low-gravity BDs and low-mass stars to build empirical photometric sequences, and we then investigate the physical properties of young BDs. We summarize and conclude in Section~\ref{sec:conclusion}.

\section{THE \emph{BASS} SURVEY}\label{sec:cand}

\emph{BASS} is a systematic all-sky search for later-than-M5 candidate members to nearby YMGs that was the focus of an earlier publication \citep{2015ApJ...798...73G}. In this work, we undertake a spectroscopic follow-up of the \emph{BASS} sample, which we briefly summarize in this section. We refer the reader to \citep{2015ApJ...798...73G} for an extensive description of the \emph{BASS} survey.

We cross-matched the \emph{2MASS} and \emph{AllWISE} catalogs outside of the galactic plane and crowded regions ($\geq 2.5$ objects per square arcminute) using a cross-match radius of 25\textquotedbl\ and applied color, confusion and photometric quality cuts to produce a starting sample of 98\,970 targets with NIR colors consistent with $\geq$ M5 spectral types and proper motion measurements larger than 30\,\masyr\ at $\geq$\,5$\sigma$ (see \citealt{2015ApJ...798...73G} for the detailed cross-match and selection algorithm). Astrometry provided in the \emph{2MASS} and \emph{AllWISE} catalogs as well as the mean epochs of observation for both surveys (\emph{JD} keyword in \emph{2MASS}; \emph{W1MJDMEAN} keyword in \emph{AllWISE}) were used to calculate proper motions. We used \emph{W1MJDMAX}-\emph{W1MJDMIN} as a conservative measurement error on the \emph{AllWISE} astrometric epoch, which typically corresponds to $\sim$\,6 months to one year, compared to a $\sim$\,11\,yr baseline between \emph{2MASS} and \emph{AllWISE}. This uncertainty as well as those on astrometric measurements themselves were propagated to the proper motion measurement errors. We obtain a typical proper motion precision of $\sim$\,15\,\masyr.

We used the BANYAN~II tool to select only objects that have a Bayesian probability $>$\,10\% of belonging to any YMG considered here (this threshold ensures that known bona fide members are recovered; see Paper~V). The BANYAN~II tool takes the sky position, proper motion and $J$, $H$, $K_S$, \emph{W1} and \emph{W2} photometry as input quantities. It then uses a naive Bayesian classifier to compare those measurements with spatial and kinematic models (SKMs) of YMGs, as well as with old and young color-magnitude diagram (CMD) sequences in both $M_{W1}(J - K_S)$ and $M_{W1}(H - W2)$ spaces. Those CMD sequences were chosen because they were found as the most efficient independent sequences to distinguish between young and field M6--L4 dwarfs. Probabilities generated from a naive Bayesian classifier can be biased when the input parameters are not independent (which is the case here); however, the relative ranking of hypotheses for a given object overcomes this bias \citep{Hand:2001tr}.

It is known that there is a large scatter in the NIR colors of young BDs even though they are redder than field dwarfs on average (e.g. \citealt{2012ApJ...752...56F}). The inclusion of the CMD sequences described above in BANYAN~II will systematically bias our sample towards red NIR colors, and decrease our sample completeness for YMG members that are not especially red. However, this effect is likely less important than the color criteria that were applied in selecting the 98\,970 objects that were input to BANYAN~II. Furthermore, a total of only two independent photometric observables (corresponding to the color-magnitude diagrams) are used in BANYAN~II, compared to four kinematics observables when no RV or parallax is available; the relative weight of kinematics is thus twice that of photometry in the calculation of probabilities. Parallax motion was not accounted for in our proper motion measurements or in the BANYAN~II tool; the maximal relative importance of this effect will become  as large as our typical \emph{2MASS}--\emph{AllWISE} proper motion precision only for objects closer than $\sim$\,10\,pc (considering the 11\,yr baseline between \emph{2MASS} and \emph{AllWISE}). This correction will properly be accounted for in a future version of the BANYAN~II tool.

We performed a Monte Carlo simulation based on the \besancon\ galactic model (A.~C.~Robin et al. in preparation, \citealp{2012A&A...538A.106R}) and the SKMs of YMGs to obtain a field contamination probability for each individual target in our sample, which allows for a more absolute interpretation in terms of the expected contamination fraction. We used the results of this simulation to reject any candidate member with a $> 50$\% probability of being a field contaminant. Note that the contamination probability from this Monte Carlo analysis is not necessarily complementary with the YMG Bayesian probability (see Paper~V for more detail). We refer the reader to Paper~V for an extensive description of all filters that were used to build the \emph{BASS} sample (e.g., minimal proper motion, color and quality filters, etc.).

There are three samples that are referred to in this Paper: (1) \emph{PRE-BASS} consists of targets that were initially selected as potential members and followed up with spectroscopy, but that were later rejected as we modified our selection criteria to reject contaminants; (2) \emph{Low-Priority BASS} (\emph{LP-BASS}) consists of targets that have NIR colors only slightly redder than field dwarfs; and (3) \emph{BASS} is the final sample presented in Paper~V that contains targets at least 1$\sigma$ redder than field dwarfs and that has a lower fraction of contaminants. As discussed in Paper~V, the statistical distance associated with the most probable YMG of a candidate member can be used to place it in two CMDs ($M_{W1}$ versus $J - K_S$ and $H - W2$) and compare its position to known field and young BDs and low-mass stars.

%Table : Observing Log
% Table : Observing Log
\clearpage
\onecolumngrid
\LongTables
\tabletypesize{\scriptsize}
\begin{deluxetable}{llllllcccc}
\tablecolumns{10}
\tablecaption{Observing log.\label{tab:obslog}}
\tablehead{\multicolumn{2}{c}{\emph{2MASS}} & \colhead{} & \colhead{Observing} & \colhead{Telescope} & \colhead{Instrument} & \colhead{Slit} & \colhead{Resolving} & \colhead{Tot. Exp.} & \colhead{Num.}\\
\cline{1-2}
\colhead{Designation} & \colhead{$J$} & \colhead{} & \colhead{Date (UT)} & \colhead{} & \colhead{} & \colhead{Width (\textquotedbl)} & \colhead{Power} & \colhead{Time (s)} & \colhead{Exposures}}
\startdata

\multicolumn{10}{c}{Candidate Members from \emph{BASS} or \emph{LP-BASS}}\\[2.5pt]\hline\\[-6pt]
00011217+1535355 & $15.52$ & & 2014 Aug 06 & Magellan & FIRE & 0.6 & 450 & 720 & 6\\
00065794-6436542 & $13.39$ & & 2014 Jan 16 & Gemini-South & Flamingos-2 & 0.72 & 500 & 840 & 4\\
00182834-6703130 & $15.46$ & & 2013 Sep 28 & Gemini-South & Flamingos-2 & 0.72 & 500 & 2000 & 8\\
00191296-6226005 & $15.64$ & & 2013 Oct 29 & Gemini-South & Flamingos-2 & 0.72 & 500 & 6200 & 20\\
00192626+4614078 & $12.60$ & & 2012 Oct 26 & IRTF & SpeX & 0.8 & 750 & 1200 & 8\\
00274534-0806046 & $11.57$ & & 2013 Oct 02 & Gemini-South & Flamingos-2 & 0.72 & 500 & 240 & 8\\
\enddata
\end{deluxetable}
\clearpage
\twocolumngrid

%Figure : Templates showcase
\begin{figure*}
	\centering
	\subfigure[Intermediate-Gravity ($\beta$) Templates]{\includegraphics[width=0.995\textwidth]{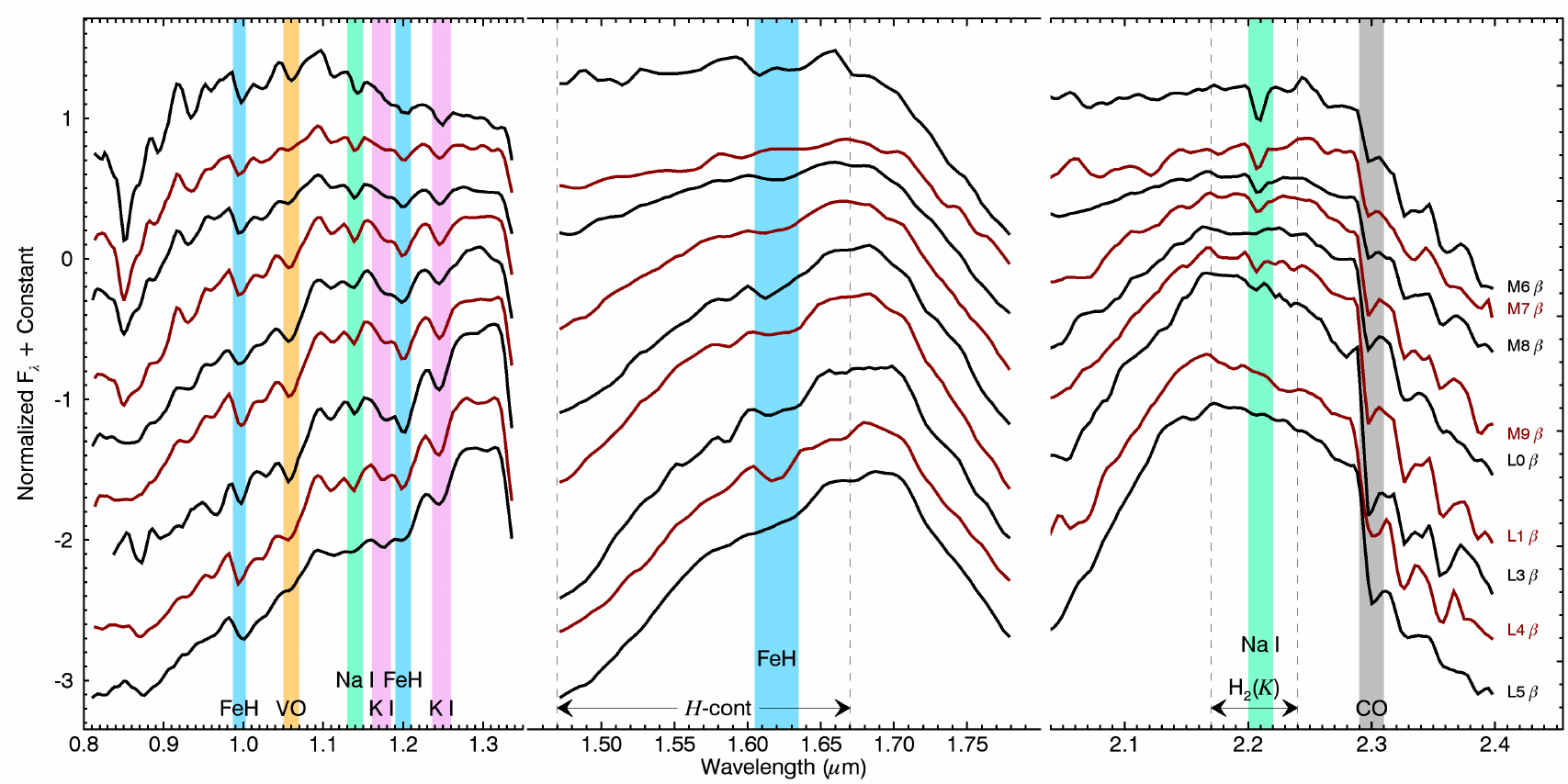}\label{fig:spectra1a}}
	\subfigure[Very Low-Gravity ($\gamma$) Templates]{\includegraphics[width=0.995\textwidth]{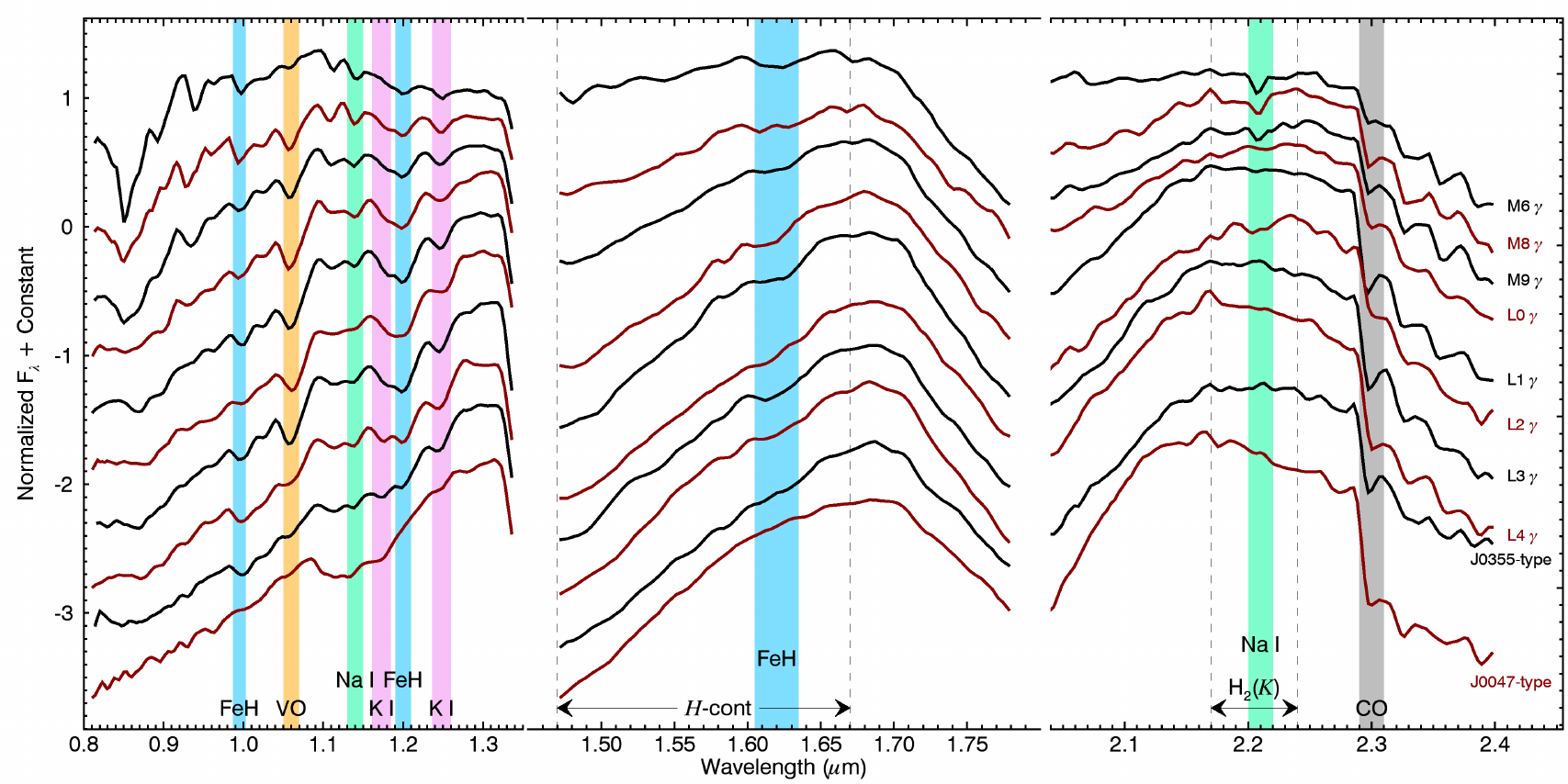}}\label{fig:spectra1b}
	\caption{NIR spectral average of templates of intermediate gravity (Panel~a) and very low gravity (Panel~b). All spectra were normalized to their median across the full wavelength range in each band, resampled at the same resolution ($R \sim 120$) and shifted vertically for comparison purposes. The vertical color bars indicate the location of gravity sensitive features.}
	\label{fig:spectra1}
\end{figure*}

All candidate members that were placed blueward of the field sequence in any of the two CMDs were rejected from \emph{BASS} and \emph{LP-BASS}. Those that were not at least 1$\sigma$ redder than both field sequences were grouped into the \emph{LP-BASS} sample, which is expected to be more contaminated by field objects and young M dwarfs with spectral types earlier than M5. We note that the \emph{PRE-BASS} sample does not necessarily consist of erroneous YMG candidate members; however, it likely suffers from a higher contamination rate from field interlopers or members of moving groups not considered in BANYAN~II.

\section{OBSERVATIONS}\label{sec:obs}

Because of their recent formation, young, low-mass objects have inflated radii compared to their field counterparts and are warmer for a given mass. As a consequence, they have a lower surface gravity at a given temperature (and spectral type). It is well known that these low-gravity dwarfs display weaker alkali and molecular absorption lines (\ion{K}{1} at 7665 \& 7669\,\AA\ in the optical and 1.17 \& 1.25\,$\mu$m in the NIR; \ion{Na}{1} at 8183 \& 8195\,\AA\ in the optical and 1.14 \& 2.21\,$\mu$m in the NIR; \ion{Rb}{1} at 7800 \& 7948\,\AA; \ion{Cs}{1} at 8521 \& 8943\,\AA; FeH at 8692\,\AA\ in the optical and 0.99, 1.20 \& 1.55\,$\mu$m in the NIR; TiO at 8432\,\AA; and CrH at 8611\,\AA). This is due to a lower-pressure in their photosphere, which is a direct consequence of their lower surface gravity.
Collision-induced absorption (CIA) of the $H_2$ molecule is also decreased in this lower pressure environment, causing a flatter $K$-band plateau at 2.18--2.28\,$\mu$m (see the H$_2$($K$) index of \citealt{2013MNRAS.435.2650C}), leaving the effect of water vapor to become apparent from the triangular-shaped continuum of the $H$ band (\citealp{2001MNRAS.326..695L,2006ApJ...639.1120K,2007ApJ...657..511A,2010ApJ...715L.165R}; see the $H$-cont index of \citealt{2013ApJ...772...79A}).
Furthermore, VO condensate clouds get thicker in the external layers of low-pressure atmospheres, causing deeper absorption bands at 7300-7550 and 7850--8000\,\AA\ in the optical and 1.06\,$\mu$m in the NIR (These effects are discussed in more detail by \citealp{2003ApJ...593.1074G, 2004ApJ...600.1020M, 2006ApJ...639.1120K, 2008ApJ...689.1295K, 2009AJ....137.3345C, 2013ApJ...772...79A} and \citealt{2013MNRAS.435.2650C}). Gravity-sensitive features were initially identified by comparing the optical spectra of M-type giants and M-type dwarfs \citep{1986ApJS...62..501K, 1998AJ....116.2520J}, and it was later demonstrated that the same features could be used to identify young, inflated M-type dwarfs by observing members of star-forming regions \citep{1996ApJ...469..706M, 1997ApJ...489L.165L, 2004ApJ...610.1045S, 2001MNRAS.326..695L, 2007ApJ...657..511A, 2008MNRAS.383.1385L}.

A number of low-gravity features (CIA effects of $H_2$ on the continuum, weaker FeH absorption and stronger VO absorption) can be measured in low-mass stars and BDs with spectral types later than M6 using low-resolution ($R \sim 75$) NIR spectroscopy, providing an efficient way of identifying field interlopers in a set of YMG candidates. A higher spectral resolution ($R \sim 1000$) allows for a more robust determination of low gravity features through the measurement of the pseudo-equivalent width (EW) of the atomic lines listed above. We thus obtained low-resolution NIR spectra of 241 candidate YMG members from the \emph{BASS}, \emph{LP-BASS} and \emph{PRE-BASS} samples. We describe in this section all observations and the individual instrumental configurations that were used. A description of individual observations is included in Table~\ref{tab:obslog}.\\%TMP

\subsection{FIRE at Magellan}\label{sec:fire}

We obtained NIR spectroscopy for 17 targets with the Folded-port InfraRed Echellette (FIRE; \citealp{2008SPIE.7014E..0US,2013PASP..125..270S}) at the Magellan Telescopes in April and December 2013, as well as May, June, August and September 2014 and February 2015. We used both the cross-dispersed and high-throughput prism modes to obtain respective resolving powers $R \sim 450$ (prism mode) and $R \sim 6000$ (echelle mode) across the 0.8--2.45\,$\mu$m range. Total exposure times ranged from 200\,s to 1800\,s, depending on source brightness, instrument configuration and weather conditions. This allowed us to obtain a typical S/N $>$\,100 per resolution element. Science targets were observed in an ABBA pattern along the slit, and a standard A0-type star was observed immediately before or after each of them at a similar airmass to ensure a proper telluric correction. We obtained ThAr (prism mode) or NeNeAr (echellette mode) lamp exposures between every science target to perform wavelength calibration, as well as high- and low-illumination flat fields that were combined to obtain a flat-field image with a large S/N across all orders while avoiding saturation. We reduced all data using the Interactive Data Language (IDL) pipeline FIREHOSE, which is based on the MASE \citep{2009PASP..121.1409B} and SpeXTool \citep{2003PASP..115..389V, 2004PASP..116..362C} packages. We supplemented the list of Ar atomic lines with those listed in \cite{1973PhyS....8..249N} to allow a more robust wavelength solution in the $K$ band in the case of prism data.

The six echellette spectra that we obtained here have a sufficient resolution to measure radial velocities down to a precision down to a few \kms. These measurements will be presented in a future publication along with a significant number of additional FIRE echellette spectra.

\subsection{SpeX at IRTF}\label{sec:spex}

We obtained NIR spectroscopy with SpeX \citep{2003PASP..115..362R} at the IRTF telescope for 118 targets from 2007 to 2015. We used the cross-dispersed and prism modes with slits of 0\farcs6, 0\farcs8 and 1\farcs0 depending on the seeing to obtain resolving powers ranging from $R \sim 75$ to $R \sim 750$ over the 0.8--2.45\,$\mu$m range. We used ABBA nodding patterns along the slit with typical exposure times of 60\,s to 250\,s which yielded typical S/N\,$>$\,100 per resolution element. A standard early A-type star was observed immediately before or after every science target at a similar airmass to ensure a proper telluric correction. Several high-S/N quartz lamp and Ar lamp exposures were obtained immediately after every target to ensure a proper wavelength calibration and flat field correction. The data were reduced with the IDL SpeXTool package \citep{2003PASP..115..389V, 2004PASP..116..362C}.

\subsection{Flamingos-2 at Gemini-South}\label{sec:f2}

We used Flamingos-2 \citep{2004SPIE.5492.1196E} at Gemini-South to obtain NIR spectroscopy for 101 targets from 2013 to 2015. We observed each target with both the \emph{JH} and \emph{HK} low resolution grisms and the 0\farcs72 slit to obtain a resolving power of $R \sim 500$ over 0.9--2.4\,$\mu$m. Targets were observed in an ABBA pattern along the slit, with total exposure times ranging from 120\,s to 3400\,s, to obtain S/N\,$>$\,80 per resolution element. Standard A0 to A6-type stars were observed immediately before or after every science target at a similar airmass to ensure a proper telluric correction. Several high-S/N quartz lamp and Ar lamp exposures were obtained immediately after every telluric standard star to ensure a proper wavelength calibration and flat field correction. Dark exposures were obtained at the end of each night, using similar exposure times than all of the science and calibration data to ensure a proper correction of the dark current. A numbers of observations were split between a few nights when observing conditions changed before the required S/N could be obtained.

We used a custom IDL pipeline to apply dark current subtraction and flat field calibrations, correct the trace curvature, optimally extract the spectrum \citep{1986PASP...98..609H} and perform a wavelength calibration using the Ar lamp observations. A dark current subtraction is usually not needed when data are reduced in $A-B$ pairs, like is the case here; however, we found that applying this correction improved the quality of the data. This is likely due to the large exposure times that were used for some targets, which resulted in a large contribution from the dark current that must be corrected both in the data and flat field exposures before applying the flat field correction. A low-pass filter was applied to the flat field exposures to avoid contaminating data with scattered light. We observed that the spectral dispersion (and thus wavelength solution) generally varied from one exposure to another; the wavelength solutions obtained from the Ar calibrations are hence only approximate.

To address this problem, we used several telluric absorption features in the raw spectra of the science and telluric observations to refine individual wavelength solutions. The $JH$ and $HK$ blocking filters also caused significant fringing in the data (up to $\sim$\,7\%). We corrected this by adjusting a sinusoid fringing solution to the low frequencies of the raw spectra. We found that a complete fringing solution (which includes finesse as an additional parameter) did not improve the results; we thus chose the simpler sinusoid approach to have a more robust algorithm.

The extracted science and telluric spectra were combined and telluric-corrected using a modified version of the SpeXtool package adapted for Flamingos-2. We observed that the slope of the continuum in the overlapping region of both observing modes (in the $H$ band) varied in a systematic way at the edge of the detector. Hence, we removed these regions before combining the spectra.
A few objects for which we obtained Flamingos-2 data (e.g. 2MASS~J07083261--4701475, 2MASS~20414283--3506442 and 2MASS~J12042529--2806364) turned out to be field dwarfs that closely match literature SpeX-prism spectra of other known objects of the same spectral type: this is an indication that the systematics mentioned above were accurately corrected.

\subsection{GNIRS at Gemini-North}\label{sec:gnirs}

We used GNIRS at Gemini-North to obtain NIR spectroscopy for three targets in 2013. We used the 32\,l\,mm$^{-1}$ grating centered at 1.65\,$\mu$m in the cross-dispersed mode with the 0\farcs675 slit to achieve a resolving power of $R \sim 750$ over 0.9--2.45\,$\mu$m. We nodded exposures along the slit in ABBA patterns with total exposure times ranging from 120\,s to 360\,s to reach S/N\,$>$\,100 per resolution element. A0-type telluric standard stars were observed immediately before or after science targets at a similar airmass to ensure a proper telluric correction. Several high-S/N quartz lamp and Ar lamp exposures were obtained immediately after every target to ensure a proper wavelength calibration and flat field correction. The data were reduced with the XDGNIRS IRAF package provided by Gemini.

\subsection{TripleSpec at Hale}\label{sec:triplespec}

We used TripleSpec \citep{2008SPIE.7014E..0XH} at the Palomar Observatory 5\,m Hale Telescope to obtain NIR spectroscopy for one target in the cross-dispersed mode with the 1\farcs0 slit, yielding a resolving power $R \sim 3\,800$ over 1.0--2.45\,$\mu$m. We observed the science target in 4-position ABBA nodding pattern along the slit with a total exposure time of 1200\,s to reach a S/N\,$>$\,100 per resolution element. High-S/N quartz lamp and NeAr lamp exposures were obtained to ensure a proper wavelength calibration and flat field correction. We reduced the data using an adapted version of SpeXtool (see Section~\ref{sec:spex}).

\section{SPECTRAL TYPES AND LOW-GRAVITY CLASSIFICATION}\label{sec:sptclass}

We describe in this section the method that we used to assign spectral types to our new observations. Our typing scheme consists of two distinct dimensions~: the first dimension consists of the usual spectral subtypes and is mostly sensitive to \Teff. The second dimension, introduced by \cite{2005ARA&A..43..195K} and \cite{2006ApJ...639.1120K}, aims at characterizing the surface gravity with the use of a greek-letter suffix. Field-gravity dwarfs are designated with the $\alpha$ suffix or no suffix, intermediate-gravity dwarfs with the $\beta$ suffix, and very low-gravity dwarfs with the $\gamma$ suffix. The $\delta$ suffix was also introduced by \cite{2006ApJ...639.1120K} to designate objects with an even younger age  (typically less than a few Myrs) and lower surface gravity than those associated to the $\gamma$ suffix.

Optical spectral standards were used to classify NIR spectra of field K7--M9 spectral types. We used the NIR data of GJ~820~B (K7), Gl~229~A (M1), Gl~411 (M2), Gl~213 (M4), Gl~51 (M5), Gl~406 (M6), GJ~644~C (M7), GJ~752~B (M8) and LHS~2924 (M9) as field-gravity spectral standards for these respective spectral types. These standards were identified from the list maintained by Eric Mamajek\footnote{\url{http://www.pas.rochester.edu/\textasciitilde emamajek/spt/}} \citep{1976PhDT........14B,1991ApJS...77..417K,2013ApJS..208....9P} and their spectra were downloaded from the IRTF spectral library\footnote{Maintained by Michael C.~Cushing and available at \url{http://irtfweb.ifa.hawaii.edu/\textasciitilde spex/IRTF\_Spectral\_Library/}.}. We did not use any of the suggested K8, K9, M0 and M3-type standards, since none of them were available in the IRTF spectral library.

While NIR L dwarfs spectral standards have been identified by \cite{2010ApJS..190..100K}, we have opted to use optically-anchored NIR spectral average templates for classifying field L0--L9 dwarfs. Templates are constructed by median-combining all spectra of a given optical spectral type and gravity class. These templates were provided by K.~Cruz and their creation will be discussed in detail and be made public as part of a forthcoming paper (Cruz et al. in preparation). The spectral morphology of these templates is consistent with the \cite{2010ApJS..190..100K} spectral standards but since they are an average of many objects, they also reflect the diversity of spectral morphologies present in each spectral type.
%Spectral standards have been determined for low-gravity M and L dwarfs by \cite{2013ApJ...772...79A}, but we have also adopted to use optically-anchored NIR in this  and thus w

Spectral standards have been determined for low-gravity M and L dwarfs by \cite{2013ApJ...772...79A}, but we opted to use spectral average templates in this case too, for the reasons mentioned above. We generated M6--M9\,$\gamma$ templates with data published in \cite{2013ApJ...772...79A} and sent to us directly by the authors. These templates are available at the Montreal Spectral Library\footnote{\url{www.astro.umontreal.ca/\textasciitilde gagne/MSL.php}}. The optically-anchored L0\,$\beta$, L1\,$\beta$ and L0--L4\,$\gamma$ templates were provided by K.~Cruz. They will be discussed in detail and be made public as part of a forthcoming paper et al. (Cruz et al. in preparation). 

All template, standard, and target spectra were re-sampled to the same resolution and wavelength grid as SpeX prism observations with the 0\farcs6 slit ($R \sim 120$). Following the method of \cite{DisentanglingLDwar:db}, the spectra were normalized in three sections in order to minimize the effect of large NIR color variations within a given spectral type. The spectra were broken into three sections: 0.80--1.35\,$\mu$m, 1.40--1.80\,$\mu$m and 1.95--2.40\,$\mu$m, roughly corresponding to the $zJ$, $H$, and $K$ bands. 

% Table : Spectral Templates
\begin{deluxetable}{ll}
\tablecolumns{2}
\tablecaption{An extended sequence of low-gravity dwarfs.\label{tab:sptextlit}}
\tablehead{\colhead{Name} & \colhead{Spectral Type}}
\startdata
USco~J160603.75--221930.0 & L0\,$\delta$ \\
USco~J160727.82--223904.0 & L0\,$\delta$ \\
USco~J160737.99--224247.0 & L0\,$\delta$ \\
USco~J160818.43--223225.0 & L0\,$\delta$ \\
USco~J160828.47--231510.4 & L0\,$\delta$ \\
USco~J160843.44--224516.0 & L0:\,$\delta$ \\
USco~J160918.69--222923.7 & L0\,$\delta$ \\
USco~J161228.95--215936.1 & L0\,$\delta$ \\
USco~J161441.68--235105.9 & L0\,$\delta$ \\
USco~J163919.15--253409.9 & L0\,$\delta$ \\
CD--35~2722~B & L3\,$\beta$\\
2MASS~J01531463--6744181 & L3\,$\beta$\\
2MASS~J17260007+1538190 & L3\,$\beta$\\
2MASS~J00011217+1535355 & L4\,$\beta$\\
2MASS~J05120636--2949540 & L5\,$\beta$\\
2MASS~J23174712--4838501 & L5\,$\beta$\\
2MASS~J03264225-2102057 & L5\,$\beta$/$\gamma$\\
SIMP~J21543454--1055308 & L5\,$\beta$/$\gamma$\\
2MASS~J03552337+1133437 & \emph{J0355-type} (L3--L6\,$\gamma$)\\
2MASS~J16154255+4953211 & \emph{J0355-type} (L3--L6\,$\gamma$)\\
2MASS~J23433470--3646021 & \emph{J0355-type} (L3--L6\,$\gamma$)\\
WISEP~J004701.06+680352.1 & \emph{J2244-type} (L6--L8\,$\gamma$)\\
2MASS~J22443167+2043433 & \emph{J2244-type} (L6--L8\,$\gamma$)\\
PSO~J318.5338--22.8603 & \emph{J2244-type} (L6--L8\,$\gamma$)\\[-7pt]
\enddata
\tablecomments{All spectral types are from this work and are based on NIR spectra. A : symbol indicates that the spectral type is based on low signal-to-noise data and is uncertain ($\pm$ 1), and a :: symbol that it is very uncertain ($\pm$ 2 subtypes); pec indicates peculiar features; $\beta$ and $\gamma$ respectively indicate intermediate gravity and very low gravity.}
\end{deluxetable}

In a first step to estimating a spectral type, we categorized our 245 new spectra with the spectral template and standard grid described above. There were 11 objects, however, that did not have a good visual match to any standard or template in the grid; this number excludes the early-type contaminants which are discussed later in this work. We collected additional low-gravity brown dwarf spectra in the literature to identify 19 more objects that do not match our standards.

We performed a visual analysis of all of the unclassifiable spectra and identified enough objects with similar spectral morphologies to create tentative new spectral types and templates for L0\,$\delta$, L3\,$\beta$, L4\,$\beta$ and L5\,$\beta$. The objects that were used in the creation of these templates are listed in Table~\ref{tab:sptextlit}. We list the revised spectral types that we obtain for other spectra from the literature in Table~\ref{tab:retypes}. We note that our L3\,$\beta$ template includes 2MASS~J17260007+1538190, which was suggested by \cite{2013ApJ...772...79A} as a tentative template for the L3\,$\beta$ spectral type.

We could not build a template for the L2\,$\beta$ spectral type, as the only objects that were confirmed as L2\,$\beta$ from optical data have either very low signal-to-noise (S/N) ratios in the NIR or no NIR data. As we gather more high-S/N spectra of low-gravity L dwarfs, we expect to fill this gap.

The L0\,$\delta$ template was built from eight candidate members of Upper Scorpius \citep{2008MNRAS.383.1385L} and one candidate member of $\beta$PMG (2MASS~00464841+0715177) that are similar to the L0\,$\gamma$ template except that their $H$ band is even more triangular and their $K$ band has a redder continuum. It is also notable that the H$_2$O-dependent slope of the L0\,$\delta$ at 1.7--1.8\,$\mu$m is slightly steeper than what is seen in any other L-type template.

\begin{figure}
	\centering
	\includegraphics[width=0.45\textwidth]{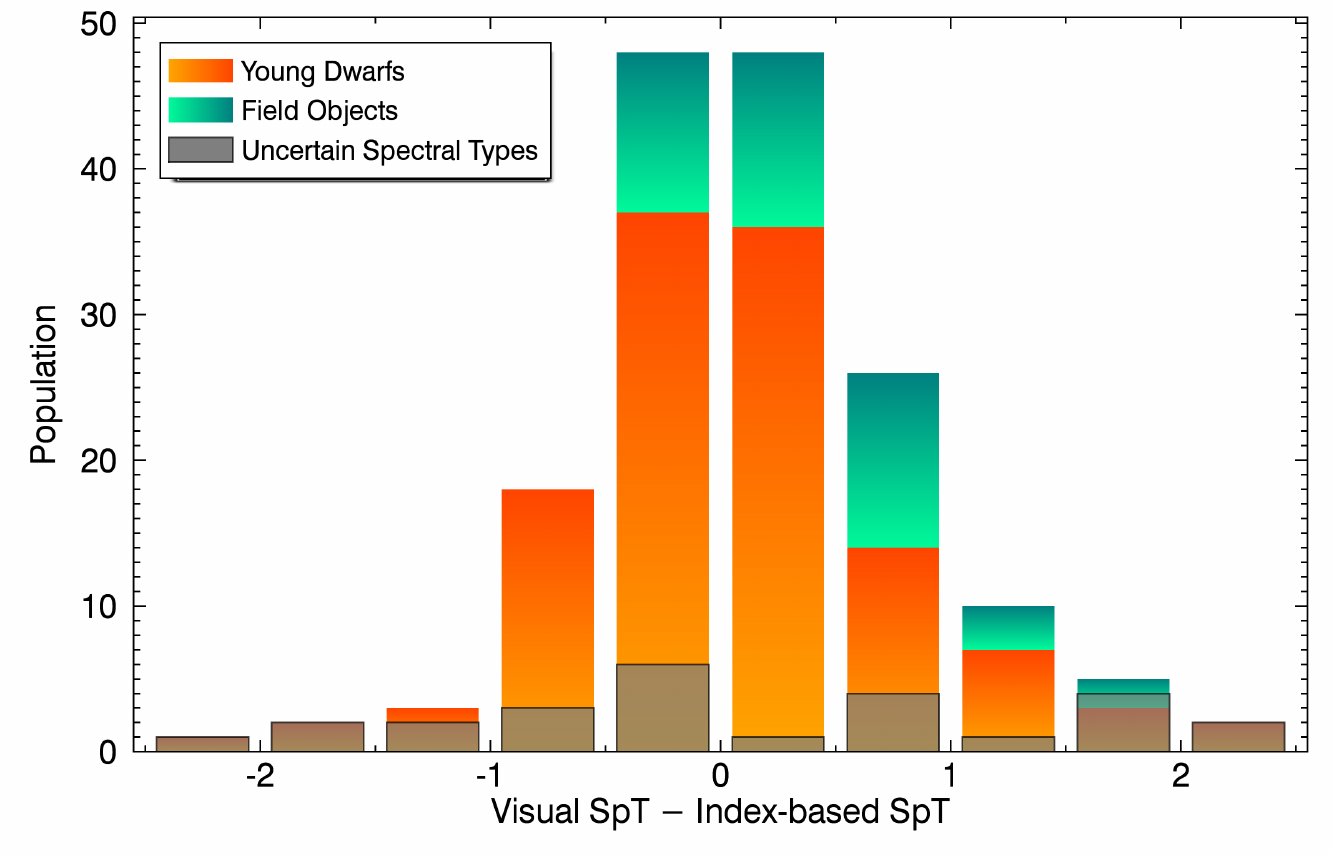}
	\caption{Distribution of the differences between our visual and index-based spectral classifications, for young dwarfs and field objects. Spectral types generally agree within one subtype, with a standard deviation of 0.7 subtypes and a reduced $\chi^2$ value of 0.8. This is indicative that our measurement errors are representative of the observed differences between the two methods. It can also be seen that most of the outliers in the distribution correspond to objects with uncertain spectral types (i.e., measurement errors of one subtype or more.)}
	\label{fig:hist_diff}
\end{figure}
% Table : Spectral Re-typings in the literature
\begin{deluxetable}{lllllll}
\tablecolumns{7}
\tablecaption{Revised NIR spectral types from the literature.\label{tab:retypes}}
\tablehead{\colhead{Name} & \colhead{} & \multicolumn{5}{c}{Spectral Type\tablenotemark{a}}\\
\cline{3-7}
\colhead{} & \colhead{} & \colhead{Optical} & \colhead{Ref.} & \colhead{NIR} & \colhead{Ref.} & \colhead{Adopted}}
\startdata
\cutinhead{Low-gravity dwarfs}
2MASS~J21324036+1029494 & & $\cdots$ & & L4.5: & 1 & L4:\,$\beta$/$\gamma$\\
2MASS~J14482563+1031590 & & L4: & 2 & L3.5 & 3 & L5:\,$\beta$\\
WISE~J174102.78--464225.5 & & $\cdots$ & & L7:: & 4 & L5:--L7:\,$\gamma$\\
G~196--3~B & & L3\,$\beta$ & 5 & L3\,$\gamma$ & 6 & L2--L4\,$\gamma$\\
2MASS~J00303013--1450333 & & L7 & 7 & L4.5:: & 8 & L4--L6\,$\beta$\\
2MASS~J20025073--0521524 & & L6 & 9 & L7:: & 10 & L5--L7\,$\gamma$\\[-3pt]
\cutinhead{Red brown dwarfs with no clear signs of low gravity}
2MASS~J08354256-0819237 & & L5 & 11 & L5 & 12 & L4\,pec\tablenotemark{b}\\
2MASS~J18212815+1414010 & & L4.5\,pec & 13 & L5\,pec & 14 & L4\,pec\\
2MASS~J21512543--2441000 & & L3 & 9 & $\cdots$ & & L4\,pec\\
2MASS~J01033203+1935361 & & L6\,$\beta$ & 7,15 & L6\,$\beta$ & 6 & L6\,pec\\
2MASS~J01075242+0041563 & & L8 & 16 & L8\,pec & 17 & L7\,pec\tablenotemark{b}\\
2MASS~J08251968+2115521 & & L7.5 & 7 & L6 & 18 & L7\,pec\tablenotemark{b}\\
2MASS~J08575849+5708514 & & L8 & 5 & L8\,$\pm$\,1 & 19 & L8--L9\,pec\\[-7pt]
\enddata
\tablenotetext{a}{All revised spectral types are from this work and are based on NIR spectra.}
\tablenotetext{b}{Candidate member of the $\sim$\,625\,Myr-old Hyades association \citep{2007MNRAS.378L..24B}.}
\tablecomments{References to this Table~: \\ (1)~\citealt{2006AJ....131.2722C}; (2)~\citealt{2008AJ....136.1290R}; (3)~\citealt{2003IAUS..211..197W}; (4)~\citealt{2014AJ....147...34S}; (5)~\citealt{2008ApJ...689.1295K}; (6)~\citealt{2013ApJ...772...79A}; (7)~\citealt{2000AJ....120..447K}; (8)~\citealt{2010ApJ...710.1142B}; (9)~\citealt{2007AJ....133..439C}; (10)~\citealt{2014ApJ...794..143B}; (11)~\citealt{2003AJ....126.2421C}; (12)~\citealt{2013AJ....146..161M}; (13)~\citealt{2008ApJ...686..528L}; (14)~\citealt{2010ApJS..190..100K}; (15)~\citealt{2012ApJ...752...56F}; (16)~\citealt{2002AJ....123.3409H}; (17)~\citealt{2011ASPC..448.1351G}; (18)~\citealt{2004AJ....127.3553K}; (19)~\citealt{2002ApJ...564..466G}.}
\end{deluxetable}

There are two sets of objects with similar spectra, each with three targets, that we identified via our visual analysis; however, we are unable to confidently assign them a spectral type that fits into our grid of templates. 
For the purposes of this paper, we label these objects as \emph{J0355-type} and \emph{J2244-type}. One set is composed of 2MASS~J03552337+1133437, 2MASS~J16154255+4953211 and 2MASS~J23433470--3646021. Their spectra are similar to the L4\,$\gamma$ template except that they have a shallower CO band at 2.3\,$\mu$m. The other set is composed of 2MASS~J00470038+6803543, PSO~J318.5338--22.8603 and 2MASS~J22443167+2043433. Their spectra display a significantly redder continuum than our templates, which might be indicative of a later spectral type. We note that two objects have previous classifications based on the index-based scheme of \cite{2013ApJ...772...79A}: 2MASS~J00470038+6803543 was classified as an intermediate-gravity L7 dwarf by \cite{2015ApJ...799..203G} and PSO~J318.5338--22.8603 was classified as a very low-gravity L7 dwarf by \cite{2013ApJ...777L..20L}. We listed these two sets of objects as well in Table~\ref{tab:sptextlit}.

We adopt a conservative estimate of L3--L6\,$\gamma$ for the spectral type range of the \emph{J0355-type}. The spectral features of the \emph{J2244-type} are indicative of a spectral type in the range L6--L8\,$\gamma$ range. For both of these new spectral types, we refrain from assigning them a more precise location in the spectral sequence until more data are available at these late low-gravity types. It is unclear at this stage whether \emph{J0355-type} and \emph{J2244-type} objects are peculiar or a simple extension of low-gravity brown dwarfs at spectral types later than L5. A larger number of late-type, low-gravity L dwarfs will need to be identified before we can assess this. Our set of low-gravity templates is displayed in Figure~\ref{fig:spectra1}.

We used the index-based classification method of \cite{2013ApJ...772...79A} to corroborate our visual classification. This method consists of measuring the slope of H$_2$O continuum features to assign a spectral type, and a combination of several gravity-sensitive spectroscopic indices to assign a gravity class. We found that spectral types obtained from the template grid system described above generally agree with index-based spectral types within one subtype (Figure~\ref{fig:hist_diff}). The standard deviation between the two methods for the 163 non-peculiar objects that we categorized is of 0.7 subtypes, with a reduced $\chi^2$ value of 0.8. A reduced $\chi^2 \approx 1$ indicates that measurement errors are representative of the discrepancies. The reduced $\chi^2$ is given by $1/(N-1)\cdot\sum y/\sigma_y$, where $N$ is the number of objects, $y$ is the spectral type discrepancy and $\sigma_y$ is the quadrature sum of the index-based and visual-based spectral type measurement errors. All cases discrepant by more than 1.5 spectral types correspond to low-S/N data, except for 2MASS~J21420580--3101162 that gets L1.5$ \pm 0.3$ from the index-based method and L3 from the visual-based method. This object does not display signs of youth or significantly peculiar features, but it has a slightly redder slope at 1.7--1.8\,$\mu$m. It unclear what is the cause of this discrepancy.

We used optical data to assign an adopted spectral type using a template-based visual classification method \citep{2009AJ....137.3345C} only for the 4 objects for which no NIR data were available. In all other cases, our adopted spectral types are based on NIR data only. Our NIR spectral types based on a visual comparison with templates show a standard deviation of 0.9 subtype with respect to optical spectral types in the literature, and the reduced $\chi^2$ of the differences is 1.5, hence slightly larger than what would be expected given the uncertainties. If we compare optical spectral types to the index-based spectral types of \cite{2013ApJ...772...79A}, we obtain a slightly larger standard deviation (1.1 subtype) and reduced $\chi^2$ (2.4). This is indicative that our visual-based classification method is more consistent with spectral types based on optical data that were reported in the literature. This should be expected, as our templates are anchored on optical data. In both cases, we observe no systematic bias (the mean of the differences is smaller than 0.1 subtype). Several objects that deserve further discussion are presented in detail in the Appendix.

We note that the index-based field-gravity, intermediate-gravity and very low-gravity classes defined by \cite{2013ApJ...772...79A} were built to correspond to the optical $\alpha$, $\beta$ and $\gamma$ classes, which is what we observe in 143/176 (81\%) of the cases. Some of the discrepancies arise for objects near the spectral type thresholds where the method of \cite{2013ApJ...772...79A} stops being applicable ($\lesssim$\,M6 or $\gtrsim$\,L6) or for data with a lower S/N. The $\delta$ gravity class does not have an equivalent in the index-based classification of \cite{2013ApJ...772...79A}, but we note that all three of the young dwarfs that we categorized as $\delta$ are assigned with the maximal index-based gravity score (2222). It does not seem that this maximal index-based gravity score always translates as a $\delta$ visual classification though, as there are four additional objects in our sample that obtained the score 2222 but that we visually categorized as $\gamma$ (2MASS~J00182834--6703130; L0\,$\gamma$, 2MASS~J01205114--5200349; L1\,$\gamma$; 2MASS~J20113196--5048112; L3\,$\gamma$; 2MASS~J22351658--3844154; L1.5\,$\gamma$; all are THA candidate members). For consistency within this work, we have adopted the visual spectral types in the remaining sections, but we list all visual and index-based spectral types in Table~\ref{tab:sptclass}. We note that this choice does not affect the conclusions presented in this work.

\section{RESULTS}\label{sec:results}

\begin{figure*}[p]
	\centering
	\subfigure[Intermediate-gravity ($\beta$) dwarfs]{\includegraphics[width=0.9\textwidth]{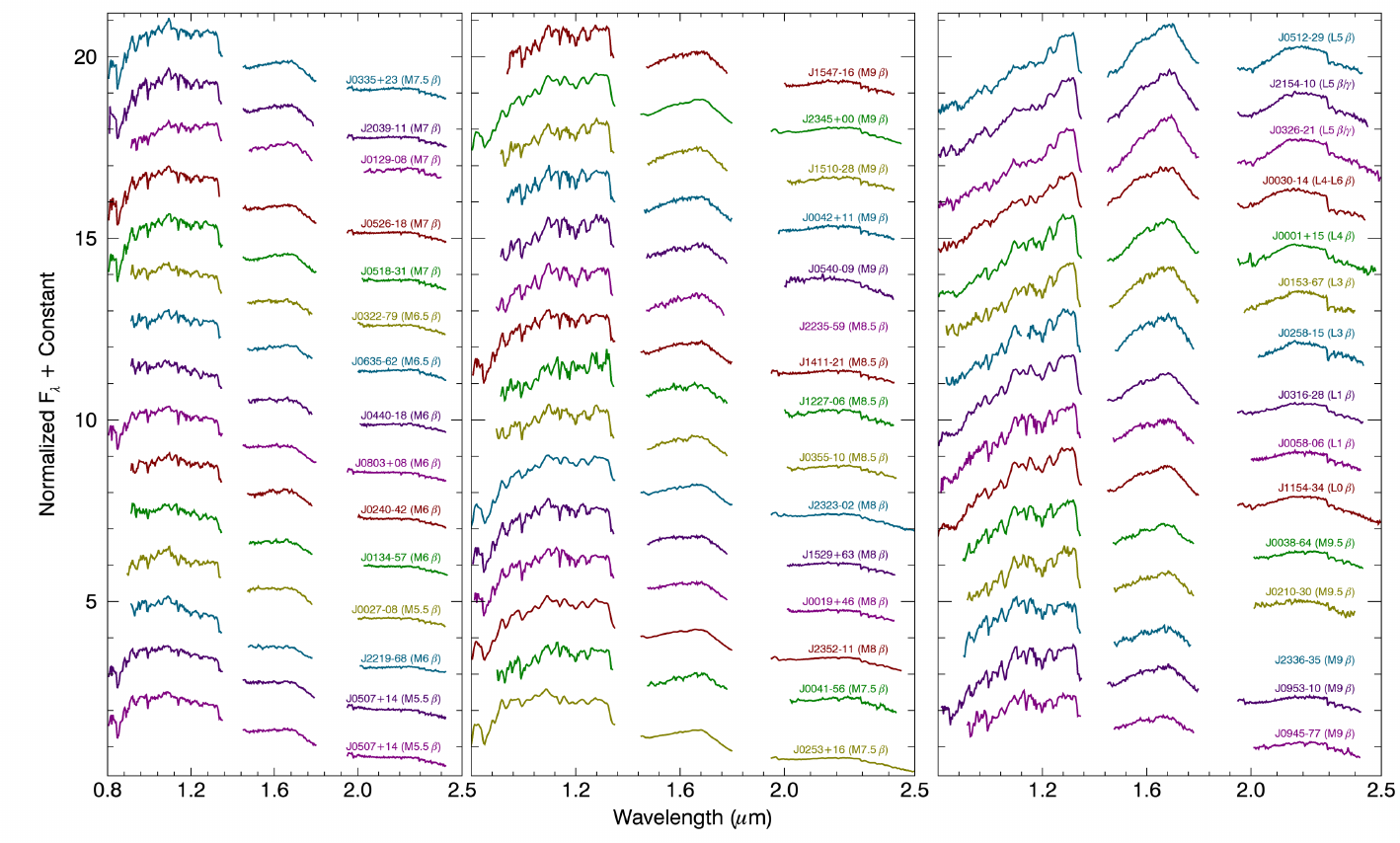}}
	\subfigure[Very low-gravity ($\gamma$) dwarfs]{\includegraphics[width=0.9\textwidth]{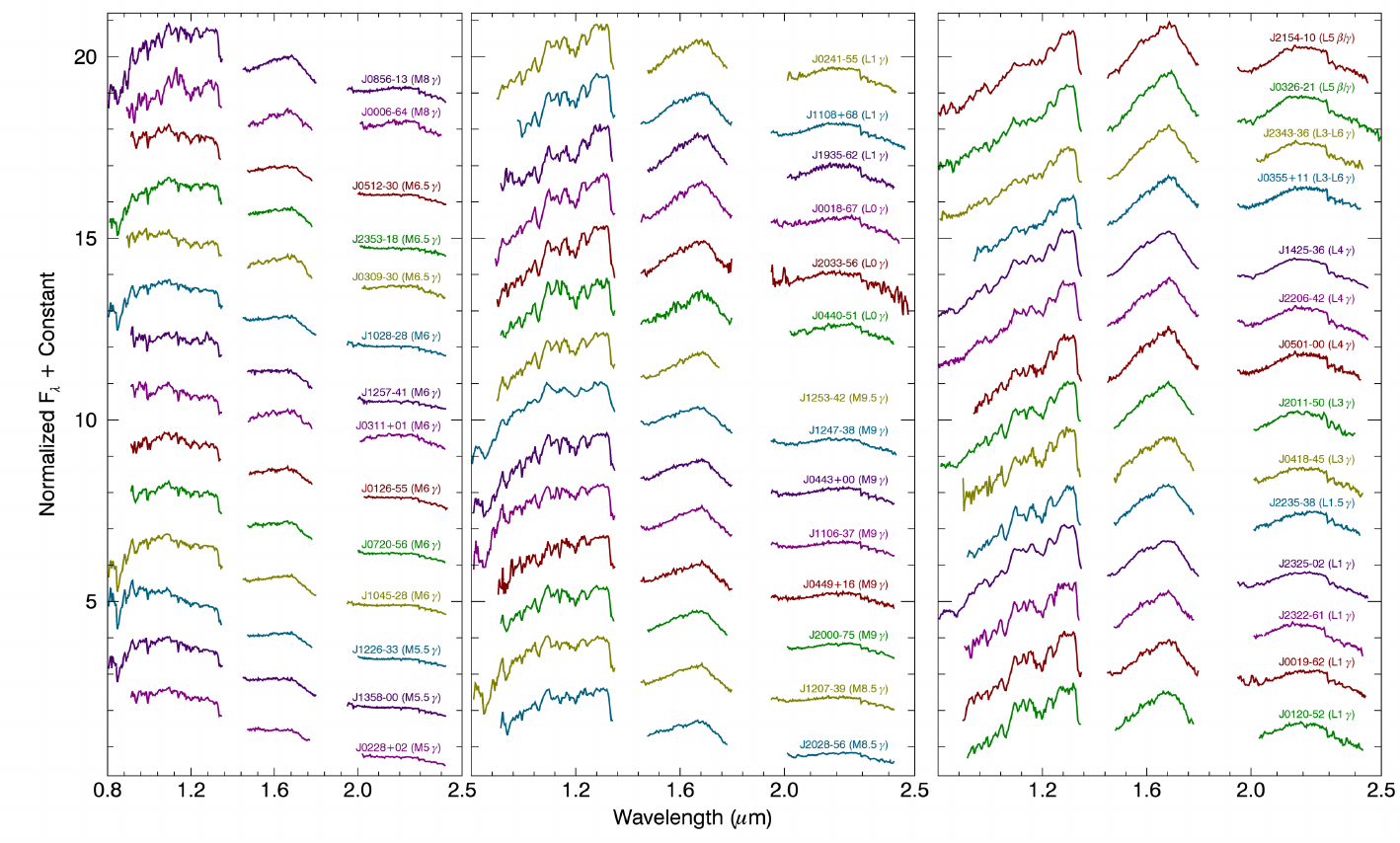}}
	\caption{NIR spectra of all new observations and objects for which spectral types were revised in this work. All spectra were re-sampled to a spectral resolution of $R \sim 120$ and a dispersion relation identical to SpeX observations in the prism mode with the 0\farcs6 slit. We used alternating colors for visibility.}
	\label{fig:allspt}
\end{figure*}

In Figure~\ref{fig:allspt}, we present the NIR spectra of several new intermediate ($\beta$) and very-low ($\gamma$) gravity dwarfs discovered in this work, as well as known dwarfs for which we have obtained new data. Several objects that were uncovered as candidate members of YMGs in \emph{BASS} had NIR or optical spectroscopy readily available in the SpeX Prism Spectral Libraries \footnote{\url{http://pono.ucsd.edu/\textasciitilde adam/browndwarfs/spexprism}} \citep{2014ASInC..11....7B} or the RIZzo Ultracool Spectral Library\footnote{\cite{2000AJ....120..447K,2003AJ....126.2421C,2007AJ....133..439C,2008ApJ...689.1295K,2008AJ....136.1290R,2009AJ....137.3345C,2010ApJS..190..100K}; see \url{http://www.astro.umontreal.ca/\textasciitilde gagne/rizzo}} with no discussion of low gravity in the literature; we included them in our present analysis and the resulting spectral classification is listed in Table~\ref{tab:sptclass} along with the new discoveries.

There are some cases where the BANYAN~II tool yields ambiguous candidate membership to more than one association (i.e., at least a second moving group shares 10\% of the total YMG Bayesian probabilities). In all such cases, we list in Table~\ref{tab:sptclass} all plausible YMGs with their relative share of the total YMG probability (i.e., excluding the field probability). An extensive RV and parallax follow-up will be required before more can be said on their YMG membership.

We have identified seven objects (Table~\ref{tab:sptclass}) that display signs of low gravity, but for which additional information was inconsistent with membership to any of the YMGs presented here (e.g., RV and distance measurements or the effect of interstellar extinction affecting the NIR spectrum which is not consistent with ages older than $\sim$\,5\,Myr). It is possible that these objects belong to YMGs or star-forming regions that are not considered here, that their RV or parallax measurements are affected by an unresolved binary companion (see the Appendix for a detailed discussion), or that other physical properties such as enhanced dust mimics a lower gravity. 

%Figure : All spectra

\clearpage
\onecolumngrid
\begin{landscape}
%Table : Spectral Classification
% Table : Spectral Classification
\LongTables
\tabletypesize{\footnotesize}
\renewcommand{\tabcolsep}{3pt}
\begin{deluxetable}{llllllllllcclllllllr}
\tablecolumns{20}
\tablecaption{Spectral Classification, Low Gravity and YMG membership.\label{tab:sptclass}}
\tablehead{\multicolumn{2}{c}{\emph{2MASS}} & \colhead{} & \multicolumn{6}{c}{Spectral Type\tablenotemark{a}} & \colhead{} & \multicolumn{2}{c}{Ind. Gravity\tablenotemark{b}} & \colhead{} & \multicolumn{5}{c}{YMG Membership} & \colhead{} & \colhead{Source}\\
\cline{1-2}
\cline{4-9}
\cline{11-12}
\cline{14-18}
\colhead{Designation} & \colhead{$J$} & \colhead{} & \colhead{Lit. Opt.} & \colhead{Ref.} & \colhead{Lit. NIR} & \colhead{Ref.} & \colhead{H$_2$O Ind.} & \colhead{Adopted} & \colhead{} & \colhead{Class} & \colhead{Score} & \colhead{} & \colhead{Lit.} & \colhead{Ref.} & \colhead{Updated \emph{BASS}\tablenotemark{c}} & \colhead{$B_P$ (\%)\tablenotemark{d}} & \colhead{$C_P$ (\%)\tablenotemark{e}} & \colhead{} & \colhead{Catalog}}
\startdata

\multicolumn{20}{c}{Bona Fide Members}\\[2.5pt]\hline\\[-6pt]
03350208+2342356 & $12.25$ & & M8.5 & 29 & M7\,VL-G & 30 & M7.3 & M7.5\,$\beta$ & & INT-G & 1n21 & & BPMG & 31 & BPMG & $84.2$ & $3.9$ & & \emph{PRE-BASS}\\
03552337+1133437 & $14.05$ & & L5\,$\gamma$ & 32 & L3\,VL-G & 30 & L2.3 & L3--L6\,$\gamma$ & & VL-G & 2122 & & ABDMG & 33 & ABDMG & $99.5$ & $1.1$ & & \emph{BASS}\\
14252798--3650229 & $13.75$ & & L3: & 15 & L5 & 52 & L3.1 & L4\,$\gamma$ & & INT-G & 11?1 & & $\cdots$ & & ABDMG & $99.9$ & $0.1$ & & \emph{BASS}\\[-5pt]
\cutinhead{Low-Gravity Candidate Members}
00011217+1535355 & $15.52$ & & $\cdots$ & & L4: & 1 & L3.7 & L4\,$\beta$ & & INT-G & 1211 & & $\cdots$ & & ABDMG & $97.4$ & $1.1$ & & \emph{BASS}\\
00065794--6436542 & $13.39$ & & L0 & 4 & $\cdots$ & & M7.4 & M8\,$\gamma$ & & INT-G & 1n12 & & THA & 5 & THA & $> 99.9$ & $< 0.1$ & & \emph{BASS}\\
00182834--6703130 & $15.46$ & & $\cdots$ & & $\cdots$ & & L0.4 & L0\,$\gamma$ & & VL-G & 2222 & & $\cdots$ & & THA & $99.9$ & $< 0.1$ & & \emph{BASS}\\
\enddata
\end{deluxetable}
\renewcommand{\tabcolsep}{6pt}

%Table : Spectral Indices.
% Table : Spectral Indices
\LongTables
\tabletypesize{\footnotesize}
\renewcommand{\tabcolsep}{6pt}
\begin{deluxetable}{llcccccccccccc}
\tablecolumns{14}
\tablecaption{Equivalent Widths and Spectral Indices.\label{tab:sptindices}}
\tablehead{\colhead{\emph{2MASS}} & \colhead{Spectral} & \colhead{} & \multicolumn{5}{c}{Equivalent Widths (\AA)} & \colhead{} & 
\multicolumn{5}{c}{Spectral Indices \citep{2013ApJ...772...79A}}\\
\cline{4-8}
\cline{10-14}
\colhead{Designation} & \colhead{Type} & \colhead{} & \colhead{\ion{Na}{1}\,1.138\,$\mu$m} & \colhead{\ion{K}{1}\,1.169\,$\mu$m}
 & \colhead{\ion{K}{1}\,1.177\,$\mu$m} & \colhead{\ion{K}{1}\,1.244\,$\mu$m} & \colhead{\ion{K}{1}\,1.253\,$\mu$m} & \colhead{} & 
\colhead{FeH$_Z$} & \colhead{VO$_Z$} & \colhead{FeH$_J$} & \colhead{KI$_J$} & \colhead{$H$--cont}}
\startdata

\multicolumn{14}{c}{Intermediate-Gravity ($\beta$)}\\[2.5pt]\hline\\[-6pt]
00011217+1535355 & L4\,$\beta$ & & $7.3 \pm 1.0$ & $5.7 \pm 1.0$ & $7.0 \pm 1.3$ & $\cdots$ & $2.9 \pm 1.4$ & & $1.22 \pm 0.01$ & $1.17 \pm 0.01$ & $1.24 \pm 0.01$ & $1.13 \pm 0.01$ & $0.92 \pm 0.01$\\
00274534--0806046 & M5.5\,$\beta$ & & $4.1 \pm 0.7$ & $1.3 \pm 0.7$ & $2.7 \pm 0.9$ & $0.6 \pm 0.6$ & $1.5 \pm 0.6$ & & $1.06 \pm 0.01$ & $1.01 \pm 0.01$ & $1.04 \pm 0.01$ & $1.03 \pm 0.01$ & $0.97 \pm 0.01$\\
00303013--1450333 & L4--L6\,$\beta$ & & $\cdots$ & $\cdots$ & $\cdots$ & $\cdots$ & $\cdots$ & & $1.10 \pm 0.03$ & $1.03 \pm 0.02$ & $\cdots$ & $1.06 \pm 0.02$ & $0.92 \pm 0.01$\\
00344300--4102266 & L1:\,$\beta$ & & $7.5 \pm 1.5$ & $6.0 \pm 1.6$ & $5.5 \pm 1.6$ & $2.9 \pm 2.1$ & $3.6 \pm 1.6$ & & $1.18 \pm 0.01$ & $1.14 \pm 0.01$ & $1.15 \pm 0.02$ & $1.09 \pm 0.01$ & $0.96 \pm 0.01$\\
00381489--6403529 & M9.5\,$\beta$ & & $8.3 \pm 1.4$ & $4.1 \pm 1.0$ & $5.2 \pm 1.3$ & $3.6 \pm 1.9$ & $2.8 \pm 1.2$ & & $1.12 \pm 0.01$ & $1.19 \pm 0.01$ & $1.18 \pm 0.01$ & $1.07 \pm 0.01$ & $1.01 \pm 0.01$\\
00425923+1142104 & M9\,$\beta$ & & $10.2 \pm 1.4$ & $4.1 \pm 1.3$ & $6.5 \pm 1.7$ & $3.0 \pm 1.0$ & $3.4 \pm 1.3$ & & $1.17 \pm 0.01$ & $1.10 \pm 0.01$ & $1.16 \pm 0.02$ & $1.08 \pm 0.01$ & $0.95 \pm 0.01$\\
\enddata
\end{deluxetable}
\renewcommand{\tabcolsep}{6pt}

%Table : Optical Spectral Indices
% Table : Optical Indices
\LongTables
\tabletypesize{\scriptsize}
\renewcommand{\tabcolsep}{1pt}
\begin{deluxetable}{llccccccccccccccl}
\tablecolumns{17}
\tablecaption{Revised Optical Spectral Types and Spectral Indices.\label{tab:sptoptindices}}
\tablehead{\colhead{\emph{2MASS}} & \colhead{Optical} & \colhead{} &
\multicolumn{12}{c}{Spectral Indices} & \colhead{} & \colhead{Low}\\
\cline{4-15}
\colhead{Designation} & \colhead{type} & \colhead{} & \colhead{K-a} & \colhead{K-b} & \colhead{Na-a} & \colhead{Na-b} & \colhead{VO-a} & \colhead{Cs-a} & \colhead{Cs-b} & \colhead{TiO-b} & \colhead{Rb-a} & \colhead{Rb-b} & \colhead{FeH-a} & \colhead{CrH-a} & \colhead{} & \colhead{Gravity\tablenotemark{a}}}
\startdata

03164512--2848521 & L1 $\beta$ & & $3.27 \pm 0.36 $\tablenotemark{b} & $2.90 \pm 0.38$ & $0.975 \pm 0.053 $\tablenotemark{b} & $0.988 \pm 0.059 $\tablenotemark{b} & $1.244 \pm 0.095 $\tablenotemark{b} & $1.125 \pm 0.036$ & $1.071 \pm 0.050$ & $1.374 \pm 0.061 $\tablenotemark{b} & $1.214 \pm 0.075$\tablenotemark{c} & $1.1089 \pm 0.080$\tablenotemark{b} & $1.254 \pm 0.043$\tablenotemark{b} & $1.196 \pm 0.045$\tablenotemark{b} & & 8 (1)\\
03264225--2102057 & L5 $\beta$/$\gamma$ & & $4.23 \pm 0.94 $\tablenotemark{b} & $8.80 \pm 3.04$\tablenotemark{b} & $1.023 \pm 0.037 $\tablenotemark{b} & $1.064 \pm 0.046 $\tablenotemark{b} & $0.993 \pm 0.030 $\tablenotemark{b} & $1.44 \pm 0.10$\tablenotemark{c} & $1.249 \pm 0.071$ & $1.058 \pm 0.022 $\tablenotemark{b} & $1.34 \pm 0.19$\tablenotemark{b} & $1.83 \pm 0.27$ & $1.345 \pm 0.053$ & $1.701 \pm 0.065$\tablenotemark{b} & & 8 (1)\\
03421621--6817321 & L4 $\gamma$ & & $3.99 \pm 0.61 $\tablenotemark{b} & $3.23 \pm 0.46$\tablenotemark{b} & $0.946 \pm 0.031 $\tablenotemark{b} & $1.026 \pm 0.035 $\tablenotemark{b} & $0.930 \pm 0.054$ & $1.011 \pm 0.028 $\tablenotemark{b} & $1.123 \pm 0.049$\tablenotemark{c} & $0.873 \pm 0.056$\tablenotemark{b} & $1.164 \pm 0.088$\tablenotemark{b} & $1.281 \pm 0.066$\tablenotemark{b} & $1.102 \pm 0.038 $\tablenotemark{b} & $1.284 \pm 0.047 $\tablenotemark{b} & & 10 (1)\\
06322402--5010349 & L3 $\beta$ & & $7.83 \pm 1.53 $\tablenotemark{b} & $5.81 \pm 1.29 $\tablenotemark{c} & $1.034 \pm 0.022 $\tablenotemark{b} & $1.079 \pm 0.036 $\tablenotemark{b} & $1.059 \pm 0.031$\tablenotemark{b} & $1.276 \pm 0.066 $ & $1.020 \pm 0.033 $\tablenotemark{b} & $1.081 \pm 0.020$\tablenotemark{b} & $1.316 \pm 0.099 $\tablenotemark{c} & $1.328 \pm 0.078 $\tablenotemark{c} & $1.214 \pm 0.023$\tablenotemark{b} & $1.555 \pm 0.028$\tablenotemark{b} & & 8 (3)\\
\enddata
\tablenotetext{a}{This column indicates the total number of spectral indices (out of 14) that are consistent with a low surface gravity. The number in parentheses indicates the total number of measurements that are consistent with an intermediate surface gravity.}
\tablenotetext{b}{This index is consistent with a low gravity at this spectral type.}
\tablenotetext{c}{This index is marginally consistent with a low gravity at this spectral type (the measurement is consistent with a low gravity but its error bars overlap with the field population).}
\end{deluxetable}
\renewcommand{\tabcolsep}{6pt}
\clearpage
\end{landscape}
\twocolumngrid

However, \cite{2013ApJ...772...79A} noted that their index-based classification scheme should not be significantly contaminated by old, dusty brown dwarfs, which makes this last hypothesis less likely.

\begin{figure}
	\centering
	\includegraphics[width=0.49\textwidth]{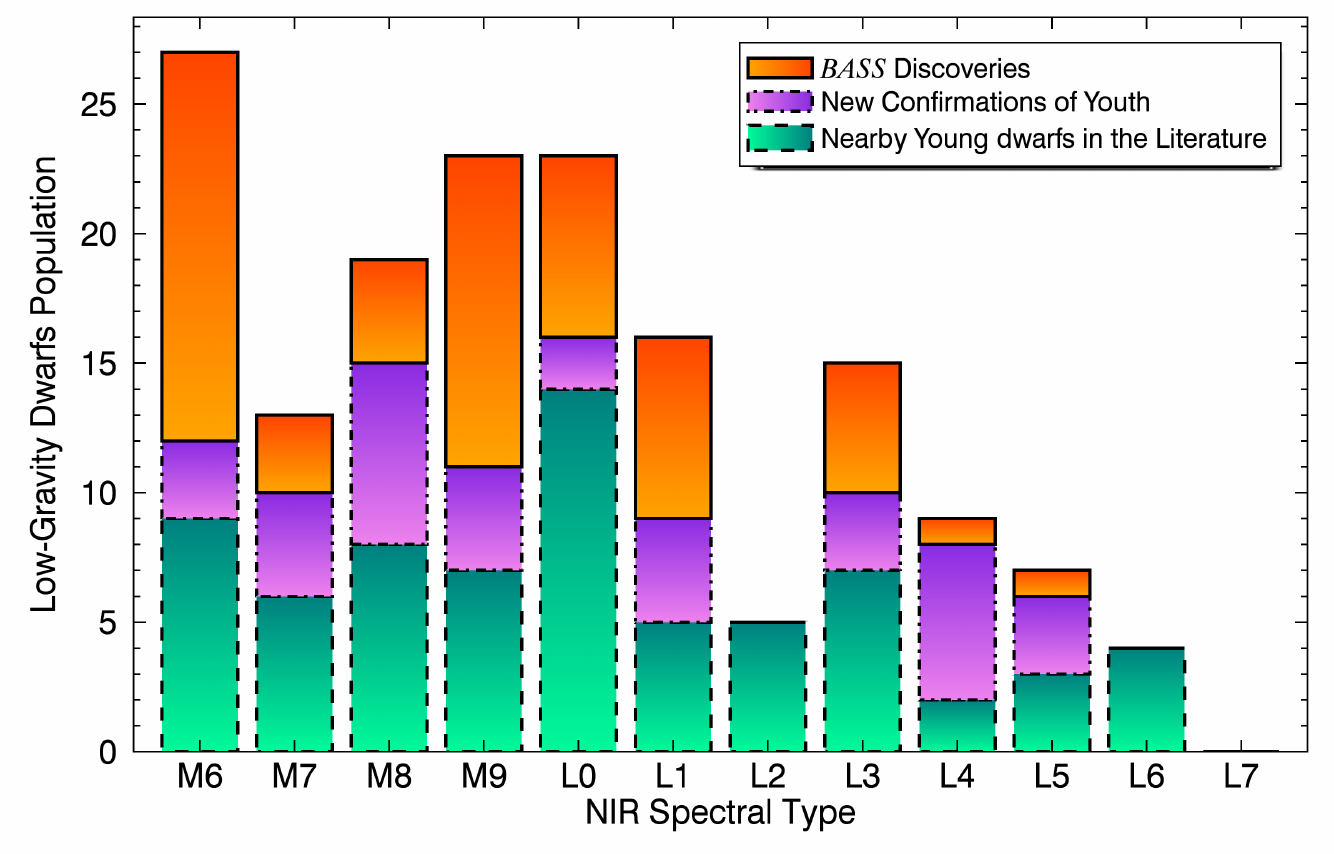}
	\caption{NIR Spectral type histogram of all known low-gravity dwarfs and those presented in this work. Green bars delimited by dashed lines represent the known population prior to \emph{BASS}, purple bars delimited by dash-dotted lines represent known dwarfs for which low-gravity features were identified here for the first time, and orange bars delimited by solid lines represent new discoveries from \emph{BASS}. The \emph{BASS} survey has contributed significantly in increasing the number of known low-gravity M6--L5 dwarfs.}
	\label{fig:spthist}
\end{figure}

In Figure~\ref{fig:spthist}, we show a histogram of all previously known low-gravity dwarfs along with new discoveries or confirmations of low gravity that are presented here. It might seem surprising that we did not identify any new low-gravity L2 dwarfs, however this is likely the effect of small number statistics and the fact that we still lack a template for the L2\,$\beta$ spectral type, e.g., some low-S/N low-gravity objects presently typed as L1: and L3: might turn out to be L2 dwarfs when more data becomes available. We anticipate our visual-based low-gravity classification scheme to improve as more data is obtained. If we account for the measurement errors on our spectral types using gaussian probability density functions (which softens the gap at L2) and use Poisson statistics to assess the significance of this lack of L2 dwarfs, we find that the differences between the number of known low-gravity L1, L2 and L3 dwarfs is insignificant (at the level of 0.2$\sigma$).

\begin{figure*}
	\centering
	\subfigure[FeH$_Z$]{\includegraphics[width=0.45\textwidth]{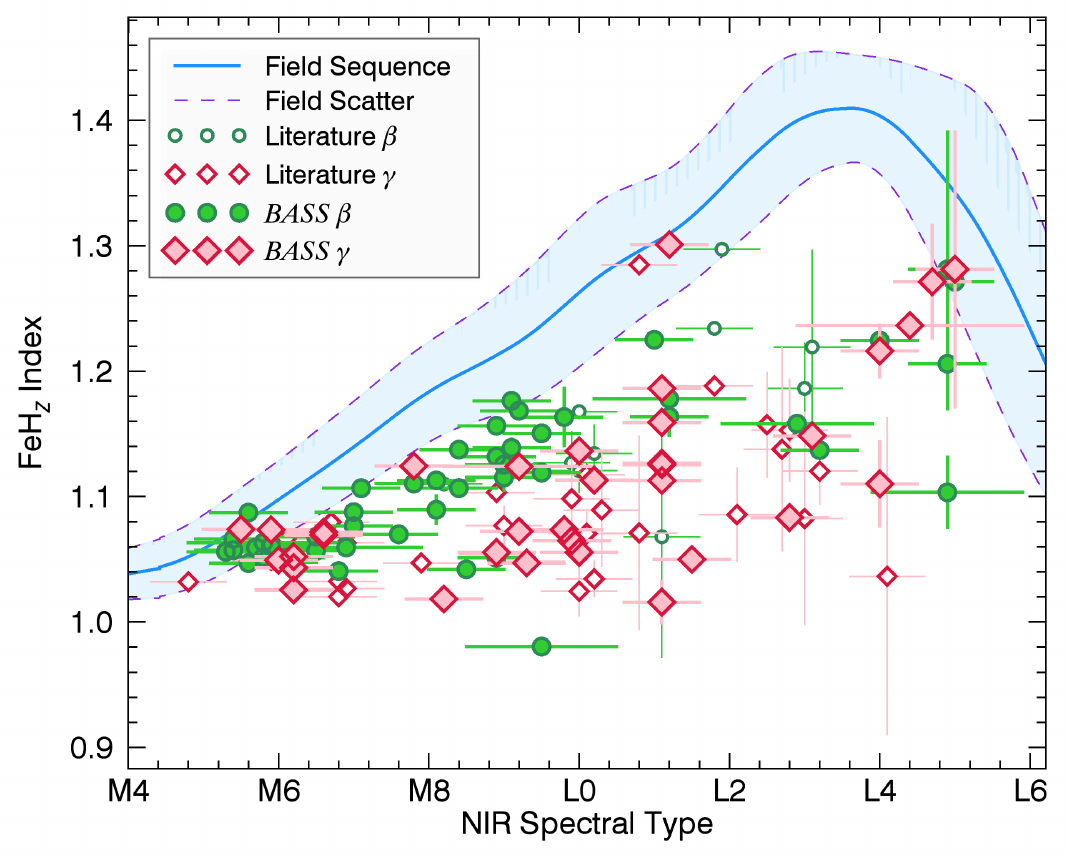}}
	\subfigure[$H$-cont]{\includegraphics[width=0.45\textwidth]{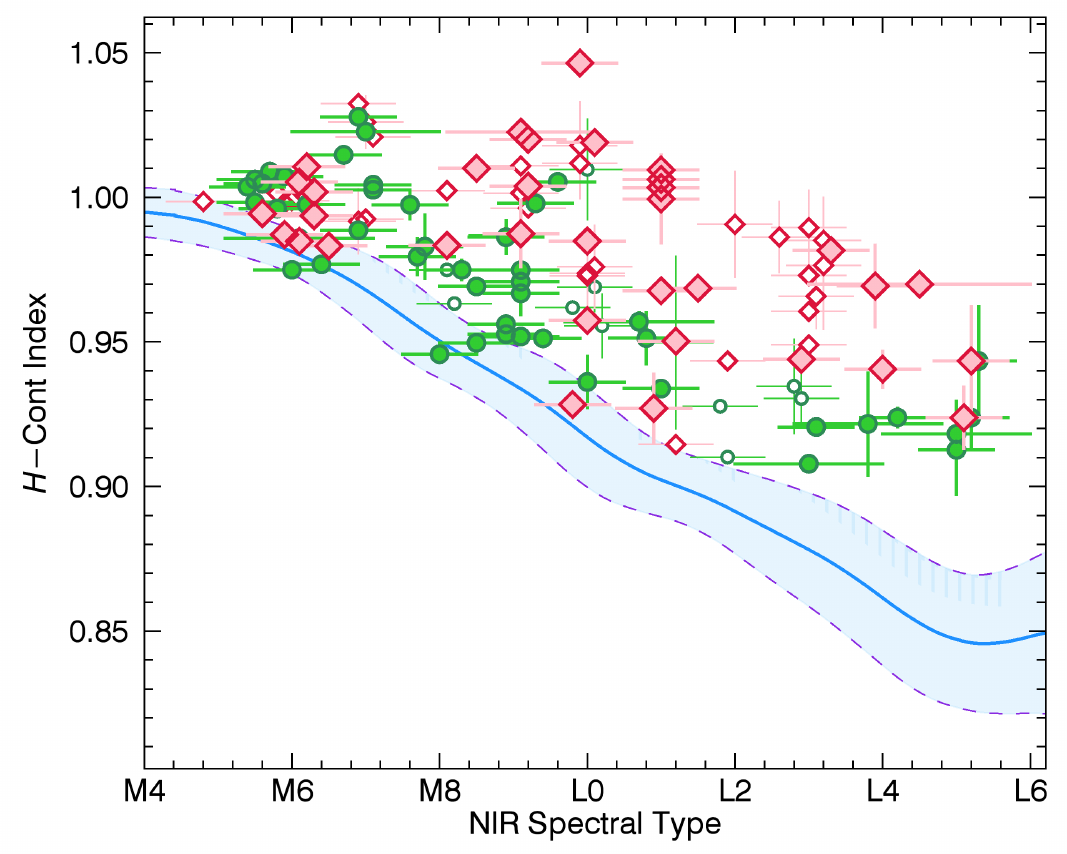}}
	\subfigure[VO$_Z$]{\includegraphics[width=0.45\textwidth]{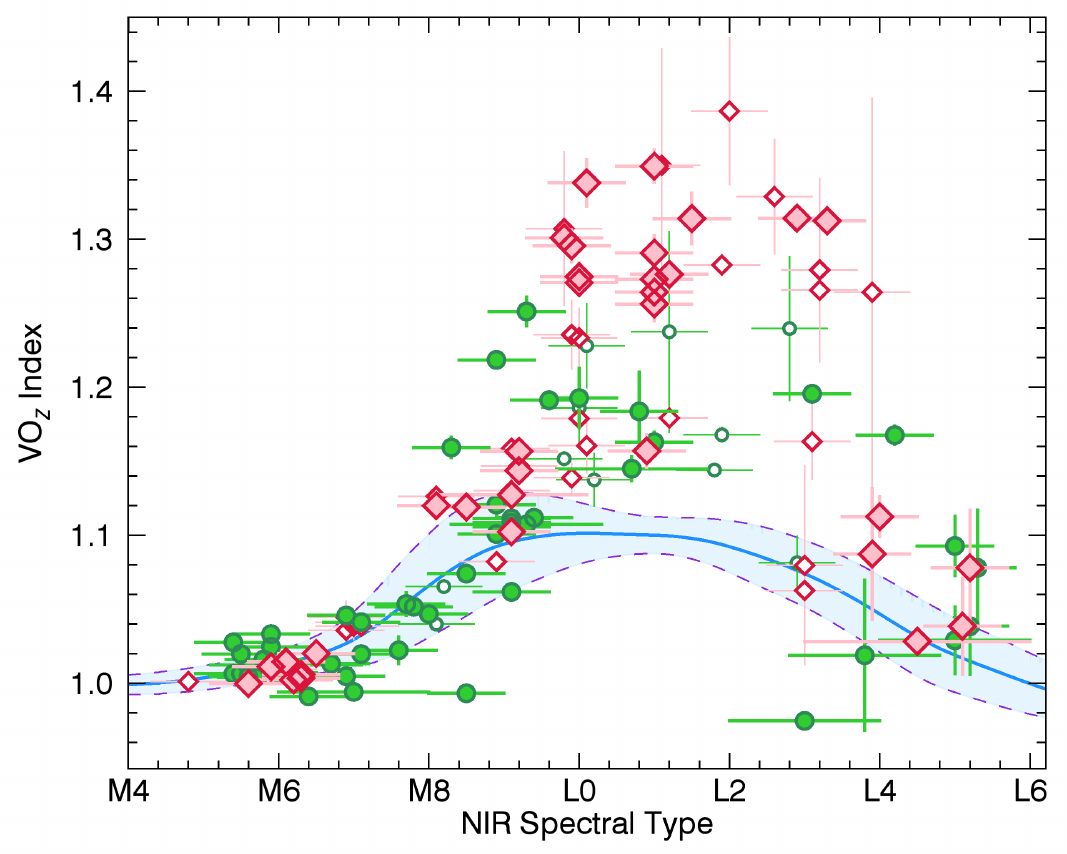}}
	\subfigure[KI$_J$]{\includegraphics[width=0.45\textwidth]{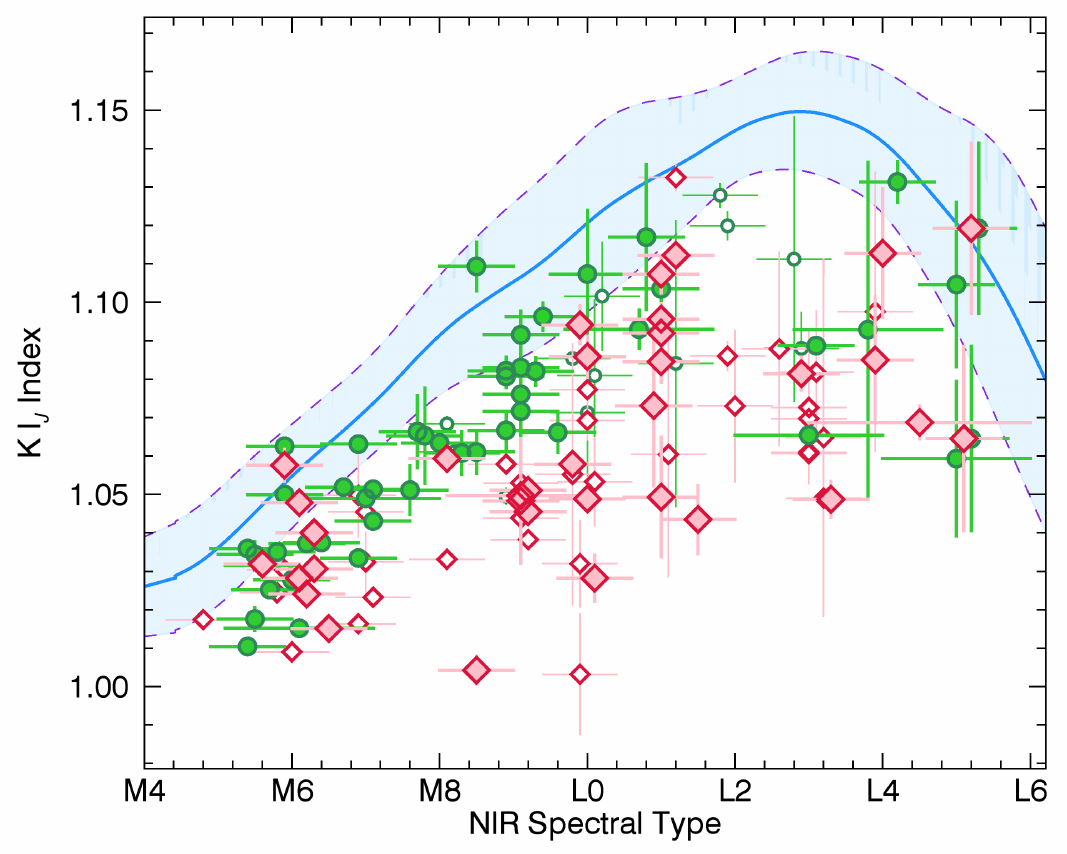}}
	\caption{Low-resolution (R\,$\gtrsim$\,75) gravity-sensitive NIR indices defined by \cite{2013ApJ...772...79A} for all intermediate-gravity (green circles) and very low-gravity (red diamonds) dwarfs from the \emph{BASS} sample. This sample consists mainly of new discoveries and known dwarfs with a new low-gravity classification. Previously known intermediate-gravity and low-gravity dwarfs from the samples of \cite{2013ApJ...772...79A} and \cite{2014A&A...564A..55M} are displayed as smaller, open symbols. The thick, blue line and the pale blue region delimited by dashed, purple lines represent the field sequence and its scatter. Random offsets smaller than 0.25 subtypes have been added to the spectral types for clarity. Lower-gravity dwarfs display (1) lower FeH$_Z$ and KI$_J$ indices at spectral types M5.5--L6; (2) larger $H$-cont indices at spectral types M5.5--L6; and (3) larger VO$_Z$ indices at spectral types L0--L4 (see text for more detail). It is readily apparent that low-gravity dwarfs of the same spectral type can display a different set of low-gravity features, which is why a classification based on multiple gravity-sensitive indices is necessary \citep{2013ApJ...772...79A}.}
	\label{fig:I1}
\end{figure*}

\begin{figure*}
	\centering
	\subfigure[FeH$_J$]{\includegraphics[width=0.45\textwidth]{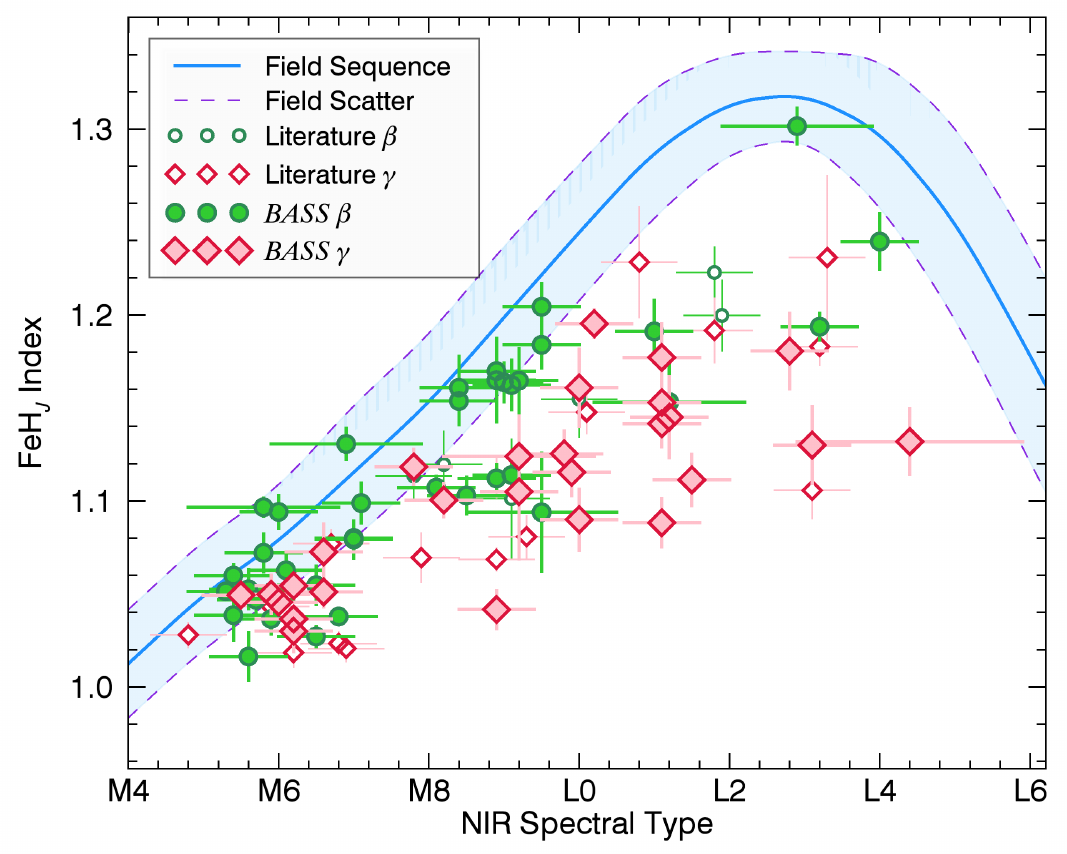}}
	\subfigure[\ion{K}{1} 1.169\,$\mu$m]{\includegraphics[width=0.45\textwidth]{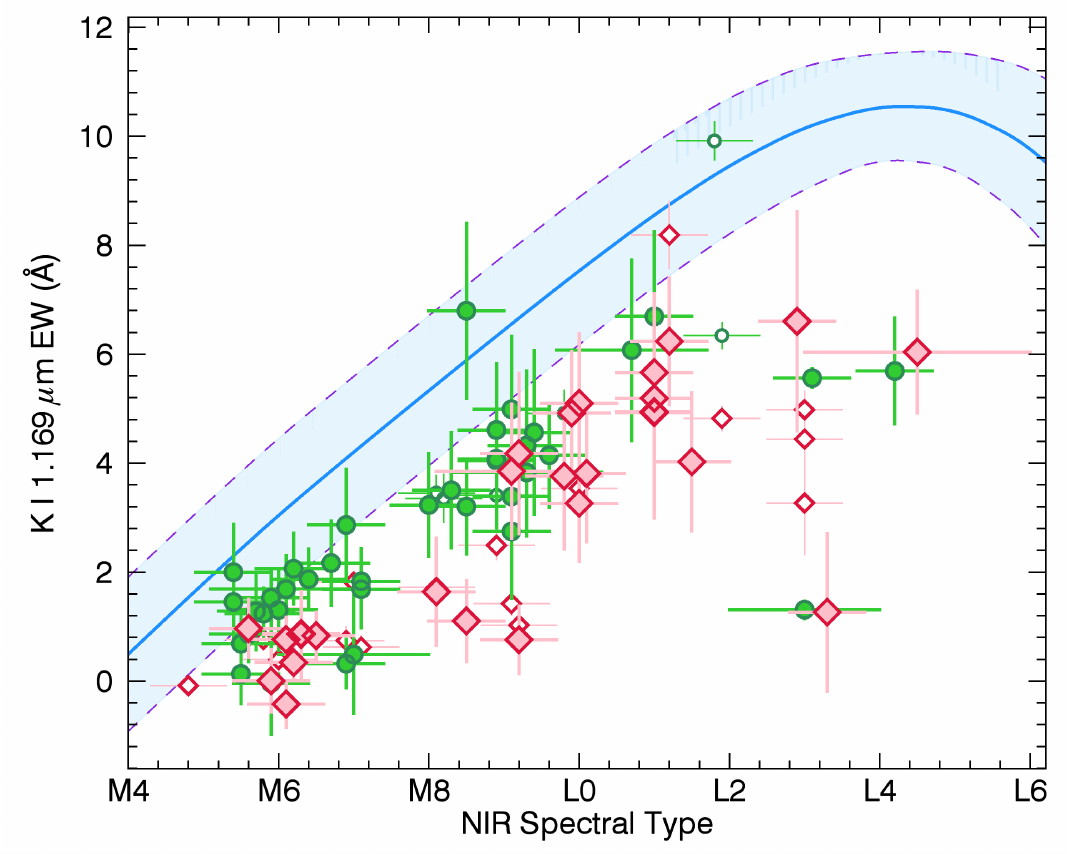}}
	\subfigure[\ion{K}{1} 1.253\,$\mu$m EW]{\includegraphics[width=0.45\textwidth]{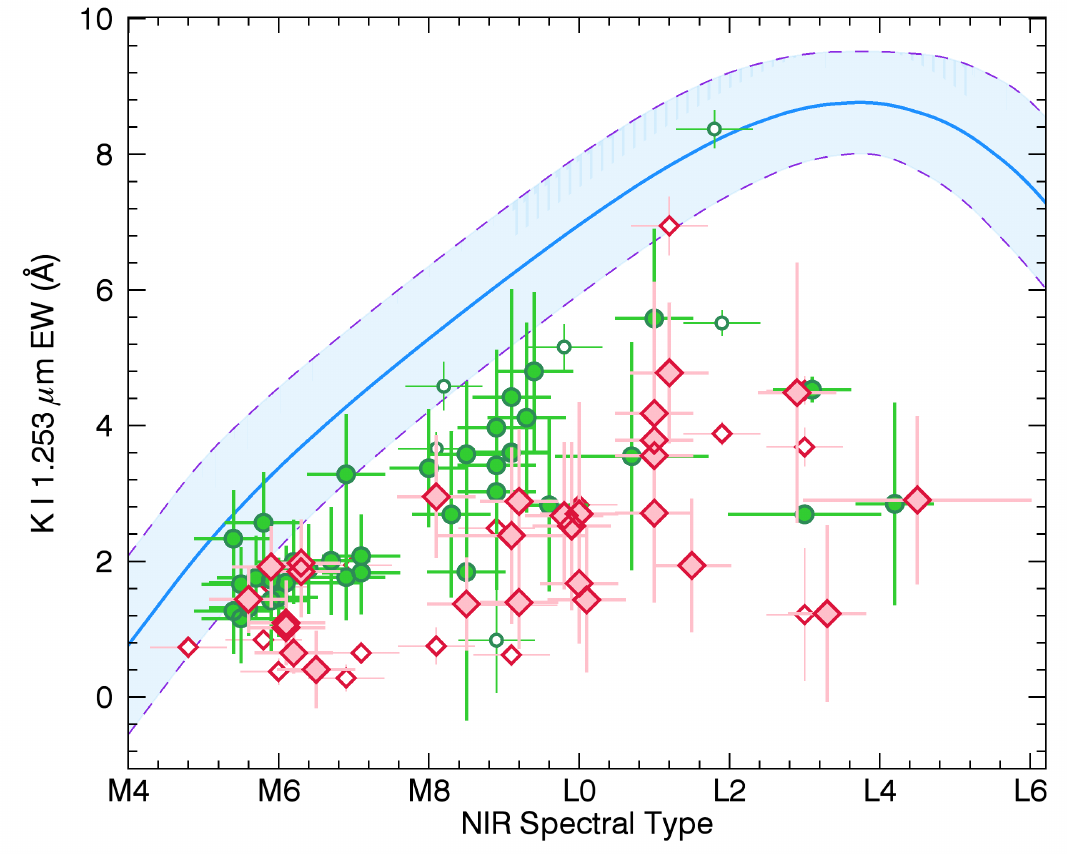}}
	\subfigure[\ion{Na}{1} 1.138\,$\mu$m EW]{\includegraphics[width=0.45\textwidth]{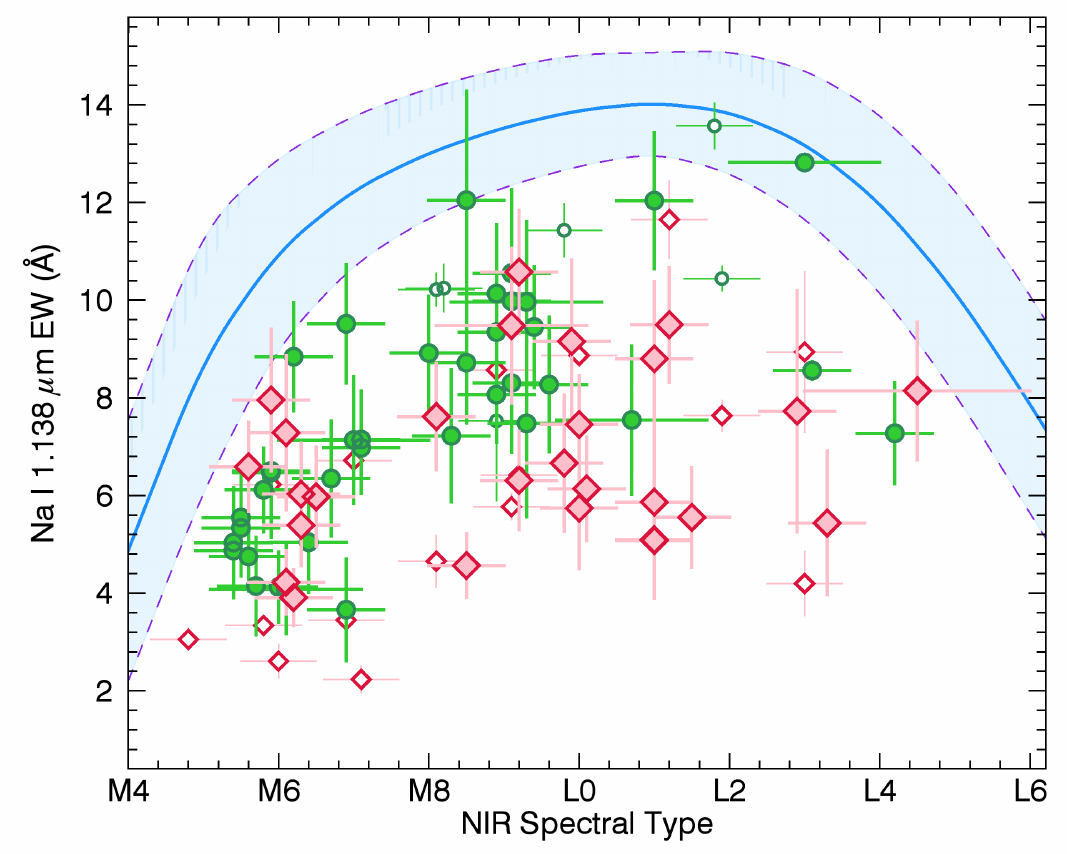}}
	\caption{Moderate-resolution (R\,$\gtrsim$\,750) gravity-sensitive NIR indices defined by \cite{2013ApJ...772...79A} for all intermediate-gravity and very low-gravity dwarfs from the \emph{BASS} sample. Symbols and color coding are identical to those of Figure~\ref{fig:I1}. Lower-gravity dwarfs display weaker alkali and FeH absorption features, which results in lower \ion{Na}{1} and \ion{K}{1} EWs and a lower FeH$_J$ index.}
	\label{fig:I2}
\end{figure*}

In Figures \ref{fig:I1} and \ref{fig:I2}, we compare all new low-gravity confirmations with the field and low-gravity sequences defined by \cite{2013ApJ...772...79A}. The individual values for these gravity-sensitive spectroscopic indices are listed in Table~\ref{tab:sptindices}. There are 7 objects in our sample that did not have a discussion of low gravity in the literature and for which optical spectra were available in the Ultracool RIZzo Spectral Library. We used them to revise their spectral types and measure gravity-sensitive optical indices defined by \cite{1999ApJ...519..802K} and \cite{2009AJ....137.3345C}. These results, based on optical data only, are presented in Table~\ref{tab:sptoptindices}. The new spectroscopic observations presented here (95 from \emph{BASS}, 26 from \emph{LP-BASS} and 120 from \emph{PRE-BASS}) allowed us to uncover a total of 108 new M6--L5 low-gravity dwarfs, doubling the number of such known objects (98 before this work).

\begin{figure}
	\centering
	\subfigure[Galactic position]{\includegraphics[width=0.49\textwidth]{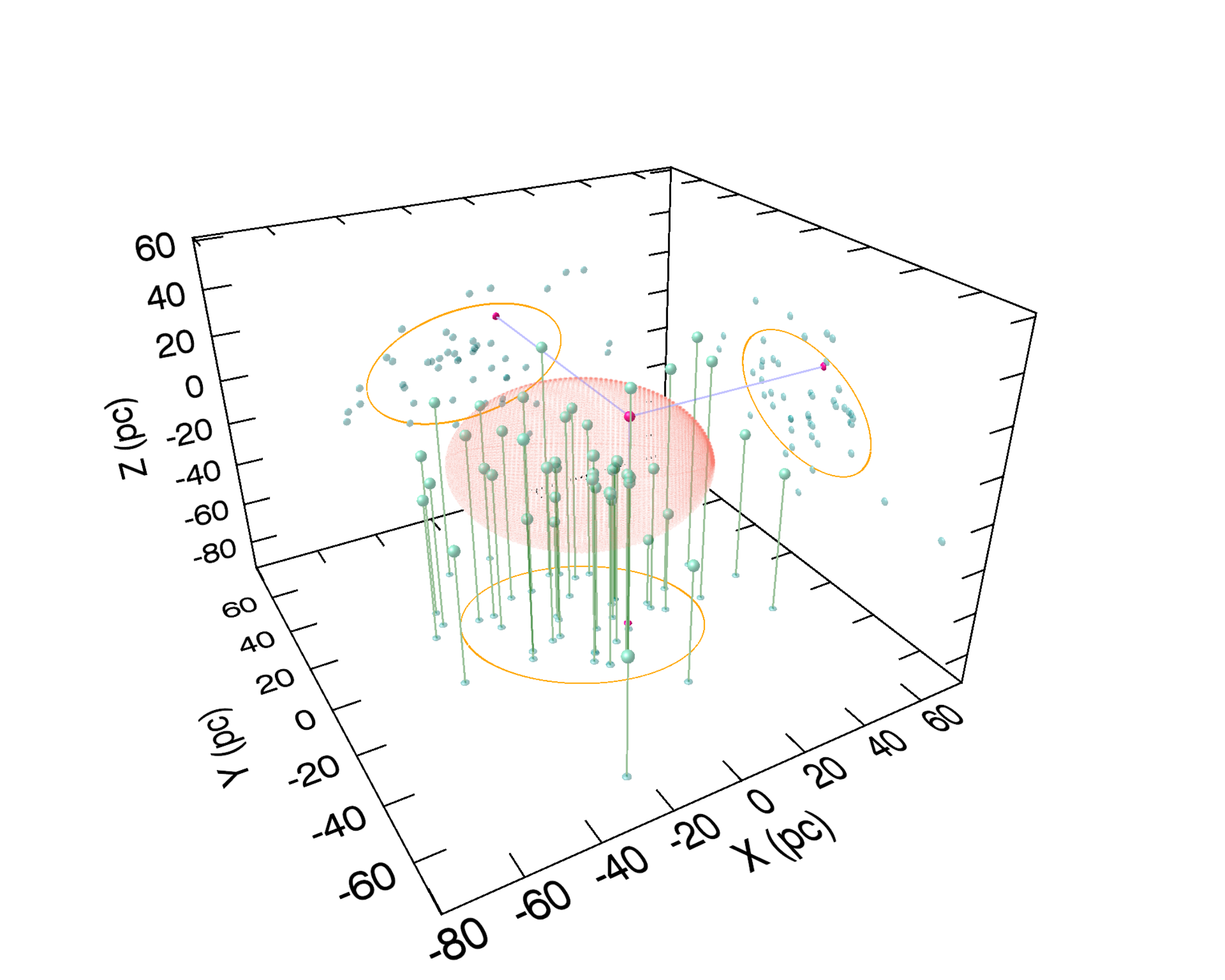}}
	\subfigure[Space velocity]{\includegraphics[width=0.49\textwidth]{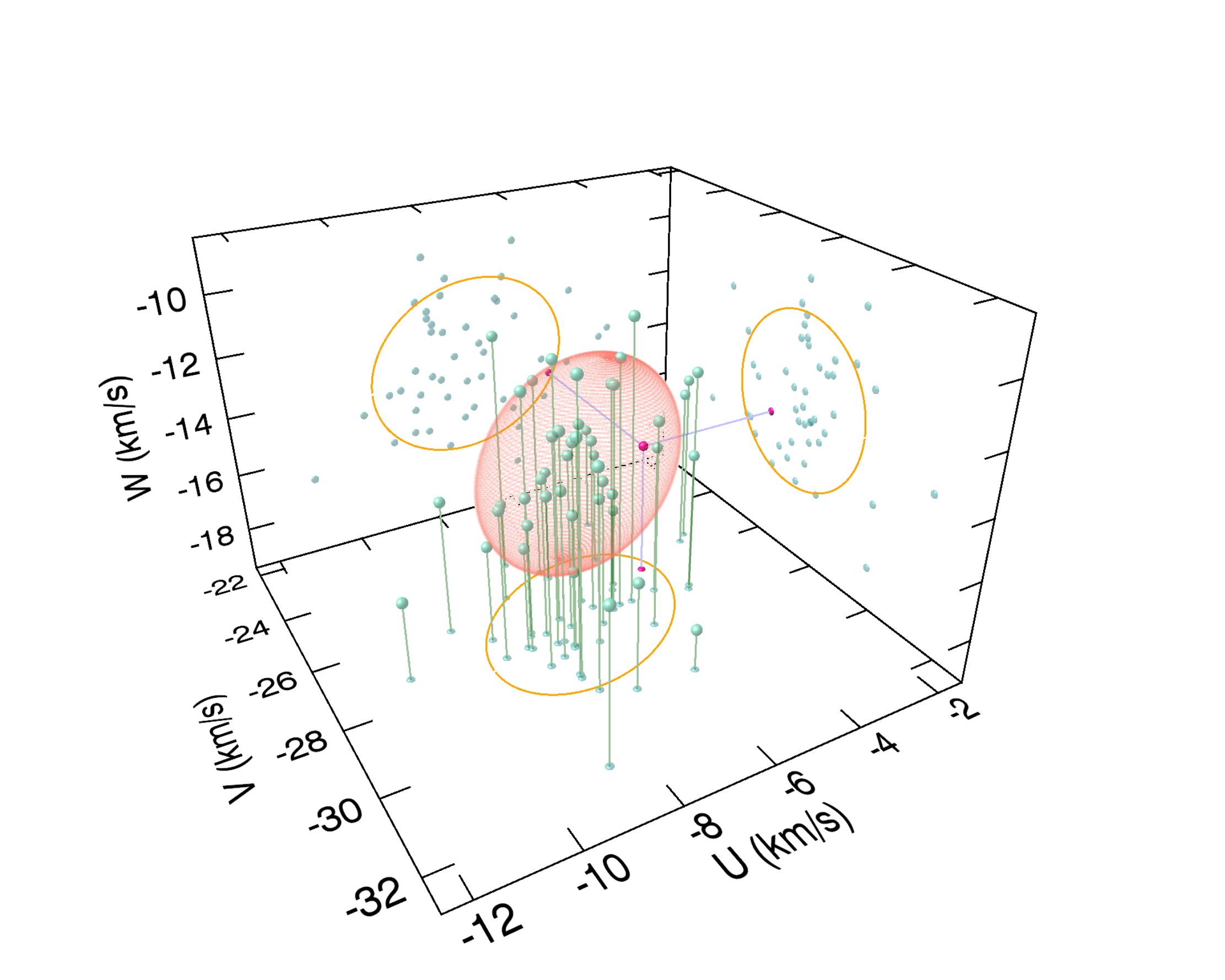}}
	\caption{Galactic position $XYZ$ and space velocity $UVW$ of the new AB Doradus bona fide member 2MASS~J14252798--3650229 (red point and its projections), compared with other bona fide members of ABDMG (green points and their vertical projections on the $XY$ and $UV$ planes) and the SKM models of ABDMG (as defined in Paper~II; orange ellipsoid and its projections).}
	\label{fig:L3XYZUVW}
\end{figure}

\begin{figure}
	\centering
	\includegraphics[width=0.49\textwidth]{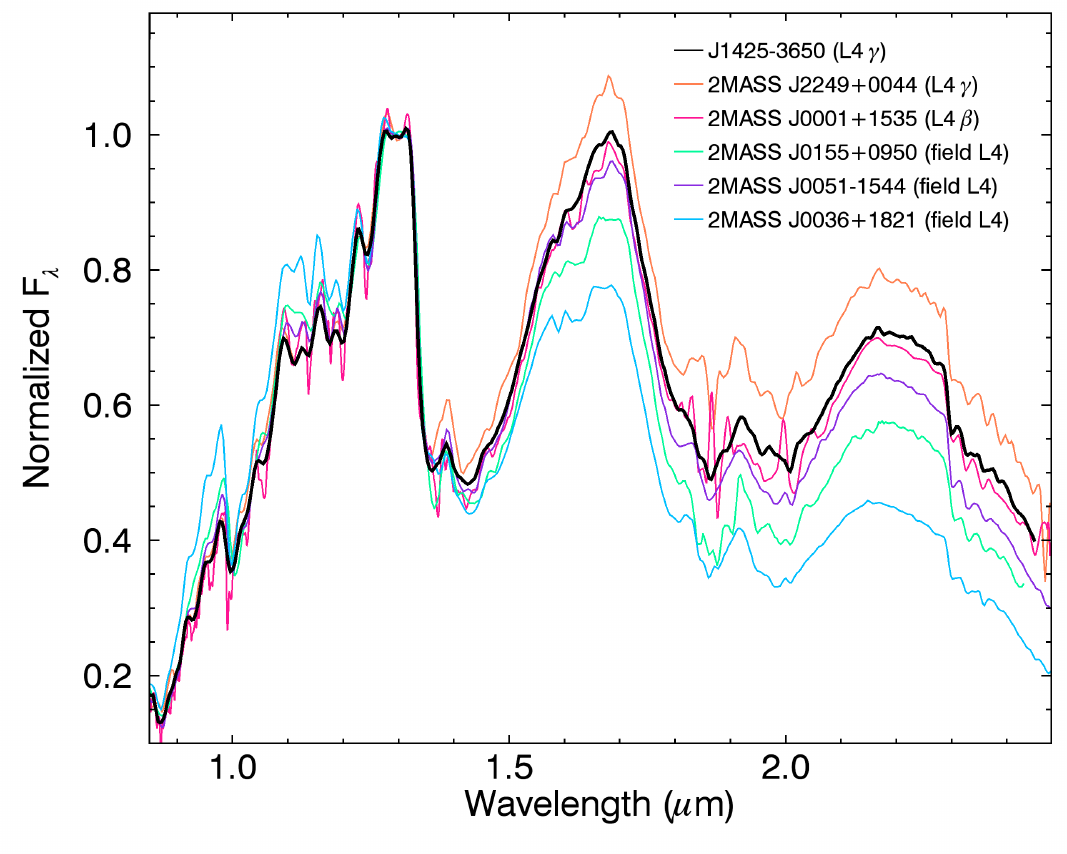}
	\caption{NIR spectrum of the new L4\,$\gamma$ ABDMG bona fide member 2MASS~J14252798--3650229 (thick black line), compared with various field and low-gravity L4 BDs. All spectra were degraded to a resolution of $R \sim 120$ and normalized at their median value in the $\sim$\,1.27--1.33\,$\mu$m range. The $H$-band continuum of 2MASS~J14252798--3650229 has a typical triangular shape and its global slope is particularly red, which are both telltale signs of low gravity.}
	\label{fig:L3ador}
\end{figure}

In addition to several new candidate members of YMGs, we report here that 2MASS~J14252798--3650229 (DENIS-P~J142527.97-365023.4) is a new low-mass BD bona fide member of ABDMG. This object was identified by \cite{2004A&A...416L..17K} as an L5 dwarf with an estimated spectro-photometric distance of $\sim$\,10\,pc. \cite{2010ApJ...723..684B} measured an RV of $5.37 \pm 0.25$\,\kms\ and \cite{2014AJ....147...94D} measured a trigonometric distance of $11.57 \pm 0.11$\,pc. \cite{2015ApJ...798...73G} reported that the galactic position and space velocities of this object are a very good match to ABDMG (Figure~\ref{fig:L3XYZUVW}), suggesting that it would be a new bona fide member if low gravity would be confirmed. They also indicated that its NIR colors are redder than those of field dwarfs of the same spectral type, which hints at low gravity. The low gravity is indeed readily apparent in the new SpeX prism spectrum that we obtained for this object (Figure~\ref{fig:L3ador}): both a visual comparison and the index-based classification of \cite{2013ApJ...772...79A} indicate that this object is an L4\,$\gamma$ dwarf. We conclude that 2MASS~J14252798--3650229 is a new bona fide member of ABDMG, making it the second latest-type confirmed member of this moving group after the L7\,$\beta$ member WISEP~J004701.06+680352.1. At the age of ABDMG, 2MASS~J14252798--3650229 has an estimated mass of $26.6^{+0.3}_{-1.0}$\,\MJup.

%Figure : Spectral Types histogram

%Figure : Allers 2013 Indices, batch 1

%Figure : Allers 2013 Indices, batch 2

\subsection{Updated YMG Membership}\label{sec:mg}

%Figure : L3 bona fide : XYZUVW

%Figure : L3 bona fide : spectrum

%Figure : Spectra showcase : Contaminants
\begin{figure}
	\centering
	\includegraphics[width=0.49\textwidth]{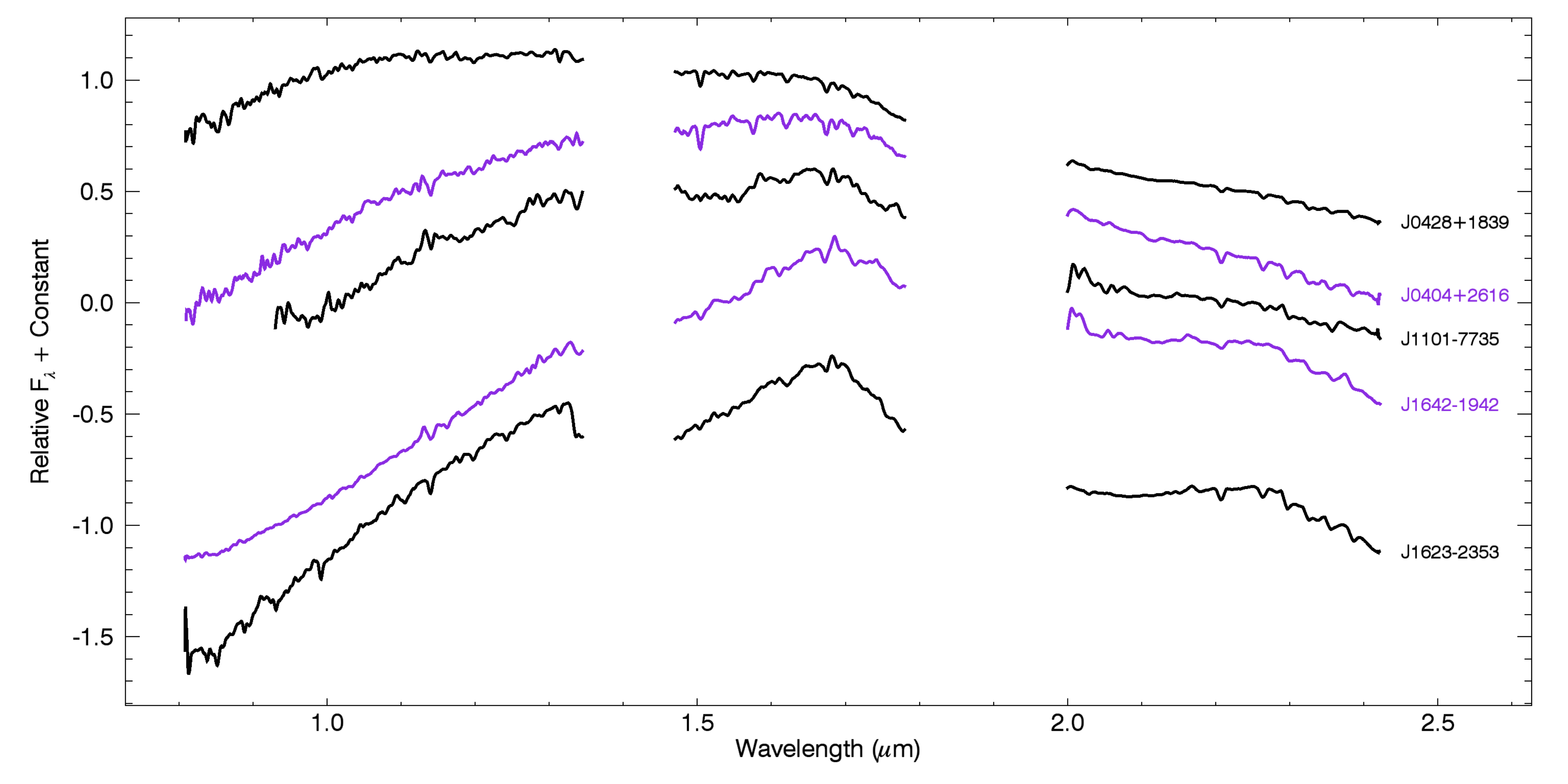}
	\caption{NIR spectra of typical contaminants in the \emph{PRE-BASS} sample. Resolution was degraded in the same way as described in Figure~\ref{fig:spectra1}. All spectra were normalized to their median across the full wavelength range and shifted vertically for comparison purposes. The contaminants presented in this figure likely correspond to background K- and M-type stars reddened by interstellar dust. We used alternating colors for visibility.}
	\label{fig:spectrac}
\end{figure}

It is possible to use the spectral type information as well as the youth of candidate members determined from the spectroscopic follow-up presented here as additional inputs in BANYAN~II to refine estimates of distance, RV and YMG membership and contamination probabilities. Spectral types are used to assess if the absolute $W1$ magnitude of a target is consistent with its spectral type at the statistical distance that corresponds to a given YMG membership (using distinct sequences for field and low-gravity dwarfs; see Paper~V), whereas prior knowledge of youth reduces the number of potential contaminants from the field and thus improves the probability that the object belongs to a YMG. We reject all objects with spectral types $\geq$\,M5 that display no signs of low gravity (17 in \emph{BASS}, 7 in \emph{LP-BASS} and 41 in \emph{PRE-BASS}), since this implies an age older than the Pleiades ($\sim$\,120\,Myr; \citealp{2009AJ....137.3345C,2013ApJ...772...79A}) and is not consistent with membership to any YMG considered here. These updated results are listed in Table~\ref{tab:sptclass}, and individual objects of interest are discussed in the Appendix.

\subsection{X-Ray Luminosity}\label{sec:xray}

We followed up several objects that turned out to have spectral types earlier than expected, some of them ($\leq$ M5) to the point where current NIR and optical index-based methods are unable to determine whether they are likely young or field objects. In this section, we take advantage of the ROSAT bright and faint source catalogs (\citealt{1999A&A...349..389V,2000IAUC.7432R...1V}; VizieR catalogs \emph{IX/10A} and \emph{IX/29}) to assess whether these objects are young candidate members of YMGs or field interlopers.

\cite{2014ApJ...788...81M} demonstrated that the distribution of absolute X-ray luminosity for M0--M5 dwarf members of ABDMG and $\beta$PMG is significantly distinct from that of field M0--M5 dwarfs. In particular, they showed that $\beta$PMG members are $\sim$\,4 times more X-ray luminous than ABDMG members, a factor that goes up to $\gtrsim$\,40 when instead compared with field dwarfs. We investigated whether any of our M0--M5 candidate members listed in Table~\ref{tab:sptclass} display X-ray emission by cross-matching their \emph{2MASS} position with the ROSAT catalogs with a 15$"$ search radius. We computed the absolute X-ray luminosity for all X-ray sources recovered this way, using trigonometric distances when possible or kinematic distances otherwise.

We have identified ROSAT entries for only three objects: 2MASS~J08540240--3051366 (M4 candidate member of $\beta$PMG; $\log L_X = 28.4 \pm 0.3$) has a low X-ray luminosity compared with M3--M5 members of ABDMG or $\beta$PMG (both have $\log L_X \approx 28.5-29.5$) and could thus be a field interloper ($\log L_X \approx 27-28.5$; see Figures 7 and 8 of \citealt{2014ApJ...788...81M}). 2MASS~J08194309--7401232 (M4.5 candidate member of COL; $\log L_X = 29.3 \pm 0.4$) and 2MASS~J21490499--6413039 (M4.5 candidate member of THA; $\log L_X = 29.3 \pm 0.3$) both have X-ray luminosities consistent with an age similar or younger than that of ABDMG, making them likely members of their respective moving groups. We note that 2MASS~J21490499--6413039 has already been reported as a candidate member of THA by \citealp{2014AJ....147..146K}, who measured its RV and found it to be consistent with other THA members. Objects that do not have a ROSAT counterpart do not necessarily have a low absolute X-ray luminosity, but might be too distant or located outside of the regions covered by the ROSAT survey.

Using the ROSAT bright catalog detection limit of $0.1$\,ct/s in the 0.1--2.4\,keV energy band and assuming a hardness ratio \emph{HR1}\,$\approx 0$, we can only put an upper limit of $\log L_X = 28-29.8$ on the remaining targets, which is generally not sufficient to reject any more candidate members. Only 3/41 of these targets (2MASS~J05484454--2942551, 2MASS~J06494706--3823284 and 2MASS~J07583098+1530146~AB) have $\log L_X < 28.5$, potentially making them less interesting candidate members. It should be noted however that one of these three objects (2MASS~J06494706--3823284) has weak \ion{Na}{1} absorption consistent with a very low surface gravity. This demonstrates how the absence from the ROSAT catalog is not a strong enough constraint to reject any of our M0--M5 candidate members. \cite{2014AJ....147..146K} has demonstrated that surveys for M-type moving group members based on either X-ray or UV-bright samples are incomplete because of the sky coverage and detection limits of current X-ray and UV catalogs.
%Limit of Rosat bright is 0.1 counts/second.

\subsection{Sources of contamination}\label{sec:contam}

In Paper~II, we demonstrated that a fraction of candidate members identified by the BANYAN~II tool are expected to be field interlopers, especially if no prior knowledge is available on age. This fraction of contaminants is dependent on the YMG considered: ARG, ABDMG and $\beta$PMG are expected to be the most contaminated, mostly due to their proximity and their overlap with the galactic plane. Counting the fraction of low-gravity dwarfs in the spectroscopic follow-up presented here allows us to estimate minimal contamination rates of 18\% and 33\% in the \emph{BASS} and \emph{LP-BASS} samples, respectively. These values are slightly larger than the estimates that we derived in Paper~V (12.6\% for \emph{BASS} and 26\% for \emph{LP-BASS}). The most likely explanation is that the kinematic distribution of field BDs is not perfectly reproduced by the \besancon\ galactic model, on which our previous estimates were based. The reason why these updated estimates correspond to a minimal contamination fraction is that some low-gravity dwarfs in our candidate sample could still be contaminants from associations not considered in BANYAN~II, e.g., the Ursa Majoris moving group (UMA; $\sim$\,300\,Myr; \citealt{2004ARA&A..42..685Z}), the Hercules-Lyrae moving group (250\,Myr; \citealt{2013A&A...556A..53E}), the $\epsilon$~Chamaeleontis association (also called Cha-Near; $\sim$\,10\,Myr; \citealt{2004ARA&A..42..685Z}), the Octans association (30--40\,Myr; \citealp{2008hsf2.book..757T,2015MNRAS.447.1267M}) and the Carina-Near moving group (200\,Myr; \citealt{2006ApJ...649L.115Z}). Measurements of RV and trigonometric distance will be helpful to identify such contaminants.
Besides field-gravity $\geq$\,M5 dwarfs, we identified other kinds of contaminants in our sample of candidates, based on our new NIR spectroscopy. We uncovered a number of objects with spectral types earlier than M5 (4 in \emph{BASS}, 4 in \emph{LP-BASS} and 28 in \emph{PRE-BASS}), for which there is no known reliable low-gravity indicators in the NIR. In addition to those, we uncovered 27 contaminants mostly in the \emph{PRE-BASS} sample (only one was found in \emph{BASS}) that correspond to K- and M-type low-mass stars reddened by interstellar dust in the line of sight (Figure~\ref{fig:spectrac}). A number of these are likely located in star-forming regions, such as $\rho$~Ophiucus ($\rho$OPH), the Scorpius-Centaurus Complex (SCC) and Taurus-Aurigae (TAU; \citealt{1978ApJ...224..857E}). These objects were all rejected from the \emph{BASS} sample, mainly because (1) we avoided star-forming regions in the final survey; and (2) the extragalactic WISE color filter defined by \cite{2011ApJS..197...19K} and the \emph{2MASS} crowding filter defined in Paper~V efficiently rejected them. 

\section{DISCUSSION}\label{sec:discussion}

%Figure : CMDs
\begin{figure*}
	\centering
	\subfigure[$M_J$ versus NIR Spectral Type]{\includegraphics[width=0.33\textwidth]{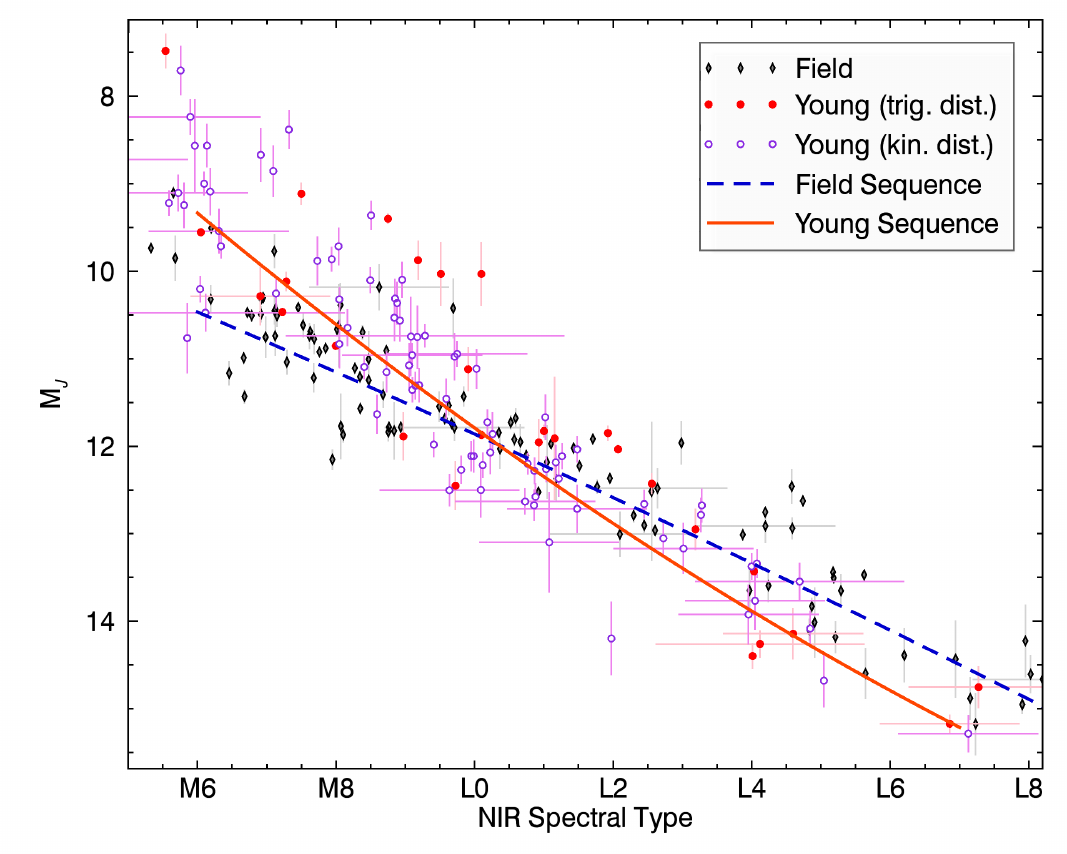}}
	\subfigure[$M_H$ versus NIR Spectral Type]{\includegraphics[width=0.33\textwidth]{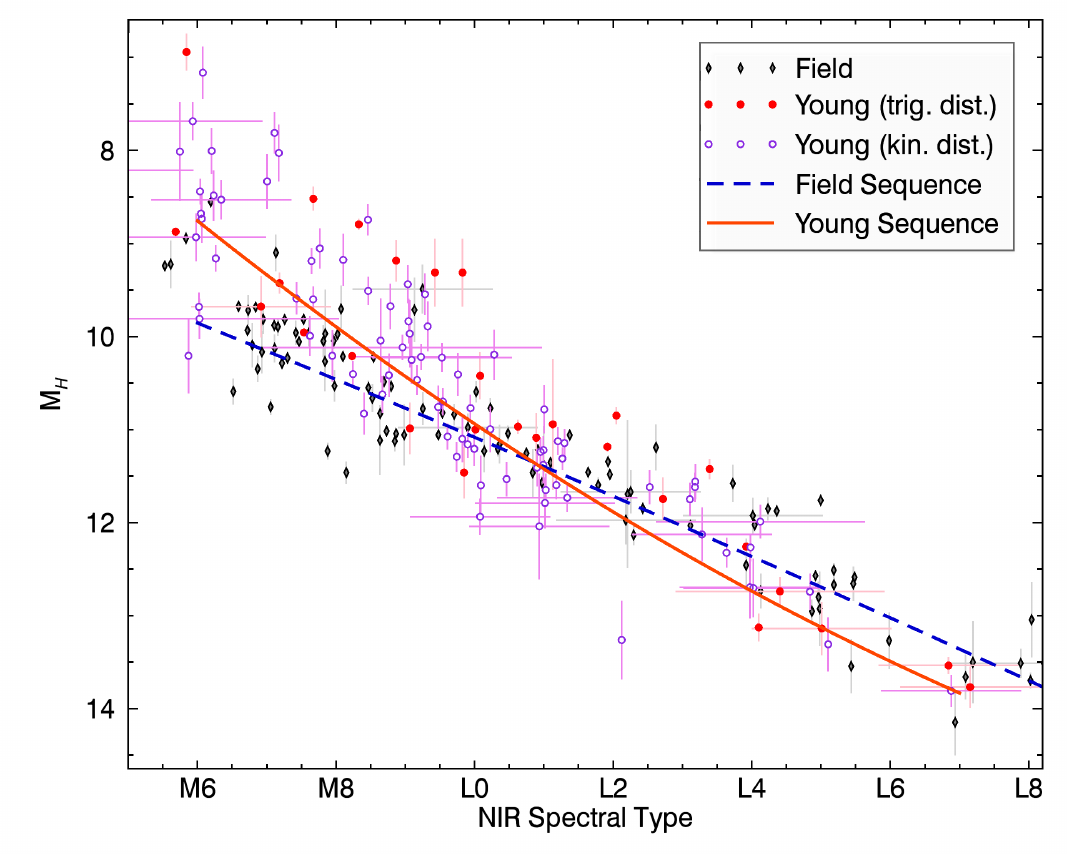}}
	\subfigure[$M_{K_S}$ versus NIR Spectral Type]{\includegraphics[width=0.33\textwidth]{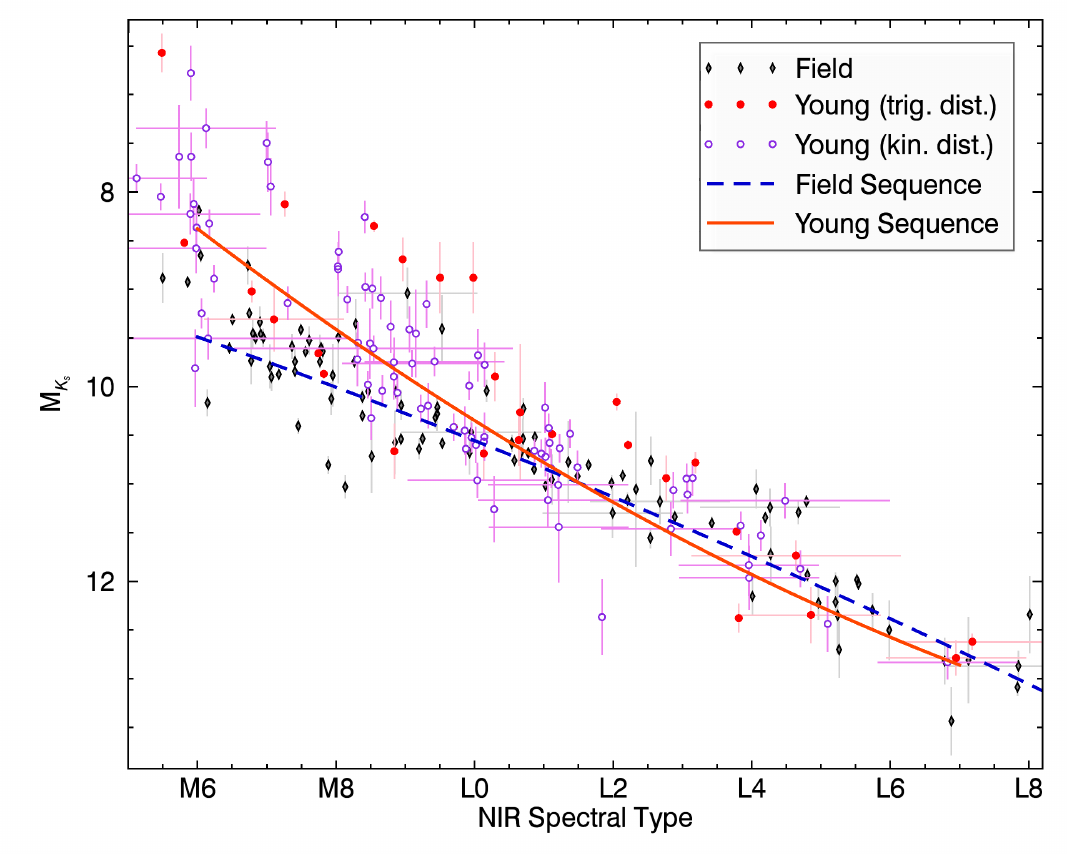}}
	\subfigure[$M_{W1}$ versus NIR Spectral Type]{\includegraphics[width=0.33\textwidth]{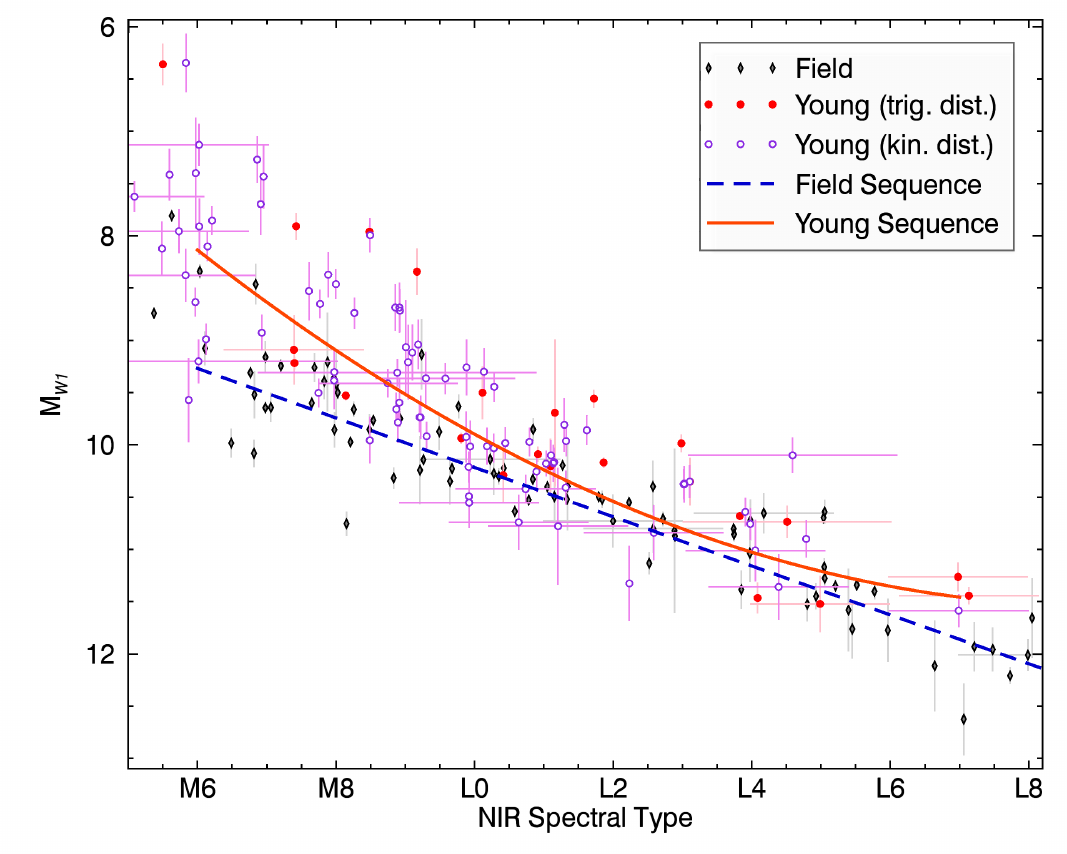}}
	\subfigure[$M_{W2}$ versus NIR Spectral Type]{\includegraphics[width=0.33\textwidth]{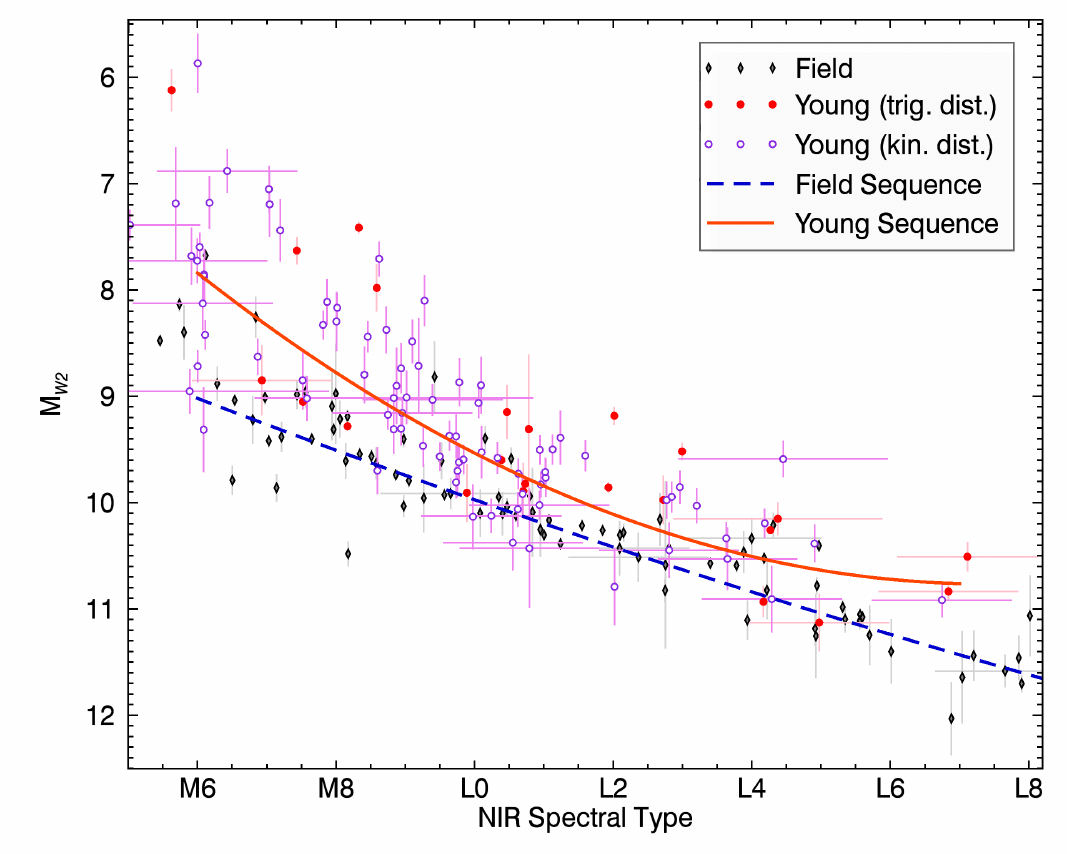}}
	\subfigure[$J-H$ versus NIR Spectral Type]{\includegraphics[width=0.33\textwidth]{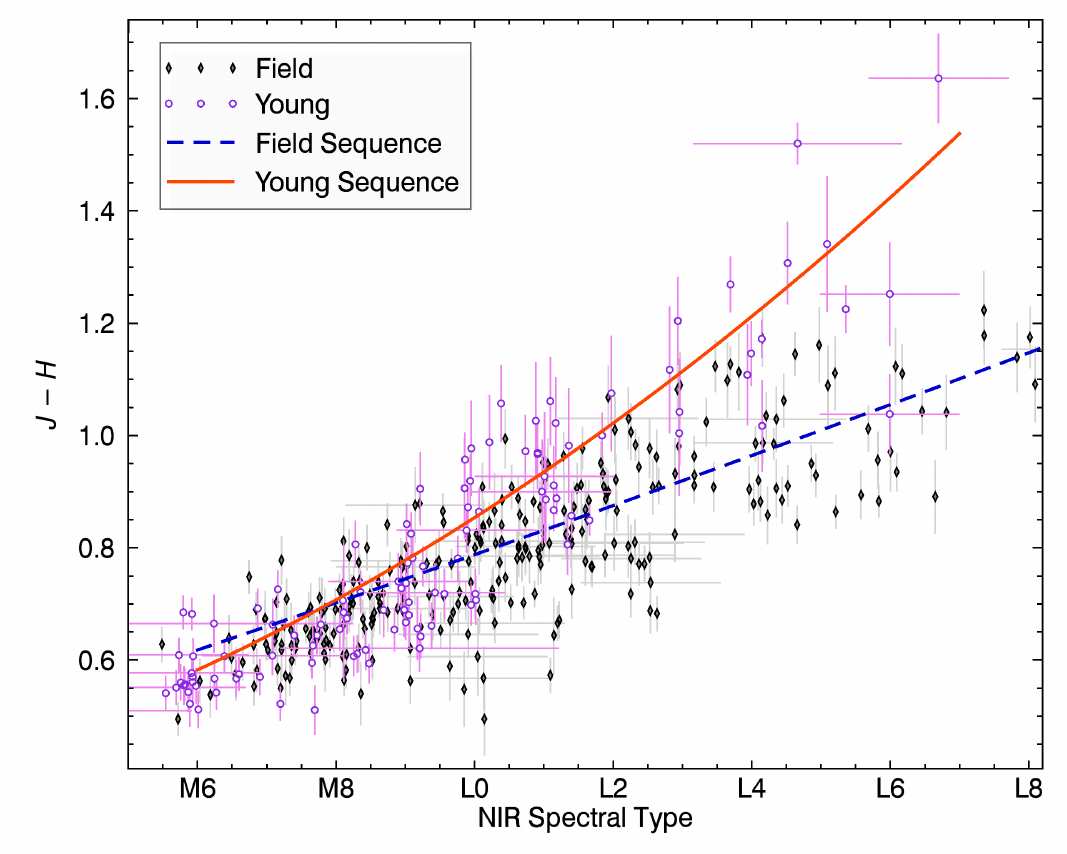}}
	\subfigure[$H-K_S$ versus NIR Spectral Type]{\includegraphics[width=0.33\textwidth]{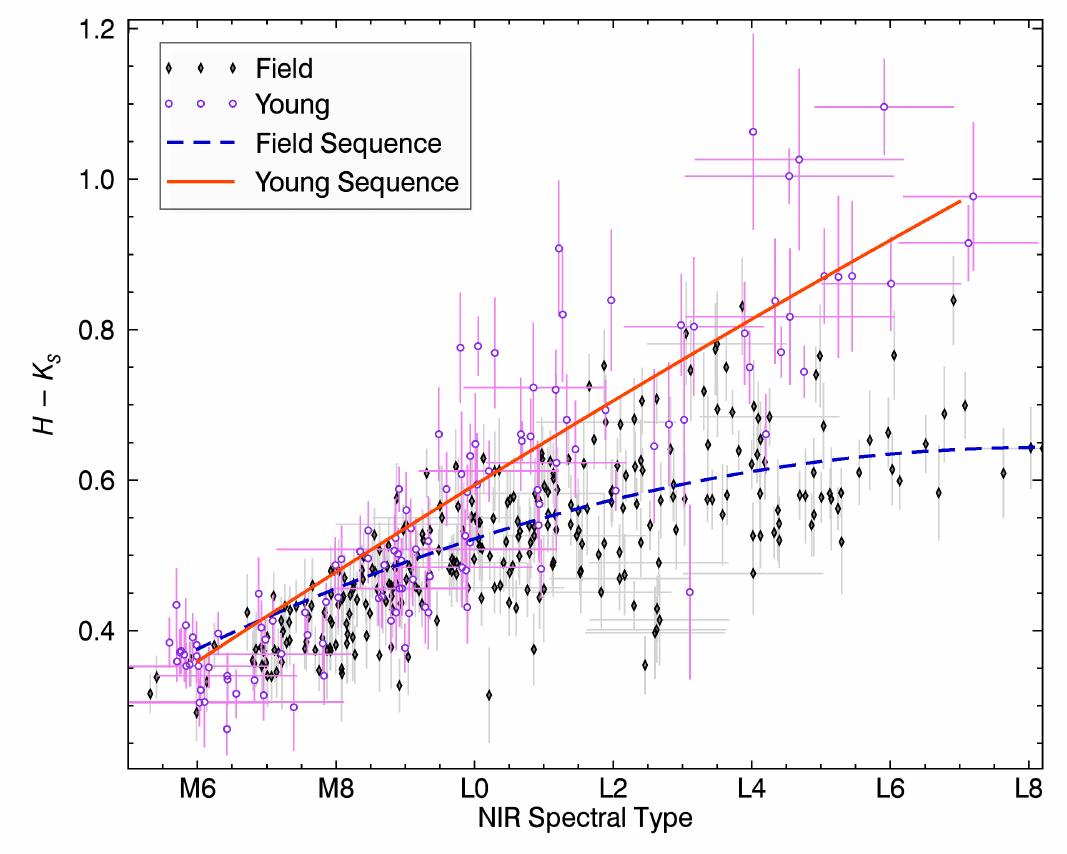}}
	\subfigure[$K_S-W1$ versus NIR Spectral Type]{\includegraphics[width=0.33\textwidth]{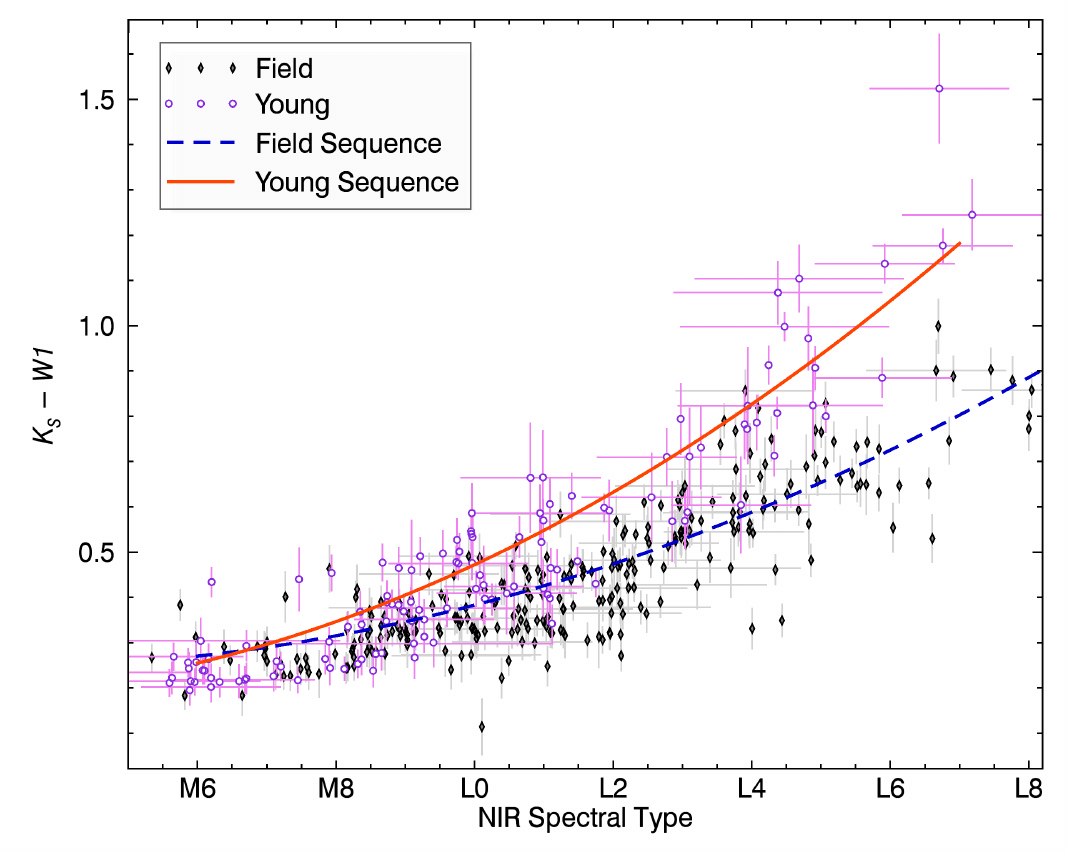}}
	\subfigure[$W1-W2$ versus NIR Spectral Type]{\includegraphics[width=0.33\textwidth]{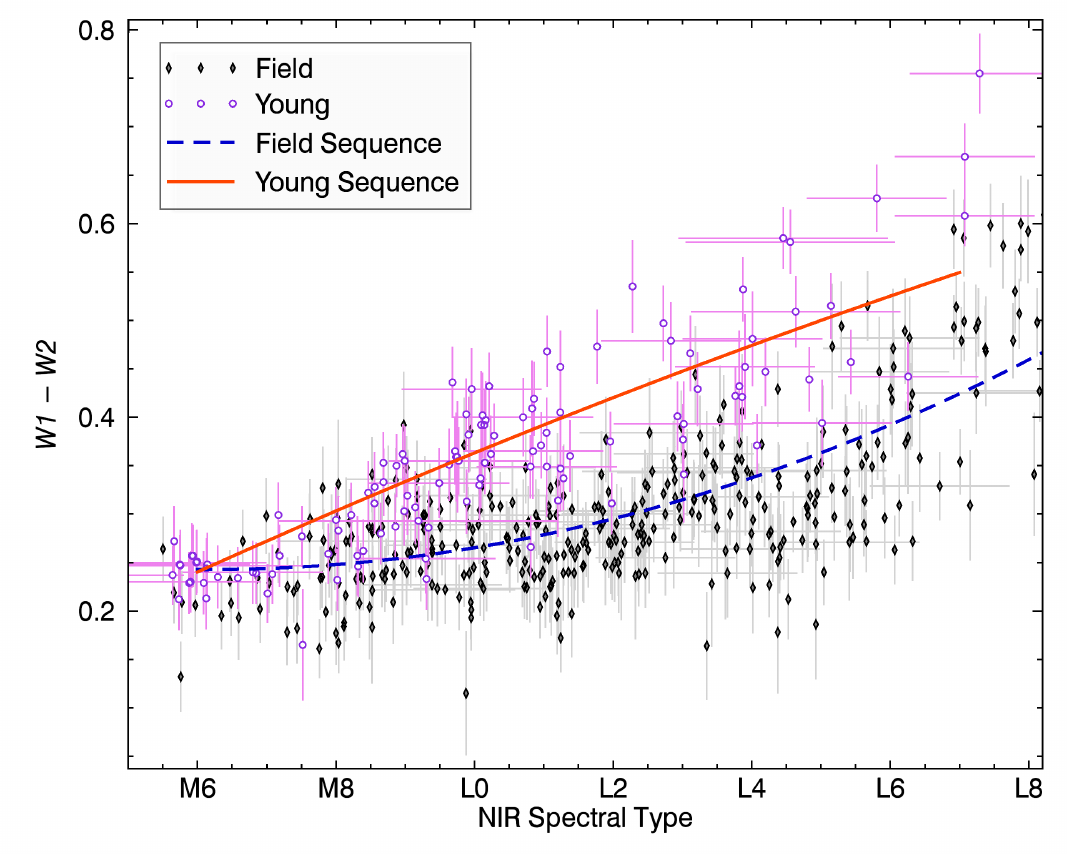}\label{fig:W1W2}}
	\caption{Absolute magnitude--NIR spectral type and color--NIR spectral type sequences for field (black diamonds) and young dwarfs (red dots when trigonometric distances were used, or purple circles when kinematic distances were used), as well as polynomial sequences (blue and orange lines, respectively) defined in Table~\ref{tab:coeff}. We used the kinematic distances obtained from the BANYAN~II tool (without photometry as input) to include low-gravity candidate members of YMGs that do not have a trigonometric distance measurement. Young dwarfs are generally brighter because of their inflated radii; however, thicker/higher dust clouds compete with this effect at spectral types L0--L7. Low-gravity L dwarfs are systematically redder than their field counterparts because of thicker/higher dust clouds in their photosphere.}
	\label{fig:CMDs}
\end{figure*}

\subsection{Updated Color-Magnitude Sequences for Young Low-Mass Stars and Brown Dwarfs}\label{sec:cmd}

%Figure : Color-Magnitude diagrams, batch 2
\begin{figure*}
	\centering
	\subfigure[$J-K_S$ versus NIR Spectral Type]{\includegraphics[width=0.33\textwidth]{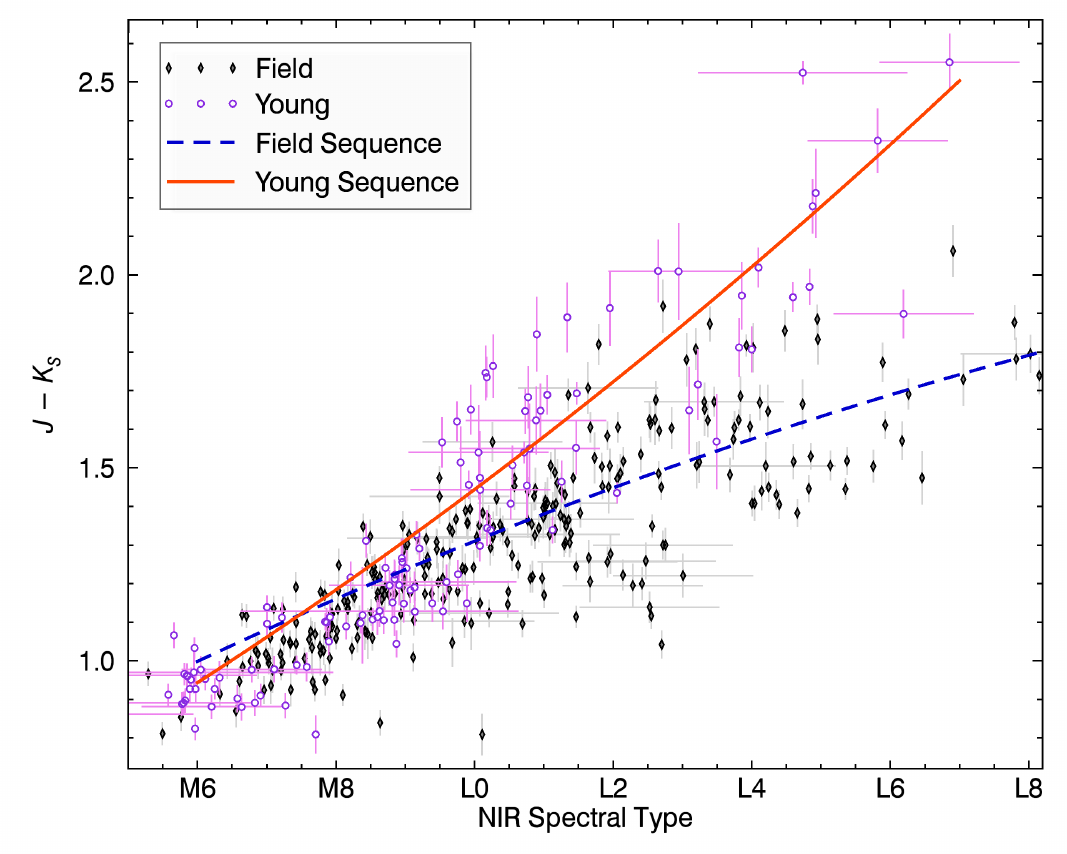}\label{fig:JK}}
	\subfigure[$J-W1$ versus NIR Spectral Type]{\includegraphics[width=0.33\textwidth]{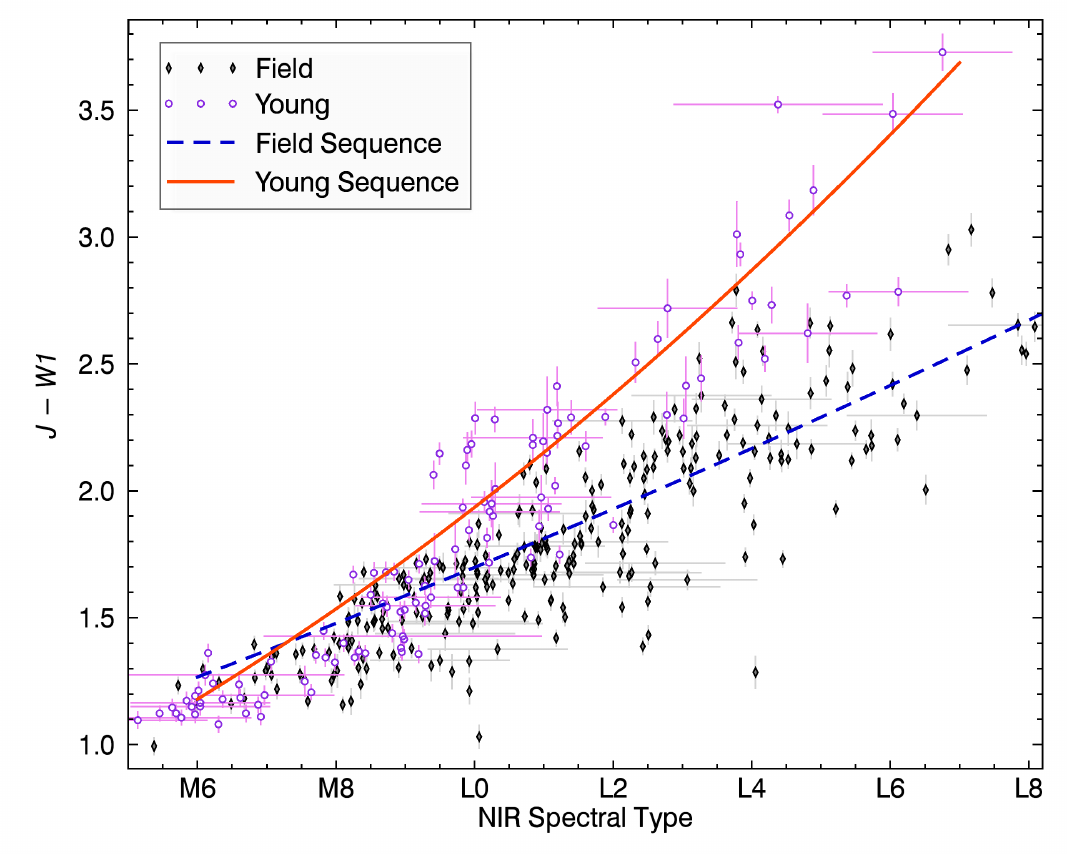}}
	\subfigure[$J-W2$ versus NIR Spectral Type]{\includegraphics[width=0.33\textwidth]{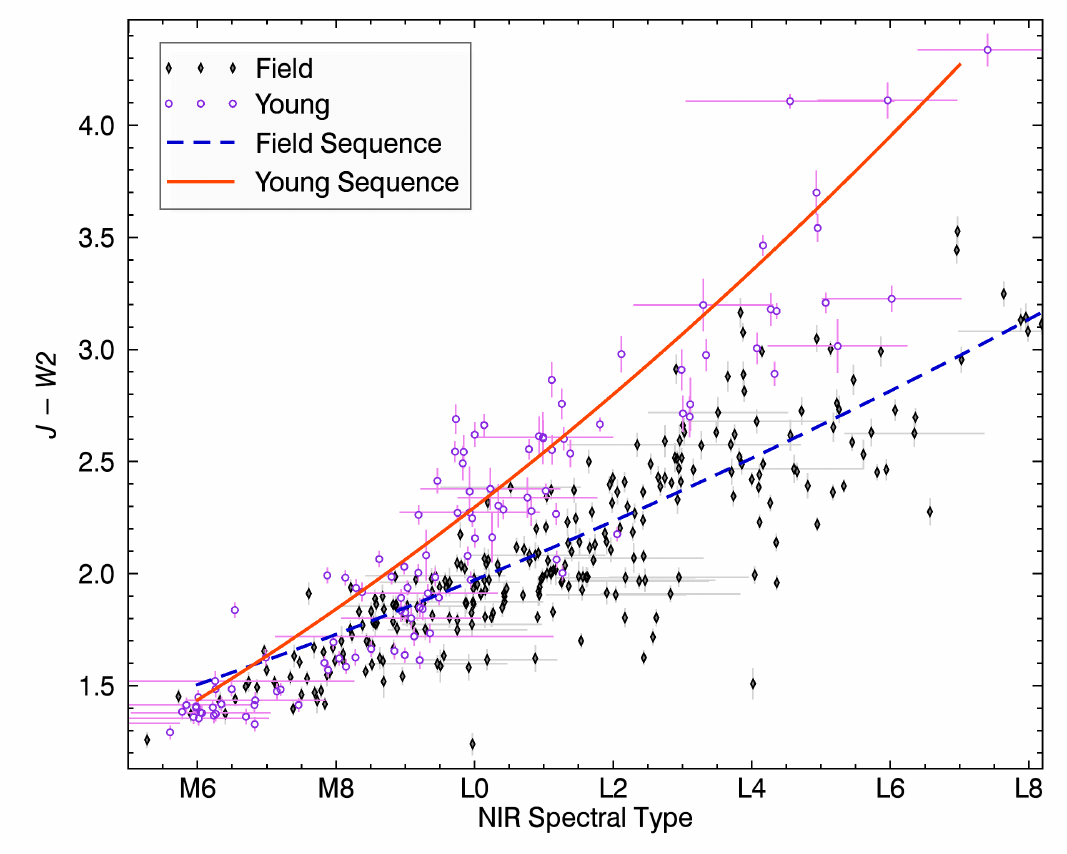}}
	\subfigure[$H-W1$ versus NIR Spectral Type]{\includegraphics[width=0.33\textwidth]{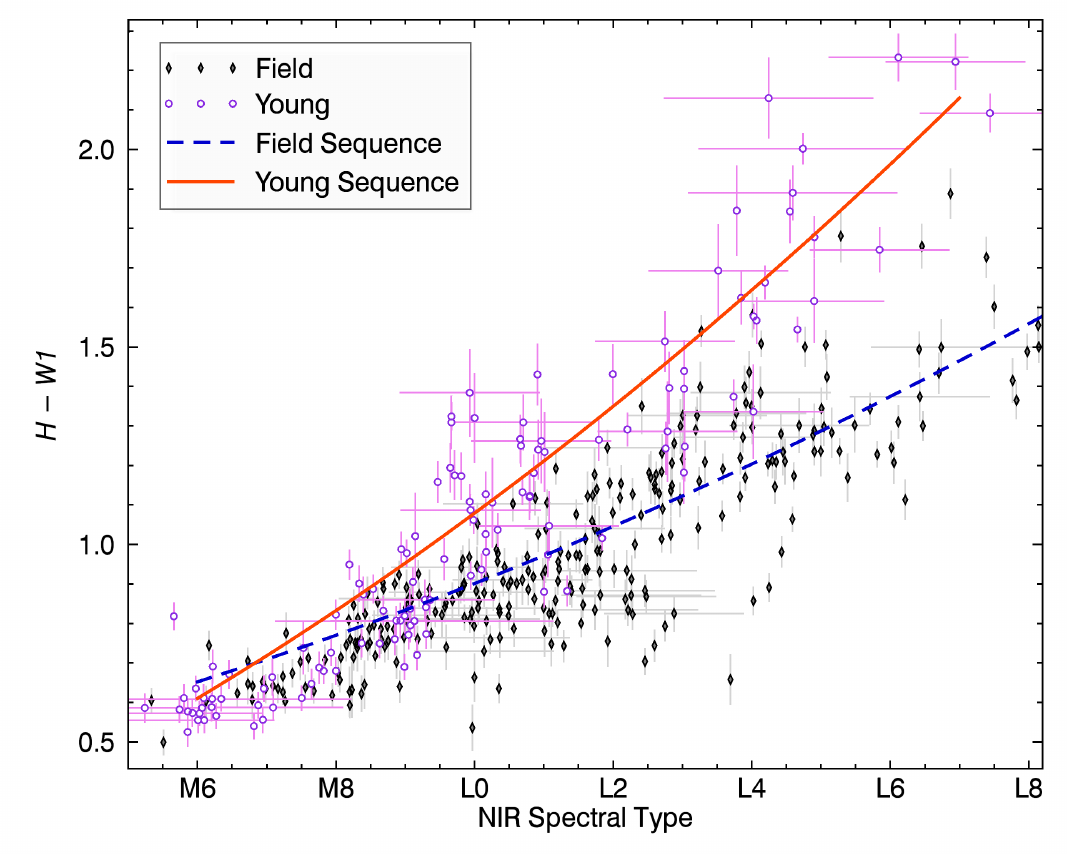}}
	\subfigure[$H-W2$ versus NIR Spectral Type]{\includegraphics[width=0.33\textwidth]{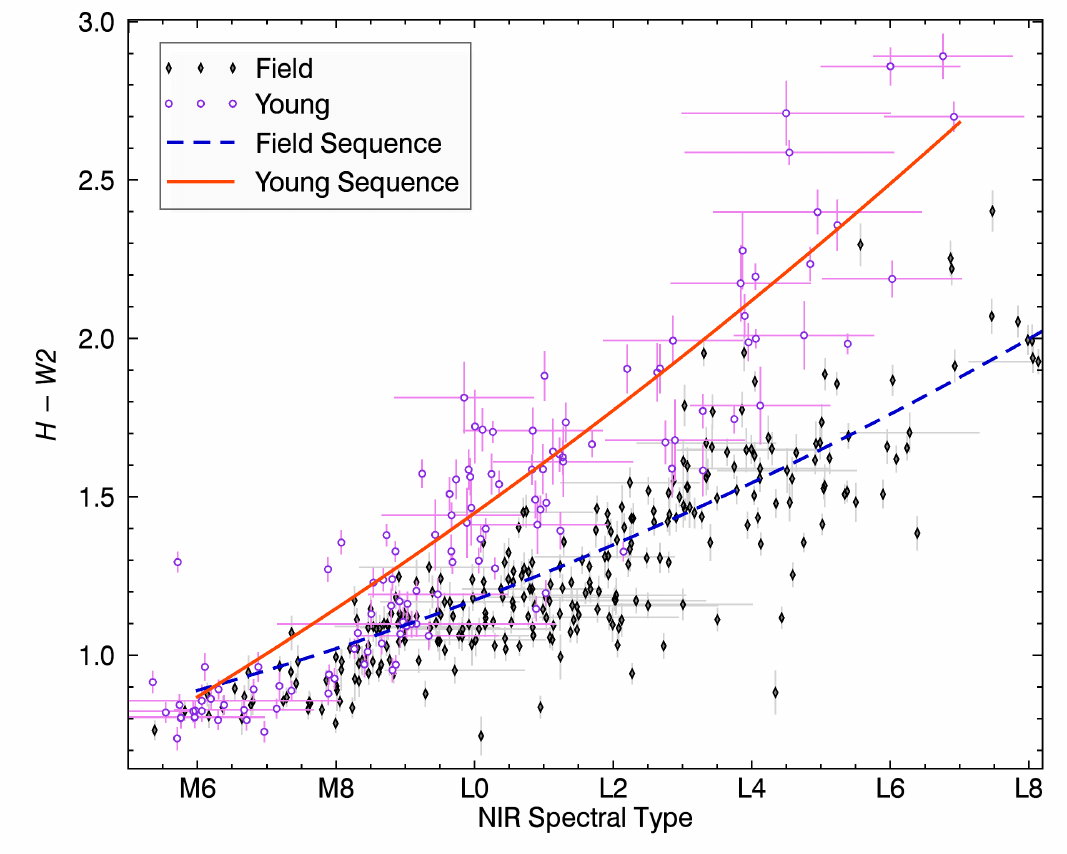}}
	\subfigure[$K_S-W2$ versus Spectral Type]{\includegraphics[width=0.33\textwidth]{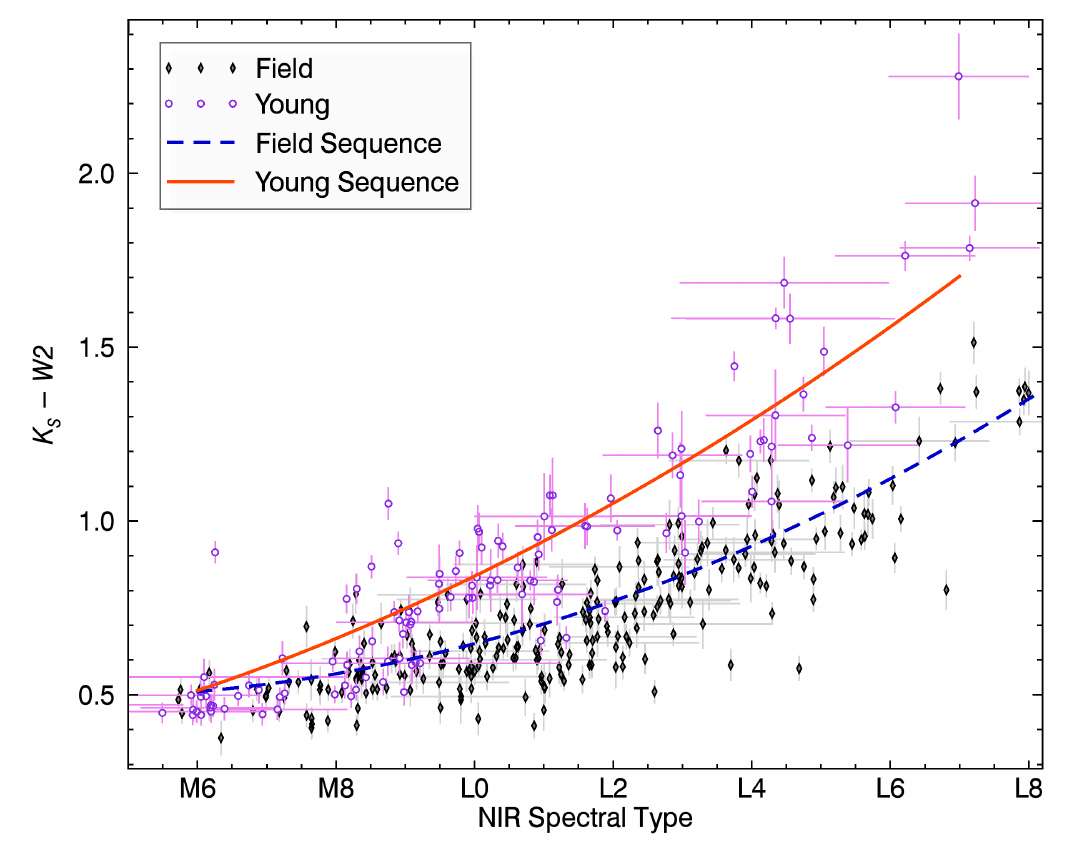}}
	\caption{Additional color--NIR spectral type sequences for young and field dwarfs as well as polynomial sequences defined in Table~\ref{tab:coeff}. The color scheme is identical to that of Figure~\ref{fig:CMDs} except that all young dwarfs are displayed with purple circles.}
	\label{fig:CC2}
\end{figure*}

%Figure : Crossing-point
\begin{figure}
	\centering
	\includegraphics[width=0.48\textwidth]{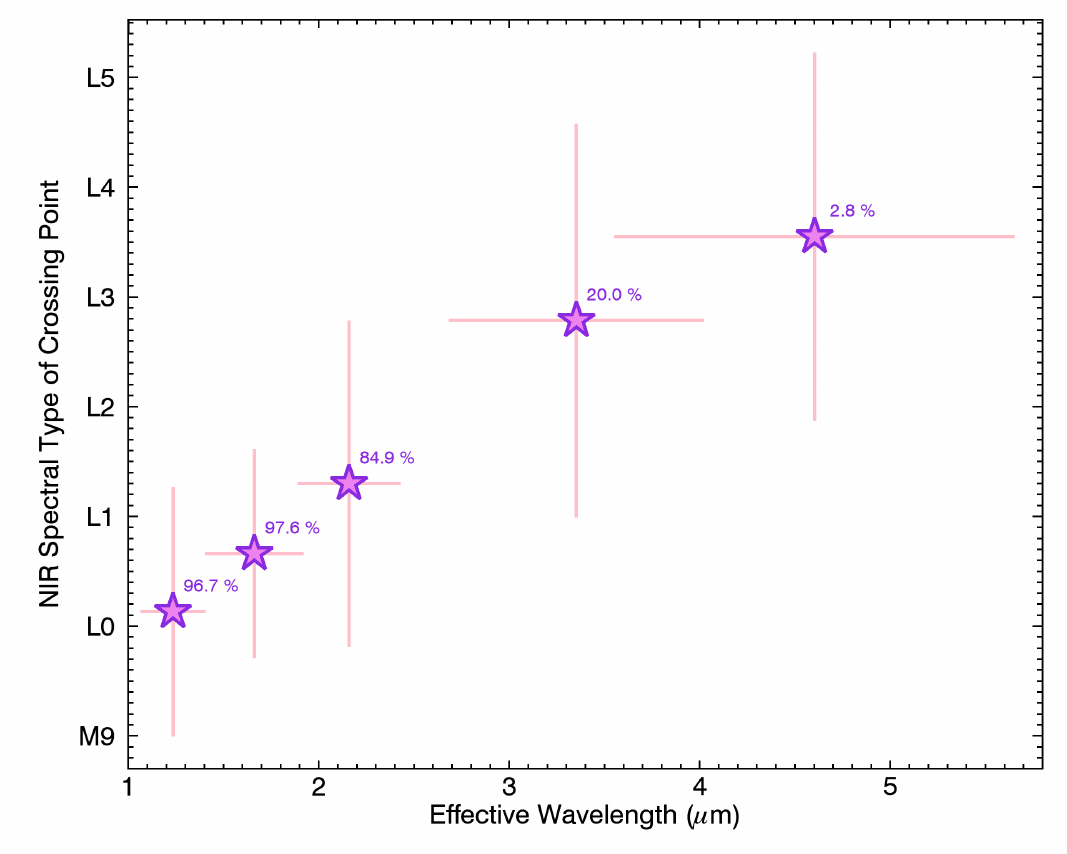}
	\caption{Spectral type at which the young and field absolute magnitude polynomial sequences cross (see Figure~\ref{fig:CMDs}), as a function of the effective wavelength in which each sequence is defined. Young dwarfs are systematically brighter than their field counterparts because of their inflated radii; however, dust clouds are thicker in the high atmosphere of young L dwarfs, which counter-balances this effect and causes the young sequence to cross the field sequence. The fraction of Monte Carlo steps where the sequences crossed is indicated next to a given data point; see text for more detail. Dust clouds are more opaque in the $J$ band ($\sim$\,1.2\,$\mu$m), hence the crossing point for this sequence happens at earlier spectral types. At longer wavelengths ($\sim$\,4.5\,$\mu$m), dust clouds do not have as much effect. This causes the sequences to cross less often and when they do, they cross at later spectral types.}
	\label{fig:crossing}
\end{figure}

%Figure : CMD.
\begin{figure*}
	\centering
	\includegraphics[width=0.995\textwidth]{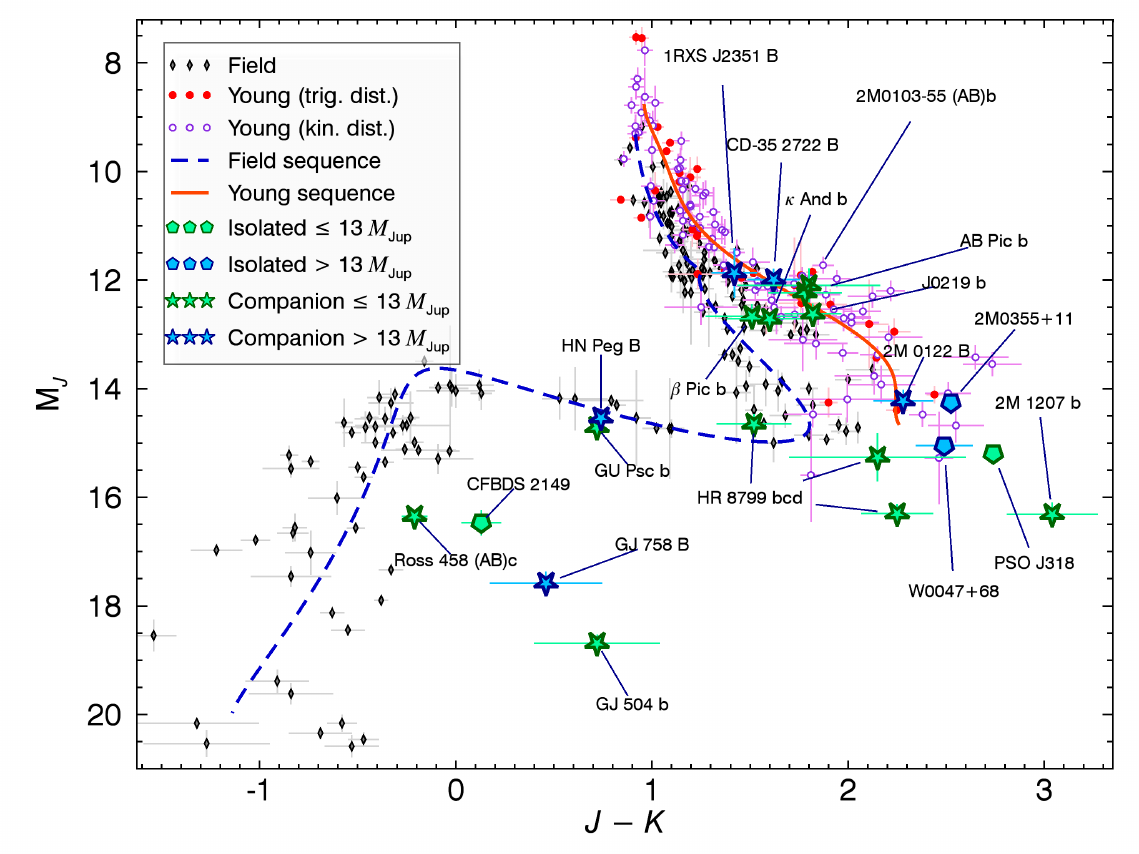}
	\caption{NIR CMD for young (red dots when trigonometric distances were used, or purple circles when kinematic distances were used) and field (black diamonds) low-mass stars and BDs. The young and field sequences are displayed with the dashed blue line and the solid orange-red line, respectively (see text for more detail). The young sequence is systematically shifted compared to field dwarfs because of the combined effect of larger radii and thicker/higher clouds. Blue stars indicate the positions of known low-mass BDs and directly imaged exoplanets (\citealp{2009ApJ...707L.123T, 2011ApJ...728...85J, 2013A&A...553L...5D, 2013ApJ...774...11K, 2012ApJ...753...14S, 2013ApJ...777..160B, 2014A&A...567L...9B, 2014ApJ...786...32M, 2013ApJ...774...55B, 2009ApJ...703..399L, 2013ApJ...763L..32C, 2005A&A...438L..29C, 2014MNRAS.439..372M, 2014ApJ...787....5N, 2008Sci...322.1348M, 2004A&A...425L..29C, 2010MNRAS.405.1140G, 2010Sci...329...57L, 2012A&A...548A..26D, 2002ApJ...567L..59G, 2007ApJ...654..570L, 2014ApJ...784...65B, 2011ApJ...729..139W, 2014ApJ...780L..30C, 2014ApJ...781...20K, 2015ApJ...806..254A}; and references therein).}
	\label{fig:JJK}
\end{figure*}

We complemented the list of all spectroscopically confirmed $\geq$ L0 dwarfs as of February 2014 \citep{2014PhDT........56M} and the DwarfArchives online library\footnote{\url{http://dwarfarchives.org}} with more recent discoveries, measurements of photometry in the literature and additional NIR photometry from a cross-match with \emph{2MASS} and \emph{WISE}, in order to build an up-to-date sequence of field dwarfs. This list currently contains $>$\,1800 published $\geq$ L0 low-mass stars and BDs\footnote{Publicly available at \url{www.astro.umontreal.ca/\textasciitilde gagne/listLTYs.php}}. We compiled a similar list of $>$\,8700 M6--M9 low-mass stars and BDs\footnote{Publicly available at \url{www.astro.umontreal.ca/\textasciitilde gagne/listMs.php}}. These two lists of dwarfs contain photometric data from articles referenced throughout the present work\footnote{In addition to the following references: \cite{2011AJ....141...54A, 2010ApJ...718L..38A, 2014ApJ...783...68B, 2006ApJ...637.1067B, 2008MNRAS.391..320B, 2013MNRAS.433..457B, 2013ApJ...776..126C, 2005AJ....130..337C, 2006AJ....132.1234C, 2011ApJ...743...50C, 2014ApJ...792..119D, 2008A&A...482..961D, 2008A&A...484..469D, 2014AJ....147...94D, 2013AJ....145....2F, 2002AJ....123.3409H, 2007A&A...466.1059K, 2007MNRAS.374..445K, 2012ApJ...753..156K, 2000ApJ...536L..35L, 2002ApJ...564..452L, 2009ApJ...695.1517L, 2010ApJ...720..252L, 2013ApJ...763..130L, 2015ApJ...799...37L, 2005A&A...440.1061L, 2007MNRAS.379.1423L, 2007AJ....134.1162L, 2007ApJ...669L..97L, 2010MNRAS.408L..56L, 2013ApJ...777...36M, 2013ApJS..205....6M, 2010A&A...524A..38M, 2013ApJ...762..119M, 1992AJ....103..638M, 2014A&A...567A...6P, 2008MNRAS.383..831P, 2008MNRAS.390..304P, 1999ApJ...522L..61S, 2013PASP..125..809T, 2003AJ....126..975T, 2014ApJ...796...39T, 2007A&A...474..653V, 2007MNRAS.381.1400W, 2003IAUS..211..197W, 2014A&A...568A...6Z}; and \cite{1995gcts.book.....V}}. In Figures~\ref{fig:CMDs} and \ref{fig:CC2}, we compare our updated population of known young low-mass stars and BDs to the field sequence in various spectral type-color and spectral type-absolute magnitude diagrams. We used data from the two aforementioned lists to build the photometric sequences. In the case of YMG candidate members that do not have a trigonometric distance measurement, we used the statistical distance from BANYAN~II, associated with the most probable YMG hypothesis. In each case, we calculated the error-weighted median sequence in bins of 1 subtype and adjusted a polynomial relation by minimizing the $\chi^2$ value. We list in Table~\ref{tab:coeff} the coefficients of these polynomial fits as well as the respective standard deviation of the data with respect to the best fit. We note that our field sequences are slightly redder than those derived from samples based on the Sloan Digital Sky Survey (\emph{SDSS}; \citealp{2000AJ....120.1579Y}) such as those presented by \cite{2008AJ....135..785W} and \cite{2015AJ....149..158S}. This is true because \emph{SDSS}-based surveys rely directly on spectra and are thus un-biased, whereas other surveys based on \emph{2MASS} and/or \emph{WISE} (e.g., \citealp{2003AJ....126.2421C,2008AJ....136.1290R,2011ApJS..197...19K}) perform a spectroscopic follow-up only on targets that were pre-selected from color cuts, which makes them biased towards detecting red objects more easily. Since \emph{2MASS}- and \emph{WISE}-based surveys dominating the population of L dwarfs identified in the literature, our field sequences are consequently redder than those based on \emph{SDSS} samples. This effect is also demonstrated in Figure 3 of \cite{2010AJ....139.1808S}.

The radii of young low-mass stars and BDs are inflated compared with old objects of the same spectral type. For this reason, it could be expected that young absolute magnitude sequences fall above the field sequences across all spectral types. However, starting at spectral type $\sim$\,L0, dust clouds form in the photosphere of BDs. Young BDs have a lower atmospheric pressure, which allows the formation of thick clouds higher in their atmosphere \citep{2006ApJ...639.1120K,2008ApJ...686..528L}. As a result, a fraction of the NIR light at $\sim$\,0.5--3\,$\mu$m gets redirected to longer wavelengths, causing young BDs to display similar absolute $J$ magnitudes to those of field BDs around spectral type $\sim$\,L0, as well as absolute $J$ magnitudes even fainter than those of field dwarfs at later spectral types \citep{2012ApJ...752...56F,2013AJ....145....2F,2013AN....334...85L,2014A&A...568A...6Z}. We could expect that this effect will eventually cease around spectral type T, where dust clouds fall below the photosphere. This has yet to be demonstrated, because there is only a very small number of young T dwarfs currently known (e.g., \citealp{2012A&A...548A..26D,2014ApJ...787....5N}). In Figure~\ref{fig:crossing}, we show the spectral type at which the young and field sequences cross as a function of spectral band. Horizontal error bars represent the effective width of the photometric filters and vertical error bars are drawn from a 10\,000-step Monte Carlo simulation, introducing noise in the data that is representative of photometric uncertainties and repeating the polynomial fit every time. Cases where the sequences do not cross are not included in the calculation of the median and standard deviation of the crossing points. We note that the fraction of Monte Carlo steps where the sequences cross significantly decreases at increasing wavelengths. This is explained by the fact that the photometric sequences become gradually disjointed in the spectral range considered; it is thus possible that in reality the sequences generally cross at spectral types $\geq$\,L7 (or not at all) in the $W1$ and $W2$ bands. This figure shows a clear correlation which indicates that flux is redistributed out to longer wavelengths in low-gravity dwarfs, a likely effect of the dust clouds (J. K. Faherty et al., in preparation). This is a known effect which is in part due to the larger opacity from the H$_2$O, CO and H$_2$ molecules at wavelengths larger than $\sim$\,1\,$\mu$m that are masking the effects of clouds \citep{2001ApJ...556..872A}. The BT-Settl isochrones \citep{2013MSAIS..24..128A,2003A&A...402..701B} do not reproduce this effect, as the young ($\leq$\,100\,Myr) and old ($\geq$\,1\,Gyr) isochrones do not cross in neither of the $J$, $H$ or $K_S$ bands over the range of effective temperatures that correspond to the M and L spectral types ($\sim$\,1300--3000\,K; \citealt{2009ApJ...702..154S}).

In Figure~\ref{fig:JJK}, we show a $M_J$ versus $J - K$ CMD in the Mauna Kea Observatories NIR filter system (MKO; \citealp{2002PASP..114..169S}) for low-gravity and field dwarfs. When MKO photometry was not available, we used \emph{2MASS} photometry with the conversion relations of \citeauthor{2004PASP..116....9S} (\citeyear{2004PASP..116....9S}; L and T dwarfs) and \citeauthor{2006MNRAS.373..781L} (\citeyear{2006MNRAS.373..781L}; M dwarfs). The combined effects of redder colors due to thicker/higher clouds \citep{2002ApJ...568..335M} and brighter absolute $J$ magnitude due to inflated radii cause a systematic shift of the low-gravity sequence to the right compared to the field sequence. This Figure brings into evidence the fact that the currently known population of young BDs does not reach a color reversal similar to the L/T transition of field dwarfs (at $J-K \sim 1.8$ and $M_J \sim 14.5$), corresponding to the temperature at which dust clouds fall below the photosphere \citep{2011ApJ...733...65B,2012ApJS..201...19D,2012ApJ...752...56F,2013AJ....145....2F,2013A&A...555A.107B,2013AN....334...85L,2014A&A...568A...6Z,2014ApJ...786...32M}. We chose this parameter space because a significant amount of data are available in these filters and it is very efficient in displaying this color reversal. It can be expected that a color reversal would eventually be reached for young dwarfs around the T spectral type, corresponding to cooler temperatures than the currently known population. The coolest known directly imaged young exoplanets and low-mass BDs (blue stars in Figure~\ref{fig:JJK}) tentatively hint at such a color reversal.

Since $J - K$ and $M_J$ are generally correlated for a given spectral type, the $(J - K)$--spectral type and $M_J$--spectral type relations listed in Table~\ref{tab:coeff} are not the best representation for the low-gravity and field sequences in this CMD diagram. In the case of the young sequence, the absence of a color reversal allowed us to simply fit a polynomial sequence to the young dwarfs directly in the $M_J$--$(J-K)$ space; however, the field sequence cannot be represented by a simple polynomial relation across the M6--T9 range. We used a Markov Chain Monte Carlo algorithm to construct a parametrized polynomial sequence that fits the field sequence across its complete spectral range. We started from a parametrized equation obtained from the combination of the $J - K$ and $M_J$ polynomial relations described in Table~\ref{tab:coeff}, and allowed the eight coefficients of each dimension to vary such that the sequence minimizes the quadrature sum of the bi-dimensional distance of all individual field dwarf positions in the CMD diagram relative to their error bars. This results in a parametrized sequence that describes $J - K$ and $M_J$ as a function of the parametric variable $\lambda$. Larger values of $\lambda$ correspond to later spectral types on average, but no relation between $\lambda$ and spectral types can be provided as the field sequence is a parametric equation that does not assign a $\lambda$ value to individual data points. We obtain:

\begin{align}
	(M_J)_{\mathrm{Young}} =\ &1.61 \times 10^{3} - 8.92 \times 10^{2}\ (J - K)\notag\\ + &2.04 \times 10^{2}\ (J - K)^2 - 2.46 \times 10^{1}\ (J - K)^3\notag\\ + &1.65\ (J - K)^4 - 5.87 \times 10^{-2}\ (J - K)^5\notag\\ + &8.59 \times 10^{-4}\ (J - K)^6
\end{align}
\begin{align}
(J - K)_{\mathrm{Field}} =\ &1.98 \times 10^{1} - 1.55 \times 10^{1}\ \lambda\notag\\ + &5.09\ \lambda^2 - 8.76 \times 10^{-1}\ \lambda^3\notag\\ + &8.68 \times 10^{-2}\ \lambda^4 - 5.12 \times 10^{-3}\ \lambda^5\notag\\ + &1.76 \times 10^{-4}\ \lambda^6 - 3.27 \times 10^{-6}\ \lambda^7\notag\\ + &2.53 \times 10^{-8}\ \lambda^8
\end{align}
\begin{align}
(M_J)_{\mathrm{Field}} =\ &-5.61 \times 10^{1} + 3.73 \times 10^{1}\ \lambda\notag\\ - &8.45\ \lambda^2 + 1.04\ \lambda^3\notag\\ - &7.64 \times 10^{-2}\ \lambda^4 + 3.56 \times 10^{-3}\ \lambda^5\notag\\ - &1.06 \times 10^{-4}\ \lambda^6 + 1.84 \times 10^{-6}\ \lambda^7\notag\\ - &1.43 \times 10^{-8}\ \lambda^8
\end{align}
\\%TMP
where the young sequence is valid in the range ${8.8 \leq M_J \leq 14.7}$ and the field sequence is valid in the range ${4.5 \leq \lambda \leq 28.5}$ (i.e., ${J \geq -1.1}$ and ${9.3 \leq M_J \leq 19.9}$).

We note that the NIR colors of young BDs discovered in the BASS and LP-BASS surveys are likely affected by a form of the confirmation bias, in the sense that we specifically looked for red objects in our survey (see Paper~V). Hence, this new photometric data should not be taken as additional evidence that young BDs are redder than field BDs. Reinforcing this result would require looking for signs of low gravity in a sample of BDs that were selected independently of their photometric colors. Directly imaged young planets and brown dwarf companions do not suffer from this potential bias however, and their colors seem consistent with those of isolated young BDs \citealp{2004A&A...425L..29C,2005A&A...438L..29C,2011ApJ...733...65B,2013A&A...555A.107B,2013A&A...553L...5D,2013ApJ...774...55B,2014ApJ...780L..30C}. This might be an indication that our confirmation bias is not significant.

%Figure : Age calibrations
\begin{figure*}[p]
	\centering
	\subfigure[Mean \ion{K}{1} EW]{\includegraphics[width=0.495\textwidth]{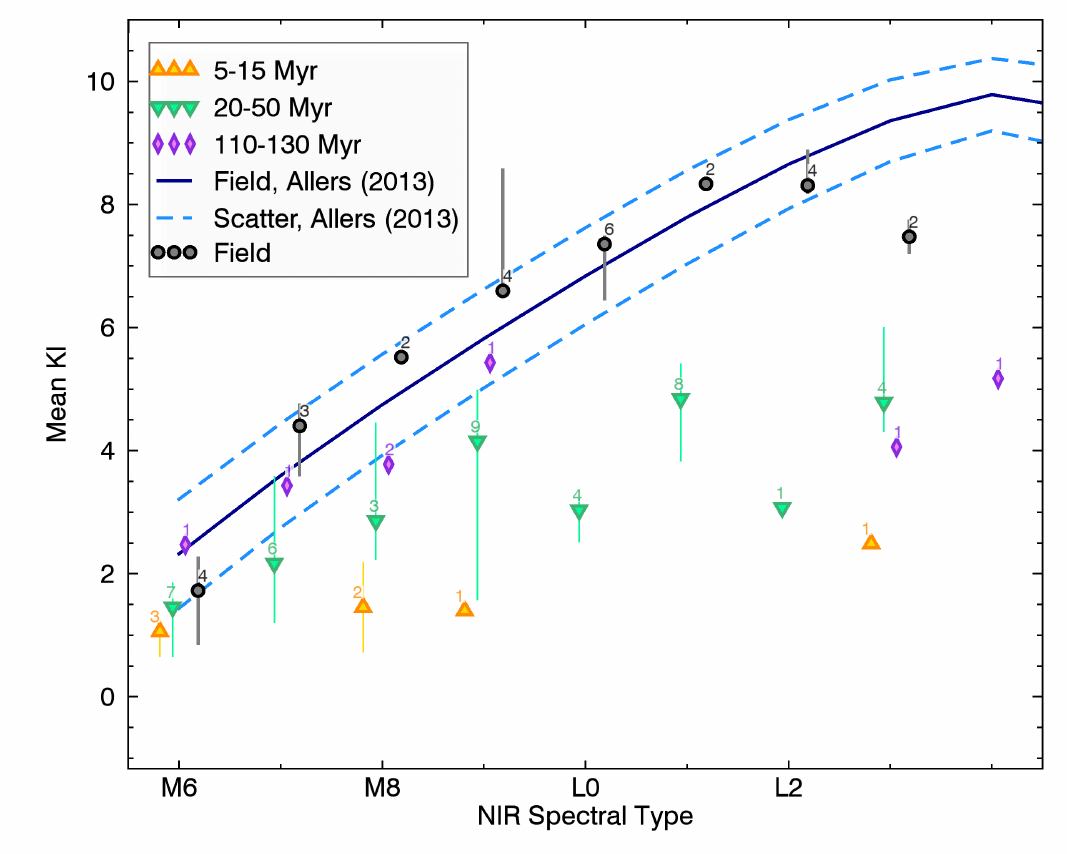}}
	\subfigure[\cite{2013ApJ...772...79A} $R \sim 75$ mean gravity score]{\includegraphics[width=0.495\textwidth]{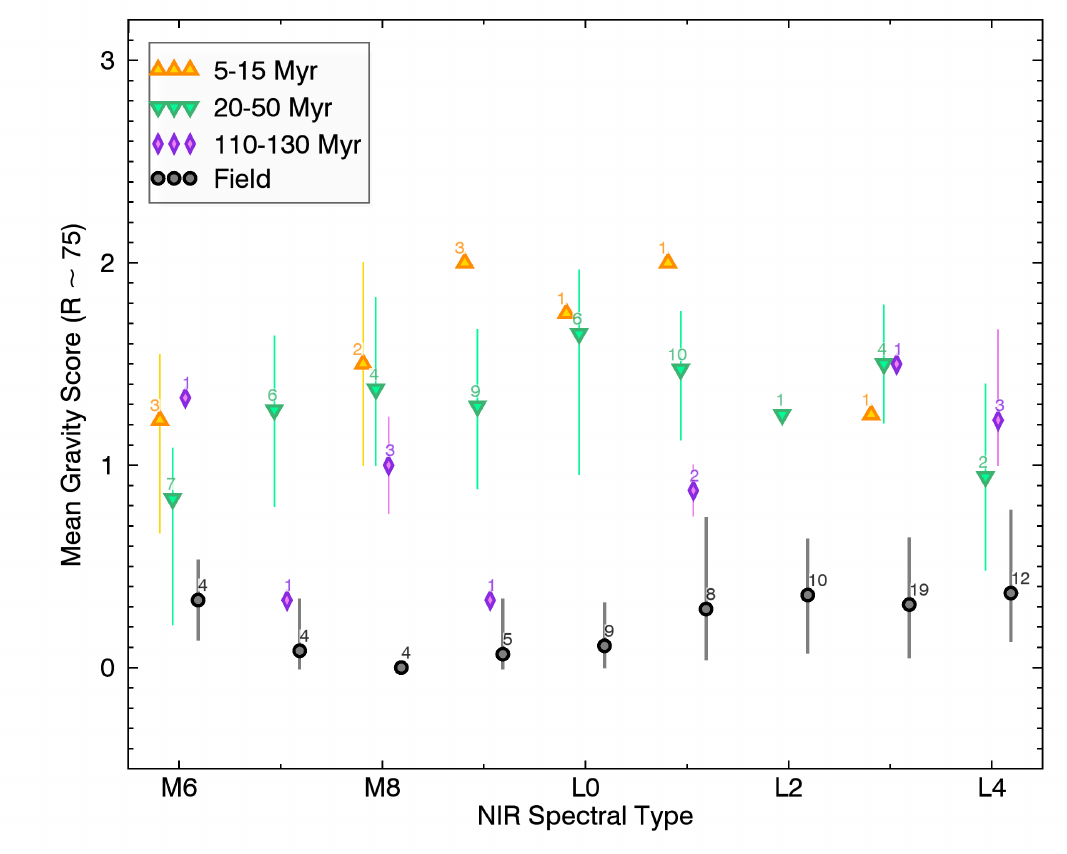}}
	\subfigure[\cite{2013ApJ...772...79A} $R \sim 750$ mean gravity score]{\includegraphics[width=0.495\textwidth]{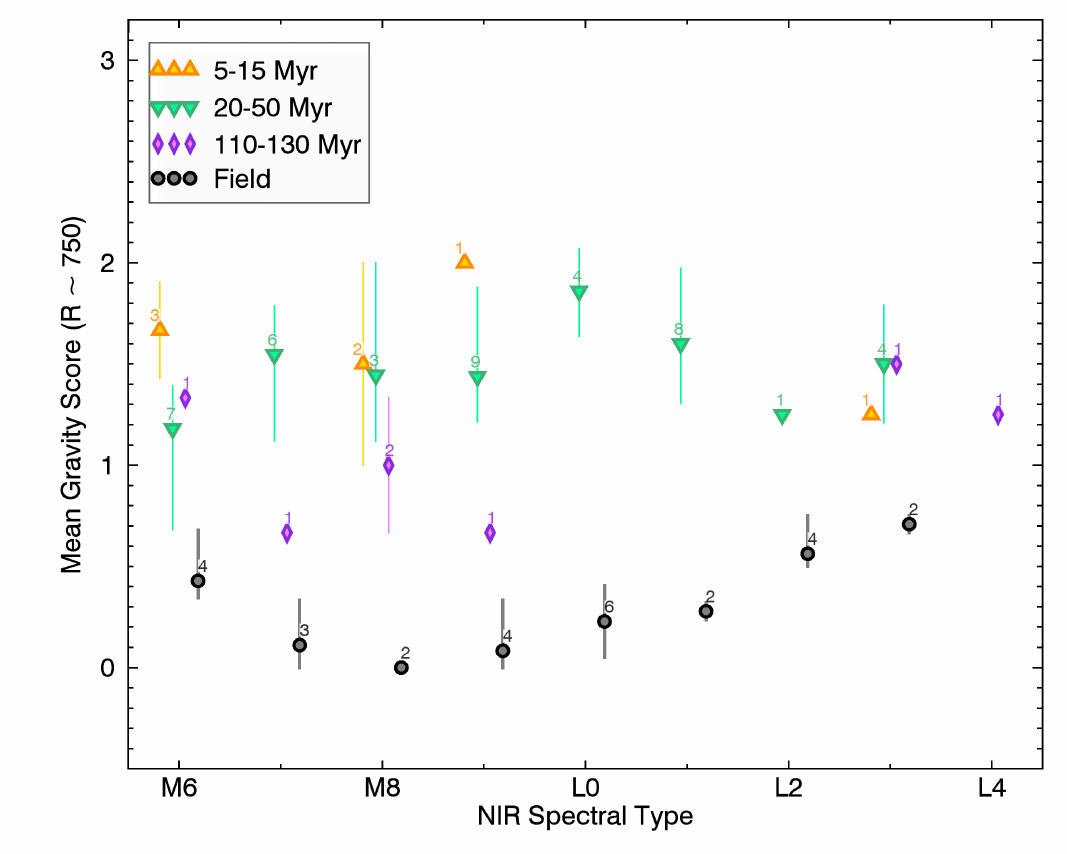}}
	\subfigure[\cite{2013MNRAS.435.2650C} H$_2(K)$\label{fig:agecalib2k}]{\includegraphics[width=0.495\textwidth]{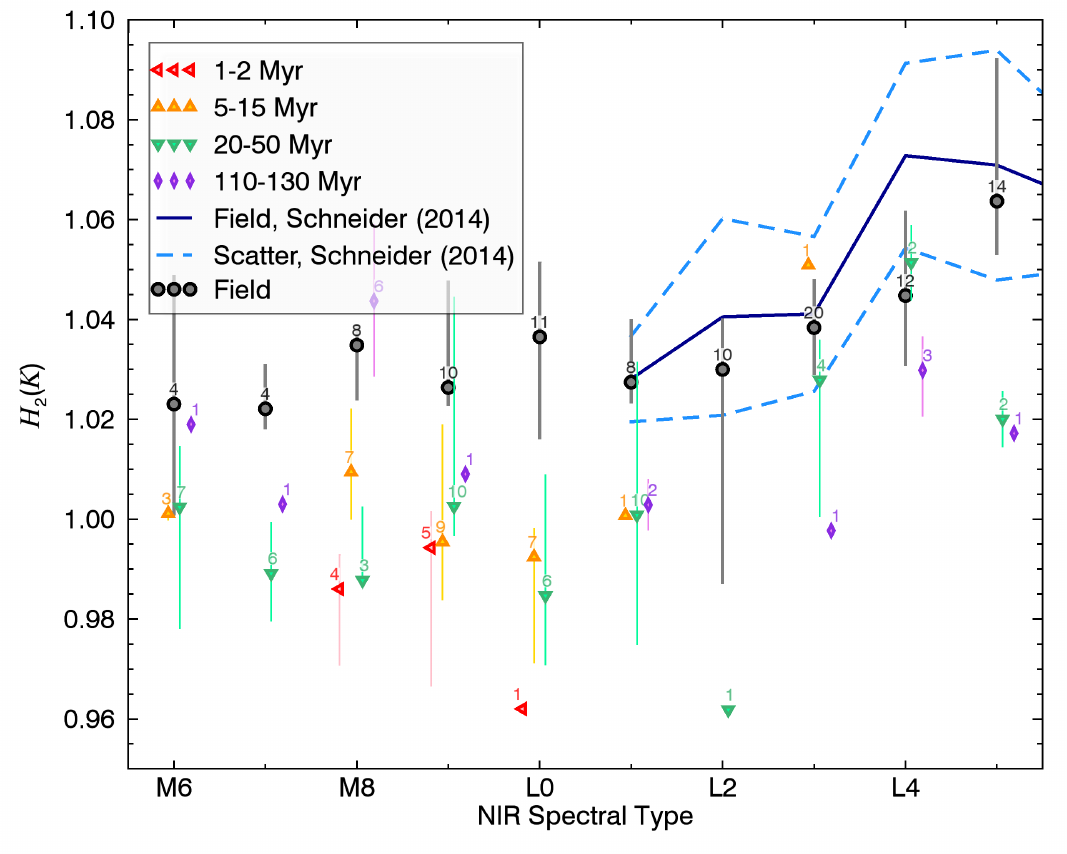}}
	\caption{Spectroscopic indices versus NIR spectral type for YMG candidates of distinct ages in our sample, binned by spectral type (see legends for color coding). We find that the mean EW of $J$-band \ion{K}{1} and the mean gravity score defined by \cite{2013ApJ...772...79A} seems to correlate with age. However, we do not see a clear correlation in the case of the H$_2$($K$) index in the 1--130\,Myr range.}
	\label{fig:agecalib}
\end{figure*}

\subsection{An Updated Investigation on the Age Dependence of Spectroscopic Indices}\label{sec:ageinv}

Since they are the only BDs with a well calibrated age, members of YMGs provide the exciting opportunity of creating a spectroscopic age calibration applicable to all young BDs. Using Pleiades members and the fact that known low-gravity BDs were located away from star-forming regions, \cite{2008ApJ...689.1295K} and \cite{2009AJ....137.3345C} estimated that the very low-gravity ($\gamma$) and intermediate gravity ($\beta$) classifications likely correspond to $\sim$\,10\,Myr and $\sim$\,100\,Myr, respectively. \cite{2013ApJ...772...79A} extended this investigation by using a restrained sample of 25 M6--L5 dwarf members of young associations. They found that very low-gravity ($\gamma$) and intermediate-gravity ($\beta$) dwarfs likely correspond to ages of $\sim$\,10--30\,Myr and $\lesssim$\,200\,Myr; however, they note that BDs with ages older than $\sim$\,30\,Myr, such as the $\sim$\,120\,Myr ABDMG member 2MASS~J03552337+1133437, can display very strong signs of low-gravity that correspond to the very low gravity ($\gamma$) classification.

We used our updated sample of low-gravity candidate members of YMGs to investigate this further. We inspected various spectroscopic index--spectral type relations of candidate members of different YMGs to identify any systematic correlation with age. We assigned the age of the most probable YMG to our candidates, while rejecting any candidate with ambiguous membership (Table~\ref{tab:sptclass}). We found that the strongest correlations with age in the $\sim$\,10--130\,Myr range resulted from: (1) the mean value of the EW of the three \ion{K}{1} doublets at $1.169$\,$\mu$m, $1.177$\,$\mu$m and $1.253$\,$\mu$m; and (2) the mean gravity score defined by \cite{2013ApJ...772...79A}. The resulting sequences are presented in Figure~\ref{fig:agecalib}. Even though they do correlate with age on average, the scatter is too large to allow a precise determination of the age of an individual system from spectroscopic indices alone. We find that the H$_2$($K$) index defined by \cite{2013MNRAS.435.2650C} does not seem to correlate significantly with age in the 1--130\,Myr range. Our results seem to be in contradiction with the findings of \cite{2013MNRAS.435.2650C} that the H$_2$($K$) index is sufficient to differentiate between objects from populations of $\sim$\,1--2\,Myr, $\sim$\,3--10\,Myr and field dwarfs in the M8--L0 range: we observe an overlap of the typical values for H$_2$($K$) in populations of $\sim$\,1--2\,Myr and $\sim$\,5--15\,Myr. However, our results are consistent with H$_2$($K$) being a good gravity-sensitive index, as it discriminates between the field population and $\lesssim$\,100\,Myr dwarfs for spectral types in the M6--L1 range, or $\lesssim$\,130\,Myr for L2--L6. It is possible that interlopers from other young associations not considered in BANYAN~II contaminate our sample, which would introduce noise in these relations. A full RV and parallax follow-up of the candidates presented here will be needed to assess this.

%Figure : Example Fitting.
\begin{figure*}
	\centering
	\subfigure[2MASS~J0126--5505 (M6\,$\gamma$)]{\includegraphics[width=0.43\textwidth]{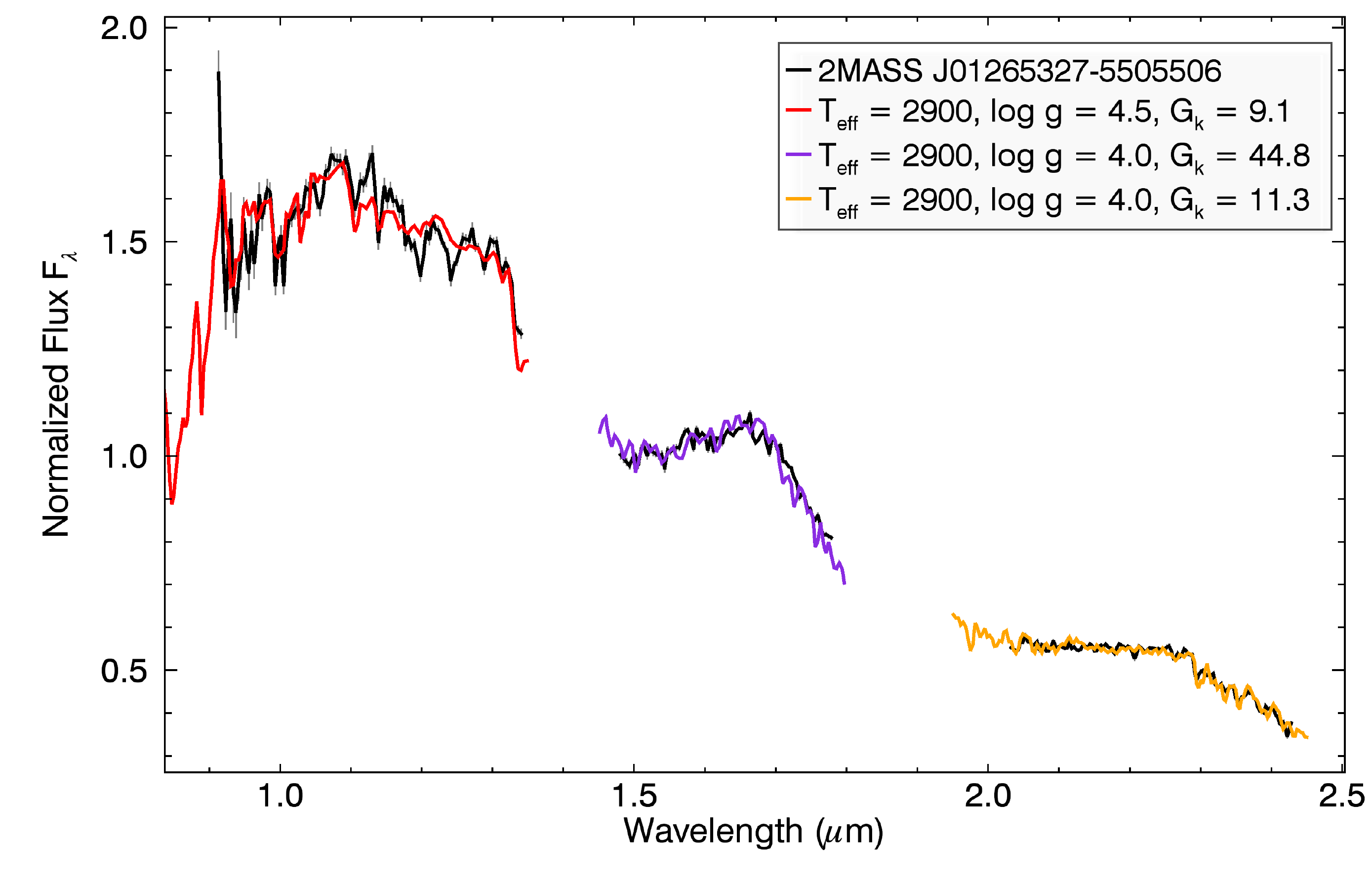}}
	\subfigure[2MASS~J0407+1546 (Field L3)]{\includegraphics[width=0.43\textwidth]{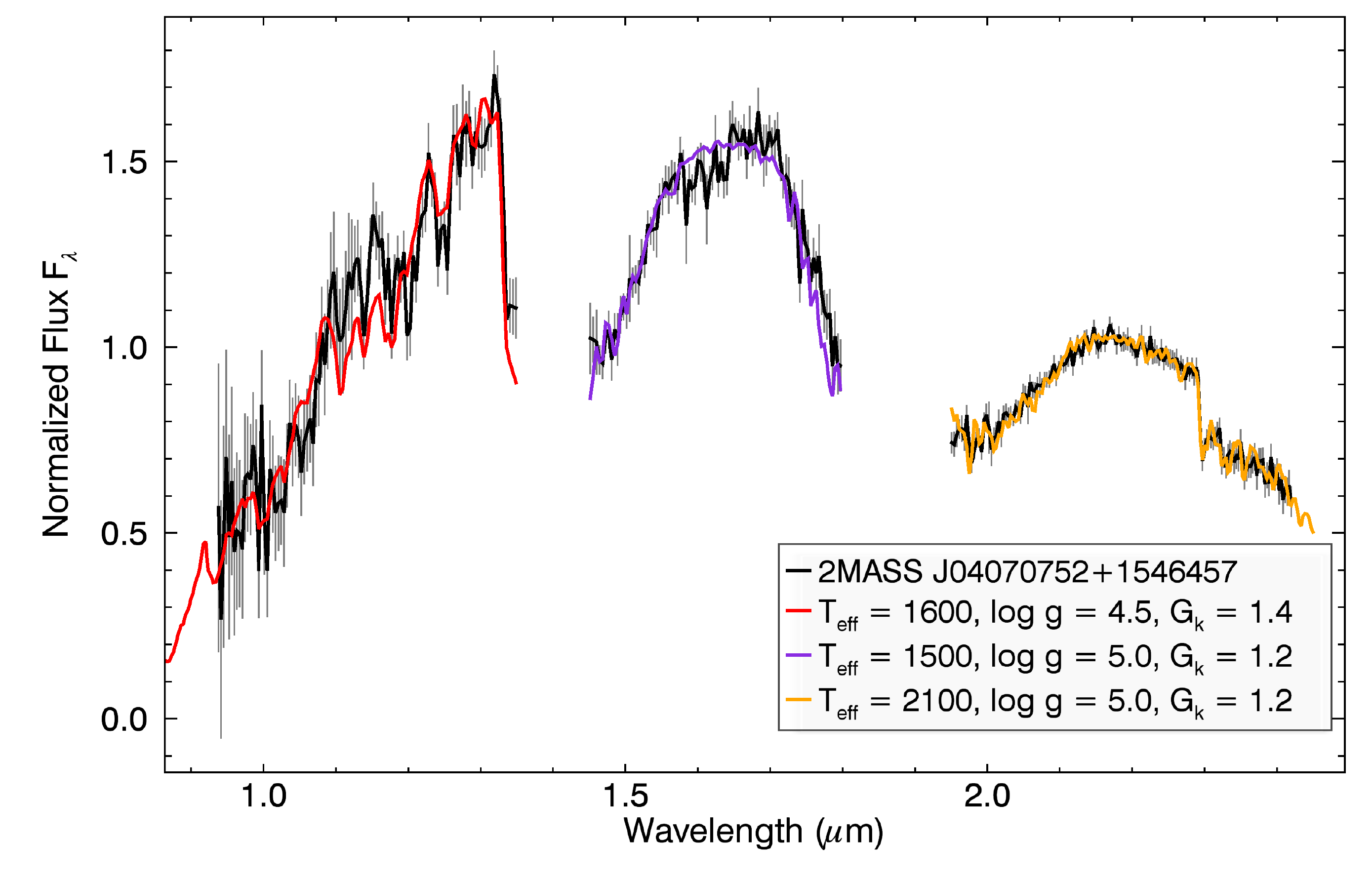}}
	\subfigure[2MASS~J0337--1758 (Field L4)]{\includegraphics[width=0.43\textwidth]{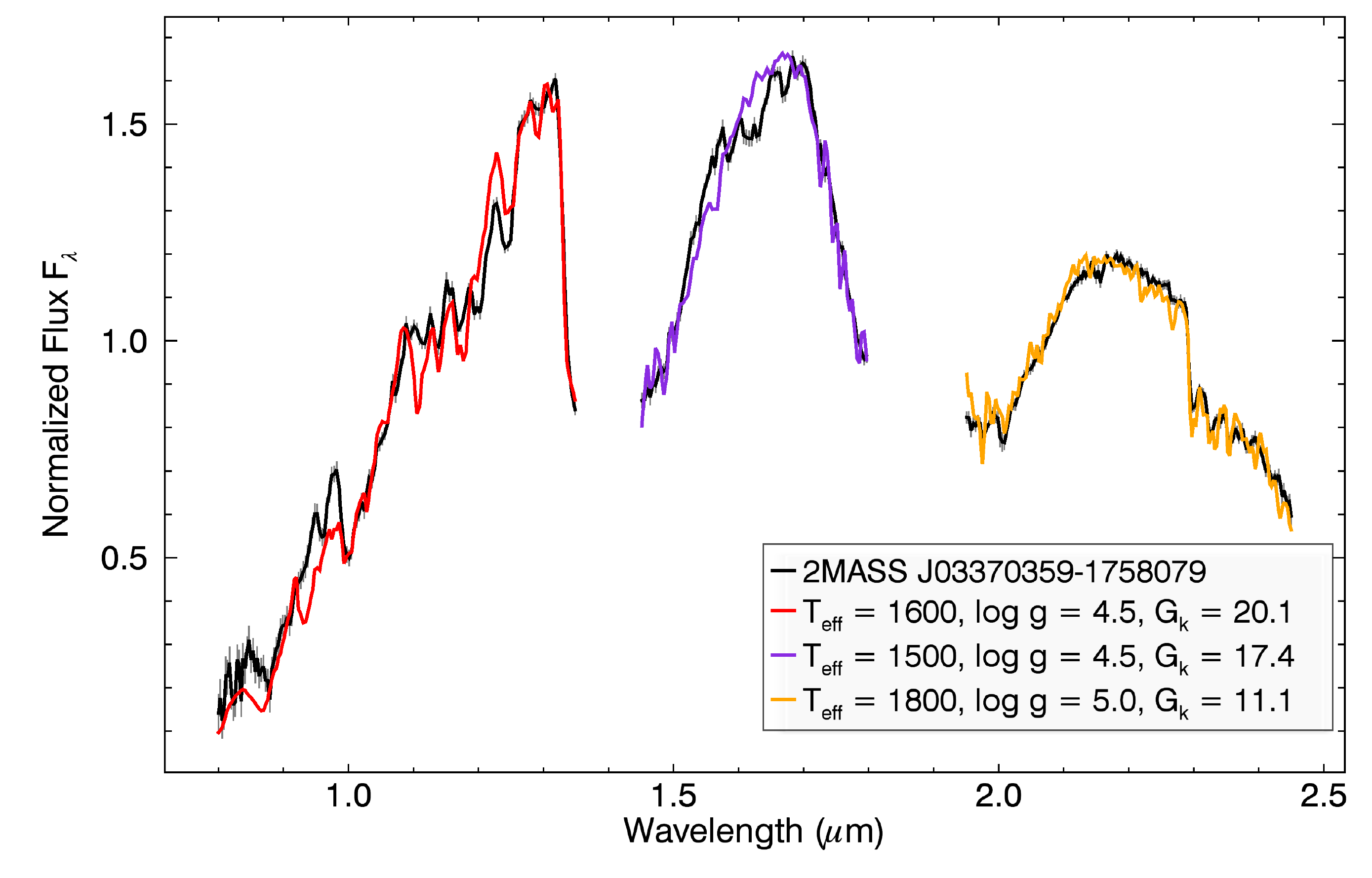}}
	\subfigure[2MASS~J0030--1450 (L4--L6\,$\beta$)]{\includegraphics[width=0.43\textwidth]{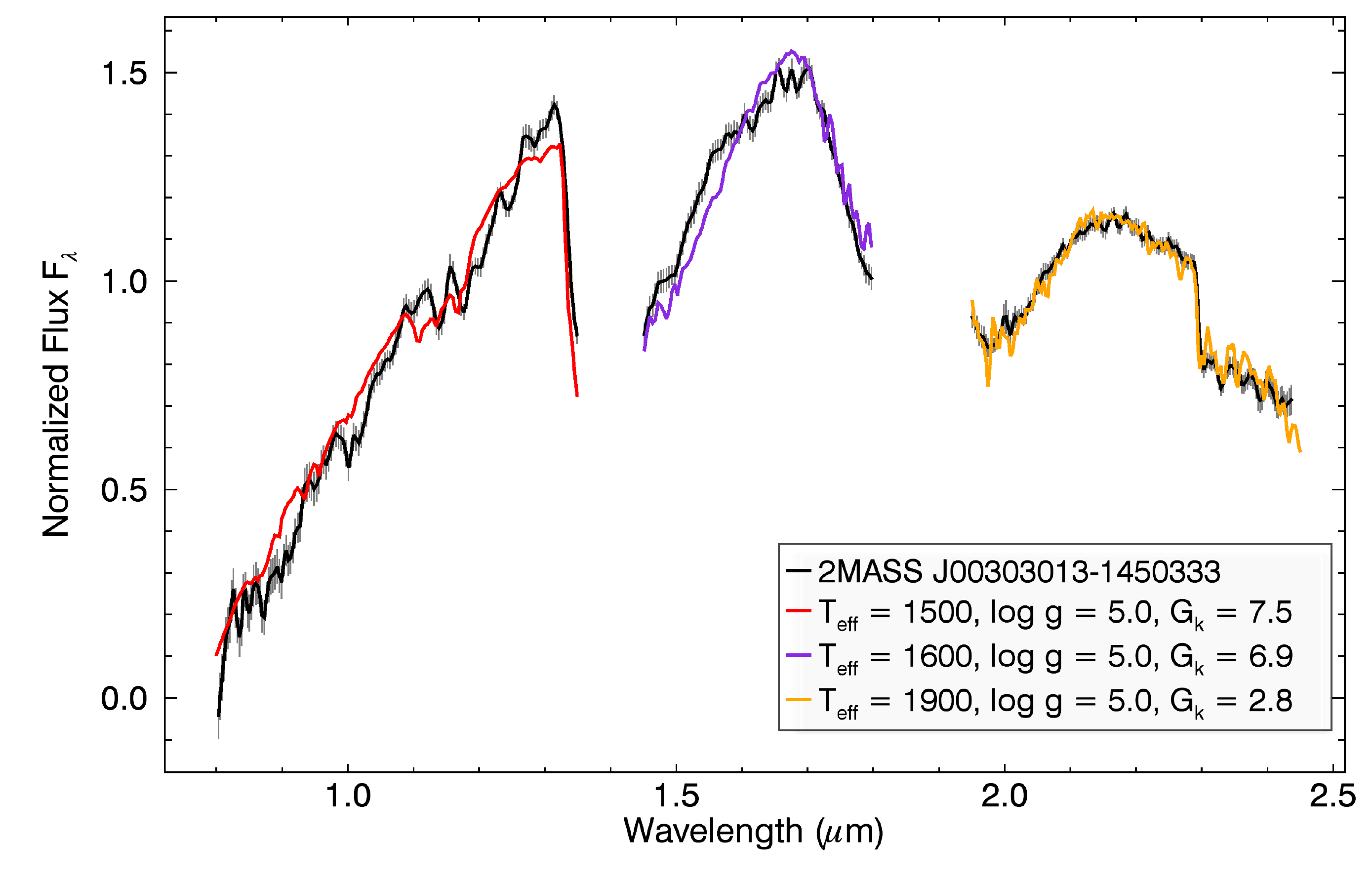}}
	\caption{Best-fitting BT-Settl atmosphere models for typical field and low-gravity BDs (thick, black line and gray error bars). The $zJ$ (red line), $H$ (purple line) and $K$ (yellow line) dilution factors were adjusted separately so that the goodness-of-fit is optimized (see text). We observe that BT-Settl models are generally unable to reproduce the $zJ$ bands or the $H$-band dip at $\sim$\,1.6\,$\mu$m that is due to FeH absorption.}
	\label{fig:spectra_fitting1}
\end{figure*}

\subsection{Model Comparison}\label{sec:models}

We used our sample of 86 new low-gravity M6--L5 dwarfs supplemented with 39 low-gravity and 131 field M6--L9 dwarfs from \cite{2013ApJ...772...79A} and the SpeX Prism Spectral Libraries to investigate the physical properties of our sample of young dwarfs, using BT-Settl atmosphere models \citep{2013MSAIS..24..128A,2003A&A...402..701B}. In Section~\ref{sec:btsettl}, we focus on effective temperatures and surface gravities obtained from a comparison of our NIR spectra with atmosphere models. In Section~\ref{sec:evmod}, we focus on the mass and radii that are obtained from a comparison of our photometry with evolution models.

\subsubsection{BT-Settl Atmosphere Models}\label{sec:btsettl}

%Figure : SpT-TEFF Relations
\begin{figure*}[p]
	\centering
	\subfigure[Adopted \Teff\ versus NIR spectral type (using \emph{WISE} data)]{\includegraphics[width=0.495\textwidth]{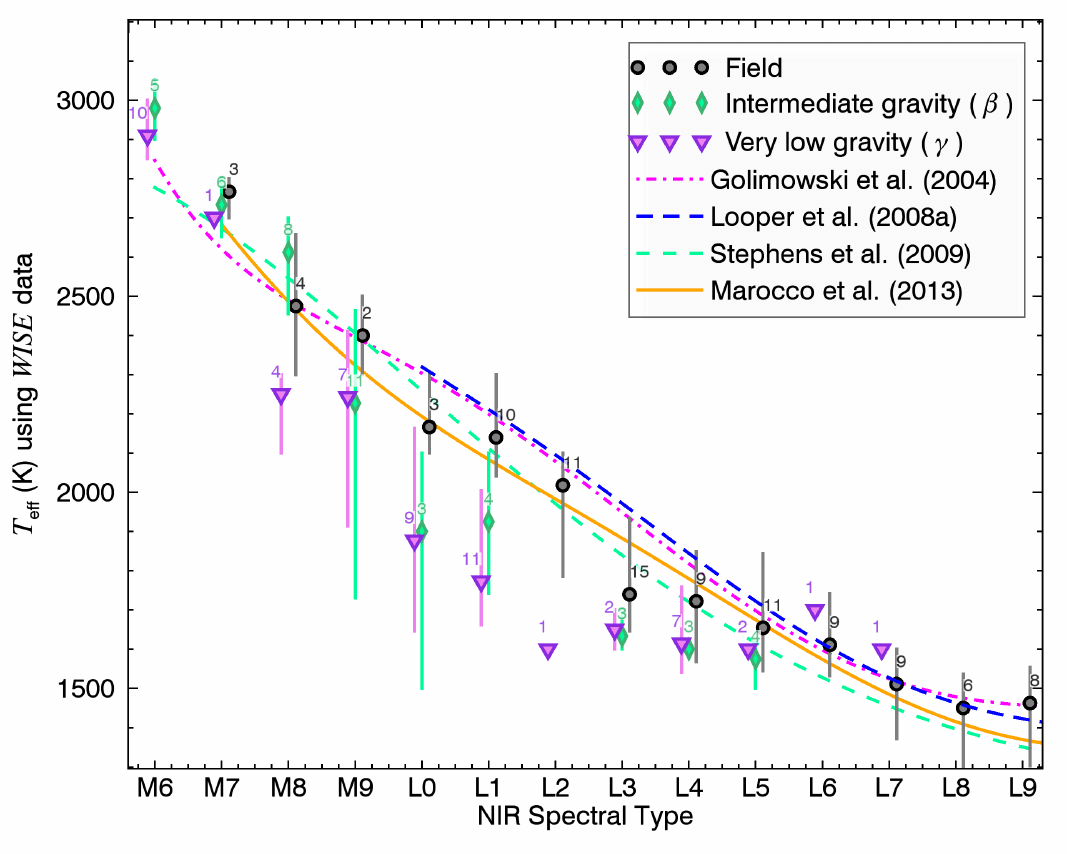}\label{fig:spt_teffk}}
	\subfigure[$\log g$ versus NIR spectral type]{\includegraphics[width=0.495\textwidth]{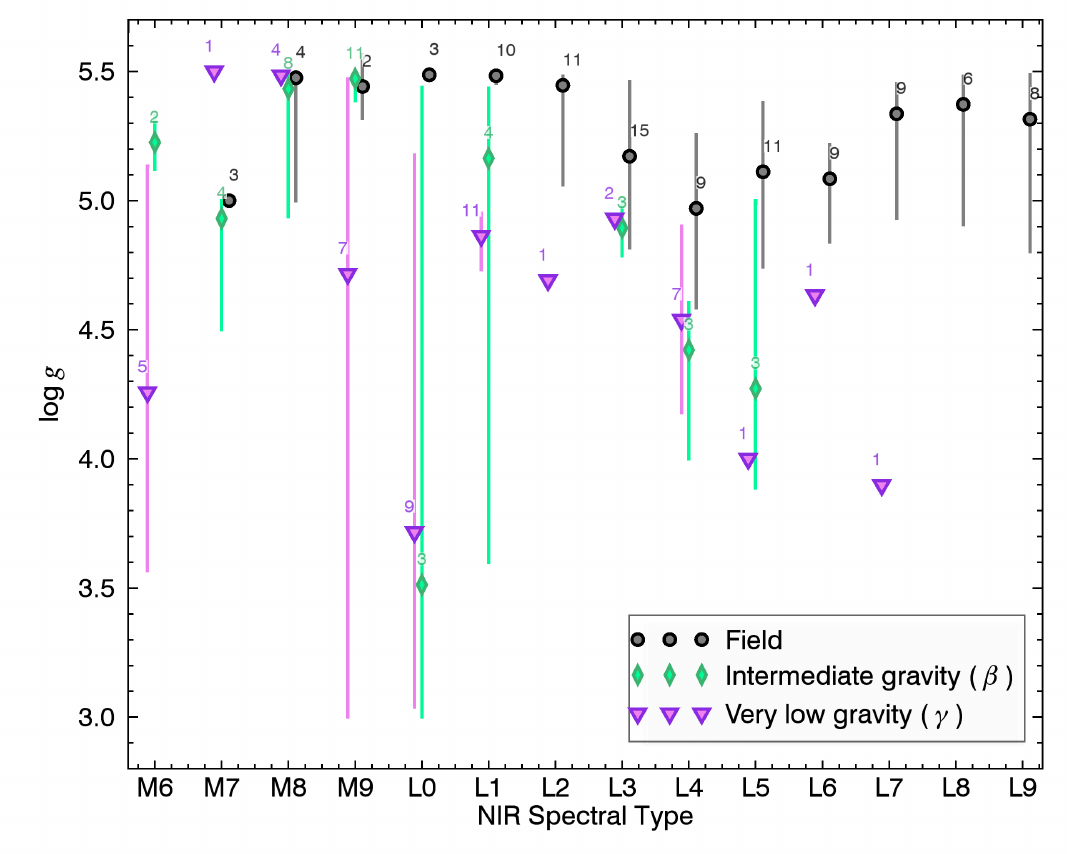}\label{fig:spt_logg}}
	\subfigure[$JH$-band \Teff\ versus NIR spectral type (without \emph{WISE} data)]{\includegraphics[width=0.495\textwidth]{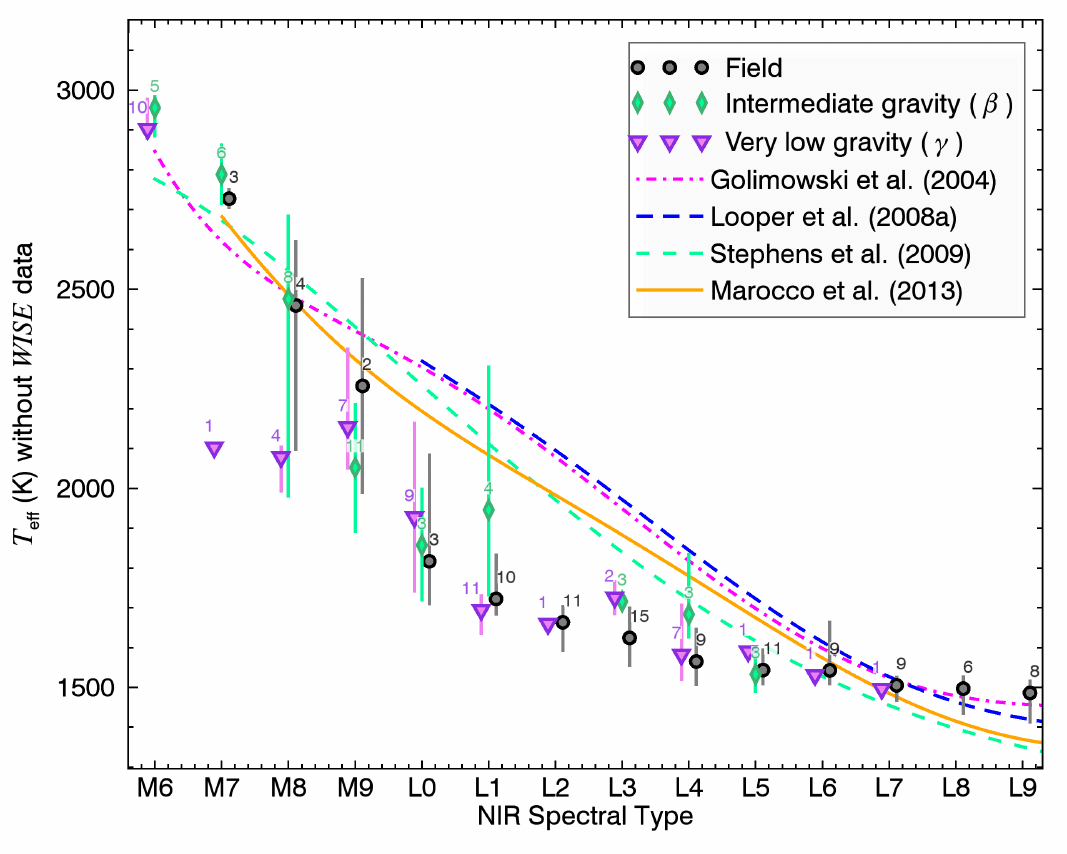}\label{fig:spt_teff}}
	\subfigure[$K$-band \Teff\ bias]{\includegraphics[width=0.495\textwidth]{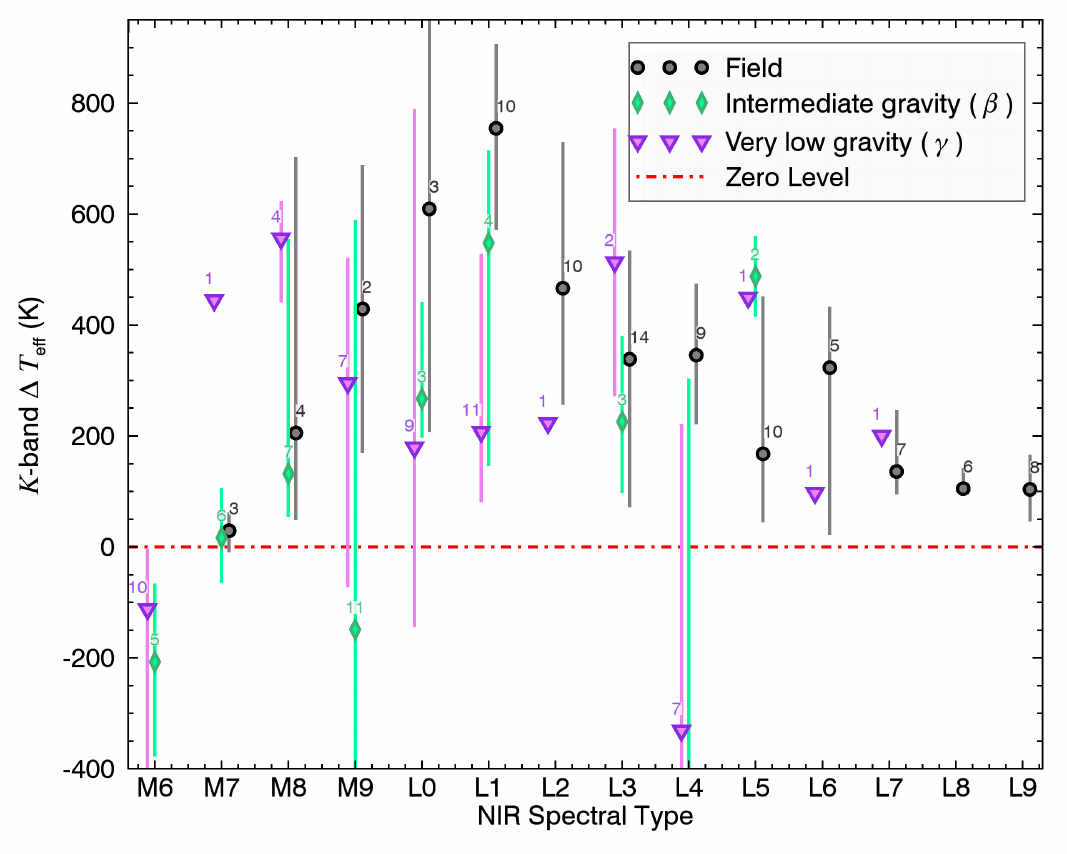}\label{fig:Kbias}}
	\caption{Panel~a: Adopted effective temperature (\Teff) derived by simultaneously fitting atmosphere models to full $JHK$ spectra and \emph{WISE} photometry as a function of spectral type for our sample of field and low-gravity dwarfs, binned by spectral type. The number of data points that were included in each bin is displayed above each symbol. Field dwarfs are represented with black circles, intermediate gravity dwarfs ($\beta$) with green diamonds and very low-gravity dwarfs ($\gamma$) with purple downside triangles. We added small systematic offsets in the spectral types of very low-gravity and field dwarfs for visibility. The solid orange line, green dashed line and fuchsia dash-dotted lines represent \Teff--spectral type relations from \cite{2013AJ....146..161M, 2004AJ....127.3516G,2008ApJ...685.1183L} and \cite{2009ApJ...702..154S}, respectively. We derive systematically cooler temperatures for young BDs. Panel~b: Adopted surface gravity ($\log g$) as a function of spectral type for our sample of field and low-gravity dwarfs, binned by spectral type. The color coding is similar to that of Panel~a. The derived $\log g$ values for dwarfs with spectroscopic confirmation of low-gravity are systematically lower than those of field dwarfs; however, the large scatter hints that model fitting alone is not an efficient way of identifying low-gravity dwarfs. Panel~c: Effective temperature, derived by fitting atmosphere models to individual $zJ$- and $H$-bands only without \emph{WISE} photometry, as a function of spectral type for our sample of field and low-gravity dwarfs, binned by spectral type. Color-coding is identical to Panel~a. We derive temperatures systematically cooler for all L0--L4 dwarfs including field dwarfs, a likely effect of dust clouds not being properly reproduced by BT-Settl models. Panel~d: Difference in the derived effective temperature from the $K$-band model fitting from that obtained by individual $zJ$- and $H$-bands model fitting (all without using \emph{WISE} photometry), binned by spectral type. Color-coding is identical to Panel~a and the red dot-dashed lines marks $\Delta$\,\Teff\,$= 0$. We derive systematically warmer temperatures when fitting atmosphere models to only the $K$ band of L0--L4 BDs.}
	\label{fig:spt_teff_all}
\end{figure*}

%Figure : Mass-Radii Relations
\begin{figure*}[p]
	\centering
	\subfigure[Mass versus Radius]{\includegraphics[width=0.6\textwidth]{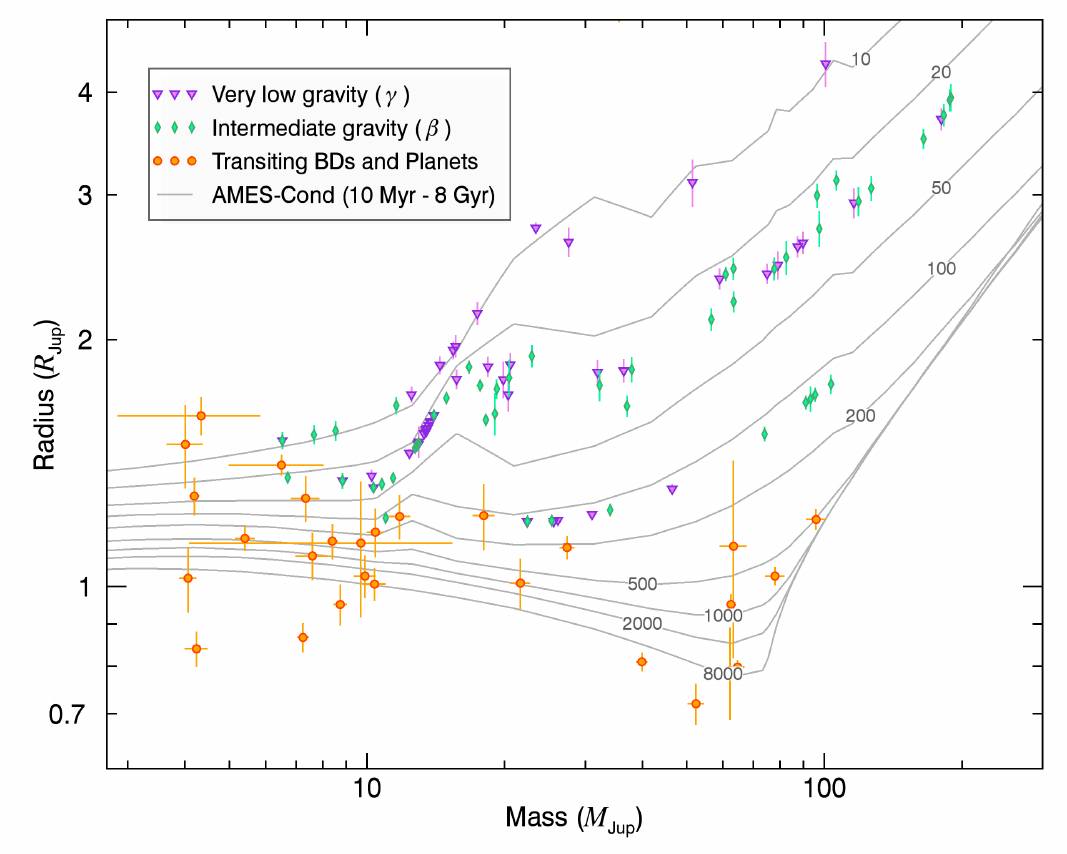}\label{fig:m_r}}
	\subfigure[Mass versus NIR Spectral Type]{\includegraphics[width=0.495\textwidth]{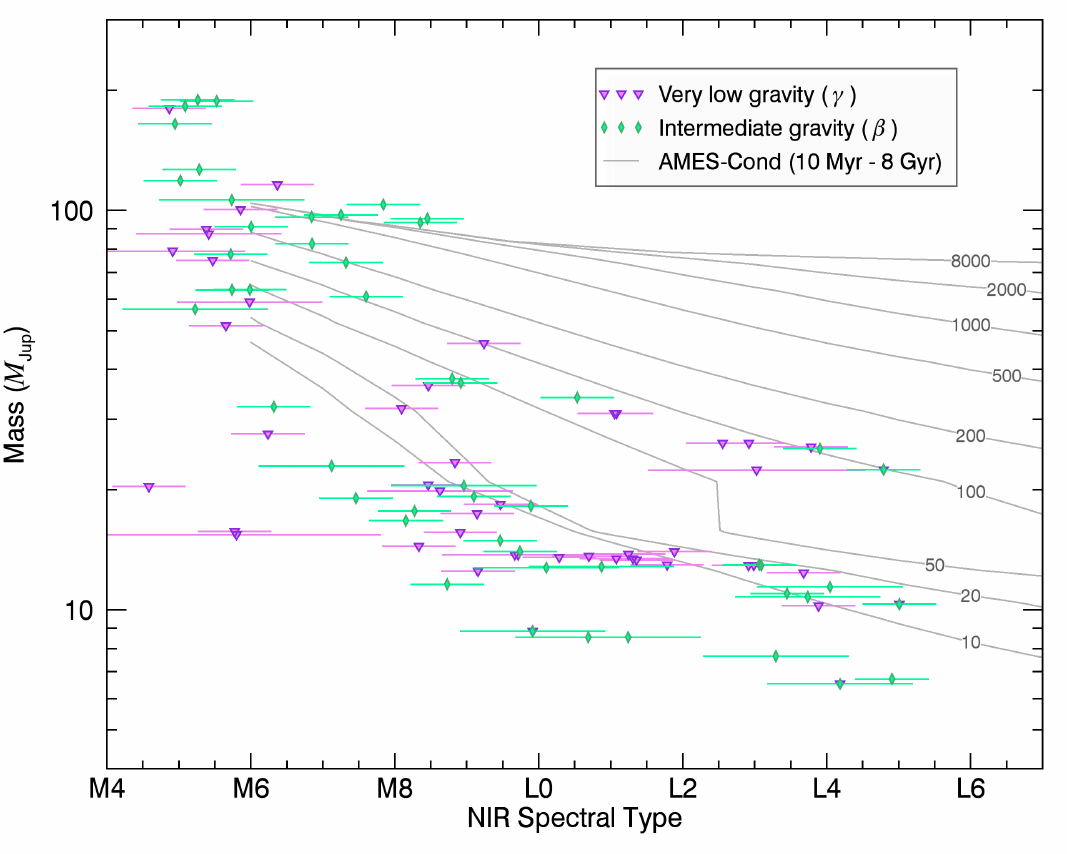}\label{fig:spt_m}}
	\subfigure[Radius versus NIR Spectral Type]{\includegraphics[width=0.495\textwidth]{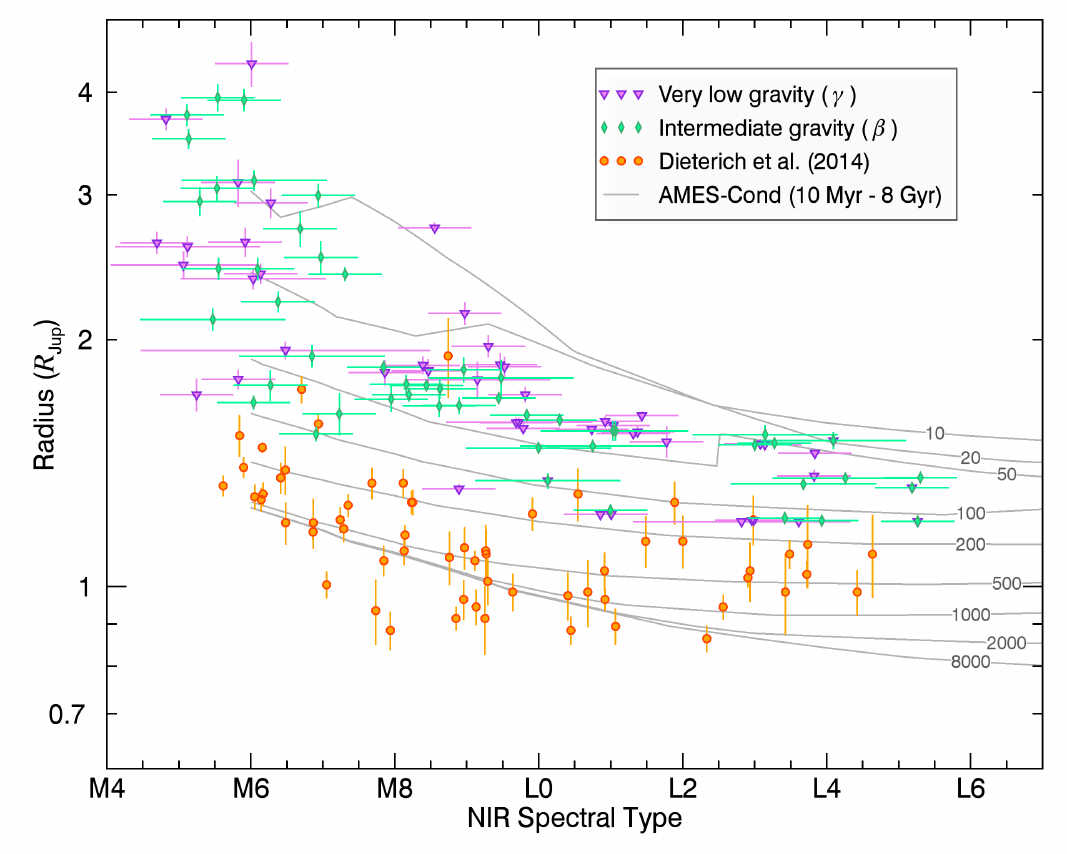}\label{fig:spt_r}}
	\caption{Panel~a: Radius as a function of mass for intermediate-gravity (green diamonds) and very low-gravity (purple downside triangles) candidate members of YMGs, derived from the BT-Settl models, compared with exoplanets and BD companions that benefit from transit and RV data (orange circles). Isochrones of various ages (gray lines; 10\,Myr to 8\,Gyr) were added for comparison. It can be seen that our sample overlaps with the planetary regime, and follow isochrones that correspond to the ages of YMGs considered here. Very low-gravity objects have a younger age on average compared with intermediate-gravity objects, which explains that they have larger radius for a given mass. Transit and RV data were obtained from \cite{2006Natur.440..311S,2006Natur.443..534S,2008A&A...491..889D,2008ApJ...683.1076W,2010ApJ...720.1118B,2010MNRAS.408.1689S,2011A&A...525A..68B,2011A&A...528A..97T,2011A&A...533A..83B,2011ApJ...742..116B,2011ApJS..197....3B,2012A&A...538A.145D,2012ApJ...761..123S,2012MNRAS.427.1877C,2013A&A...549A..18T,2013A&A...551L...9D,2013A&A...554A.114H,2013A&A...558L...6M,2013ApJ...779....5B,2014A&A...562A.140P,2014A&A...572A.109D,2014ApJ...788...92S,2014MNRAS.445.2106L,2014arXiv1411.4047M}; and \cite{2014arXiv1411.4666Q}. Panel~b: Mass as a function of spectral type for intermediate-gravity (green diamonds) and very low-gravity (purple downside triangles) candidate members of YMGs, compared with BT-Settl isochrones (gray lines; 10\,Myr to 8\,Gyr) from which the masses were derived. The isochrones were mapped on the spectral type dimension by converting effective temperatures to spectral types using the polynomial relation of \cite{2009ApJ...702..154S}. Panel~c: Radius as a function of spectral type for intermediate-gravity (green diamonds) and very low-gravity (purple downside triangles) candidate members of YMGs, compared with radii measurements from \citeauthor{2014AJ....147...94D} (\citeyear{2014AJ....147...94D}; orange circles). BT-Settl isochrones of various ages (gray lines; 10\,Myr to 8\,Gyr), which were used to derive our radii, were added for comparison. They were mapped on the spectral type dimension by converting effective temperatures to spectral types using the polynomial relation of \cite{2009ApJ...702..154S}. The radii measurements of \cite{2014AJ....147...94D} are all based on a comparison of the synthetic colors of atmosphere models with photometric measurements combined to trigonometric parallaxes. Their data consists in a majority of field objects. It can be seen that we determine radii that are consistent with the young ages of our objects when comparing to isochrones; however, the radii measurements of \cite{2014AJ....147...94D} for dwarfs later than $\sim$\,L2 yield larger values than those predicted by field-age evolution models alone (see \citealp{2014AJ....147...94D} for more detail).}
	\label{fig:mr_all}
\end{figure*}

\cite{2014A&A...564A..55M} used BT-Settl atmosphere models to determine the physical parameters of seven young L dwarfs and found that (1) low-gravity L0--L3 dwarfs fit models with similar temperatures of $\sim$\,1800\,K; (2) the continuum shape of the $H$ band is not well reproduced by solar-metallicity models; (3) the 1.1--2.5\,$\mu$m range in the $zJ$ bands is not well reproduced by models; and (4) the global continuum slope is not well reproduced by atmosphere models for L dwarfs. 

We used a method similar to that of \cite{2008ApJ...678.1372C} and \cite{2014ApJ...787....5N} to identify the best fitting solar-metallicity CIFIST2011 BT-Settl atmosphere model for our observed spectra, on a grid of effective temperature and surface gravity ranging from \Teff\ = 500--5000\,K and $\log g =$\,3.0--5.5\,dex with a grid spacing of 100\,K and 0.5\,dex, respectively. We computed the goodness-of-fit ($G_{k,j}$; \citealt{2008ApJ...678.1372C}) in each case. 

\cite{2008ApJ...678.1372C} demonstrated that \Teff\ can only be recovered efficiently by performing such a model fitting on a very large spectral range in the case of field L1--L8 dwarfs; however, while fitting a single model spectrum in this way allows recovering a good \Teff\ estimate, it does not reproduce well the general slope and the features in individual spectral bands. Since gravity-sensitive spectral features are generally narrow, this method will not yield good estimates of $\log g$. We have thus performed our model fitting in two different steps : (1) by fitting one single BT-Settl spectrum to the full 0.8--5\,$\mu$m range (\emph{WISE} $W1$ and $W2$ magnitudes were added as additional data to our spectra in order to do this); and (2) by fitting one BT-Settl spectrum to each one of the $zJ$, $H$ and $K$ spectral bands. The first method allowed us to obtain an estimate of \Teff, while the second one allowed us to obtain an estimate of $\log g$ for each object in our sample.

In order to append the \emph{WISE} photometric data to an observed NIR spectrum, we compute the synthetic $J$, $H$ and $K_S$ \emph{2MASS} magnitudes of the spectrum and determine the three corresponding normalization factors. We then use the median of these factors to bring back the two \emph{WISE} photometric data points to the same scale as the observed spectrum. The dilution factor is treated as a free parameter in our analysis so that no estimate nor measurement of distance is needed in the model fitting. We thus choose the dilution factor that minimizes $G_{k,j}$ for each fitted model. We do so in an analytical way to decrease computing time. We thus define the goodness-of-fit as :

\begin{align}
	G_{k,j} &= \sum_{i=1}^N W_{i,j} \left(\frac{F_{obs,i,j}-D_{k,j}F_{k,i,j}}{\sigma_{obs,i,j}}\right)^2\label{gof}, \\
	\mbox{where } D_{k,j} &= \frac{\sum_{i=1}^N W_{i,j} \frac{F_{obs,i,j}F_{k,i,j}}{\sigma_{obs,i,j}^2}}{\sum_{i=1}^N W_{i,j} \frac{F_{k,i,j}^2}{\sigma_{obs,i,j}^2}}\label{dilute}, \\
	W_{i,j} &= \frac{\left.\frac{\mathrm{d}\ln\lambda}{\mathrm{dx}}\right|_{x=x_{i,j}}}{\sum_{i^\prime=1}^N \left.\frac{\mathrm{d}\ln\lambda}{\mathrm{dx}}\right|_{x=x_{i^\prime,j}}},\label{eqweights}
\end{align}

where $x_{i,j}$ is the pixel number (i.e., the spectral position), $j$ is the index of the spectral band (i.e., $zJ$, $H$ or $K$, applicable only when we fit by individual bands), $k$ is the atmosphere model index (each value of $k$ corresponds to a given combination of \Teff\ and $\log g$), $N$ is the total number of pixels in the fitting range, $\lambda$ is the wavelength ($\mu$m), $W_{i,j}$ are the normalized weight factors, $D_{k,j}$ is the dilution factor that minimizes $G_{k,j}$, $F_{obs,i,j}$ and $\sigma_{obs,i,j}$ are the observed spectrum and its measurement error, and $F_{k,i,j}$ is an atmosphere model. The weights are chosen to ensure that equal wavelength ranges in log space equally contribute to the goodness-of-fit. For example, a broadband photometric measurement or one pixel of a low-dispersion spectroscopic order would be given a larger weight than one pixel of a high-dispersion spectroscopic order as it covers a larger wavelength range. \cite{2008ApJ...678.1372C} introduced this weighting method except that it was not done in log space; \cite{2014ApJ...787....5N} noted that the log space provides a more physically meaningful scale (i.e. using the log space prevents a bias that would be caused by working in wavelength space rather than frequency space).

We calculated errors on the adjusted parameters $a_l$ (i.e., \Teff\ and $\log g$) from \citeauthor{Wolberg:2006ve} (\citeyear{Wolberg:2006ve}; p.50) :

\begin{align}
	\sigma_{a_l,k,j} &= \sqrt{\frac{N}{N-2} G_{k,j}\ C_{l,l,k,j}^{-1}}\ \mbox{with}\label{erreq} \\
	C_{l,m,k,j} &= \sum_{i=1}^N W_{i,j} \frac{\partial}{\partial a_l}\left(D_{k,j}F_{k,i,j}\right) \frac{\partial}{\partial a_m}\left(D_{k,j}F_{k,i,j}\right), 
\end{align}

where $C_{l,m,k,j}$ are elements of the correlation matrix. Equation \eqref{erreq} and the equivalent expression of \cite{Wolberg:2006ve} differ by a factor $\sqrt{N}$ to compensate for our use of normalized weights in Equation \eqref{gof}. These error estimates do not take into account any systematic error in either our observations or the BT-Settl atmosphere models, and are thus only based on the variation of the goodness-of-fit with respect to each parameter. We show a few typical examples of per-band model fitting in Figure~\ref{fig:spectra_fitting1}. 

As noted by \cite{2014A&A...564A..55M}, we find that the BT-Settl models generally fail to accurately reproduce the $zJ$-bands spectra of L dwarfs, especially at wavelengths smaller than $\sim$\,1\,$\mu$m; the general slope seems to be in agreement, but a high-gravity solution is almost always preferred for all L dwarfs. Moreover, the FeH absorption features at $\sim$\,1.6\,$\mu$m are not present at all in the atmosphere models, which could be explained by missing opacity sources in the synthetic models. For this reason, we have only kept results from the $H$ and $K$ bands to determine $\log g$. The adopted $\log g$ value is thus determined from the weighted mean of the values obtained from the $H$-band and $K$-band fitting, where the weights are set to the total values of $W_{i,j}$ (see Equation~\ref{eqweights}) within the fitting range divided by the inverse square of the individual measurement errors. This corresponds to the optimal weights that account both for the measurement error and the wavelength range used in the fitting process. Both the measurements and errors were rounded to the nearest half-integer and to the nearest factor of 100\,K in the case of $\log g$ and \Teff, respectively. We imposed a floor on measurement errors that correspond to the grid size of our BT-Settl models, i.e. 0.5 and 100\,K for $\log g$ and \Teff.

Our adopted \Teff\ and $\log g$ values are listed in Table~\ref{tab:physpar} for our complete sample of low-gravity and field dwarfs. In Figures~\ref{fig:spt_teffk}, we show the spectral type--\Teff\ sequence that we obtain, compared to various sequences from the literature \citep{2009ApJ...702..154S, 2004AJ....127.3516G, 2013AJ....146..161M}. We find \Teff\ values that are consistent with the literature across the full range of spectral types, except for low-gravity objects which seem to be systematically cooler. This might be an additional indication that low-gravity brown dwarfs have cooler effective temperatures compared with field brown dwarfs of the same spectral types, an effect that was previously hypothesized and then demonstrated for the young, directly-imaged BD and exoplanet companions HD~203030~B, TWA~27~b, HR~8799~b and $\beta$~Pictoris~b \citep{2006ApJ...651.1166M,2011ApJ...733...65B,2011ApJ...735L..39B,2014ApJ...786...32M}, as well as for young brown dwarfs (\citealp{2012ApJ...752...56F,2013ApJ...777L..20L};~Joseph~C.~Filippazzo et al., submitted to ApJ).

In Figure~\ref{fig:spt_logg}, we show the spectral type--$\log g$ sequence that we obtain for low-gravity and field dwarfs. The $\log g$ values that we derive for our low-gravity sample are systematically lower than those of our field sample, as expected. However, we observe a large scatter in the $\log g$ values of low-gravity dwarfs, although they are lower on average. This indicates that the model fitting method that we described above might not be very efficient in recovering low-gravity dwarfs in an ensemble of NIR spectra. Additionally, we derive slightly lower $\log g$ values for field dwarfs with spectral types M7 and L3--L6, indicating that the false positive rate might be larger when identifying low-gravity dwarfs based solely on model fitting in this range of spectral types. Our results also tentatively indicate that M7 dwarfs are systematically better fit by low-gravity atmosphere models, however this is based on only three objects and is thus possibly an effect of small number statistics.

We tried to reproduce the results of \cite{2008ApJ...678.1372C} showing that fitting individual bands yield systematically offset \Teff\ values, and to extend this result to our full M6--L9 range as well as to low-gravity dwarfs. In Figure~\ref{fig:spt_teff}, we show the spectral type--\Teff\ sequence that we obtain if we combine the $zJ$- and $H$-band measurements in a weighted mean (using similar weights than described above for $\log g$). We show that the systematic offsets in \Teff\ values derived with this method are significant in the M9--L5 range, and independent of surface gravity. In Figure~\ref{fig:Kbias}, we compare the difference of \Teff\ values obtained from $zJ$- and $H$-band fitting to those obtained from the $K$-band fitting only. We show that \Teff\ values derived from the $K$-band only are systematically warmer in the M9--L5 range. The values of \Teff\ obtained from $K$-band fitting only are thus closer to those presented in Figure~\ref{fig:spt_teffk}, except that the scatter is much larger. These results confirm the findings of \cite{2008ApJ...678.1372C}, while extending them to earlier spectral types (down to M9) and seem to indicate that the $zJ$ and $H$ bands are the most likely cause of the systematic offset in \Teff.

It will be interesting to investigate whether fixing the \Teff\ value using a large spectral coverage, and subsequently determining the best $\log g$ value using wavelength regions significantly smaller than a spectral band that are known to be gravity-sensitive, might provide a better way to determine accurate $\log g$ values for L dwarfs. This will be the subject of a future work, along with repeating this analysis with future generations of BT-Settl atmosphere models that include a more realistic treatment of dust clouds (see \citealt{2014A&A...564A..55M} for a discussion on this topic).

\subsubsection{Evolution Models}\label{sec:evmod}

We estimated the physical parameters (mass, radius, \Teff, $\log g$) of all low-gravity candidate members presented here from a comparison of their absolute \emph{2MASS} and \emph{WISE} photometry with isochrones from CIFIST2011 BT-Settl models using a likelihood analysis. The age range of the most probable host YMG was used in each case, and statistical distances from BANYAN~II are used when a trigonometric distance is not available. These models do not account for magnetic fields and assume a hot-start formation (large initial entropy). Both effects could cause a systematic underestimation of mass \citep{2010ApJ...711.1087K,2012ApJ...756...47S,2014MNRAS.437.1378M,2014ApJ...792...37M}. However, it has been demonstrated that BD masses derived from evolution models are systematically too large when compared to dynamical mass measurements \citep{2001ApJ...560..390L,2004A&A...423..341B,2009ApJ...706..328D,2009ApJ...692..729D,2009ApJ...699..168D,2010ApJ...721.1725D,2014ApJ...790..133D,2015arXiv150306212D}. This seems in contradiction with what would be expected from the model limitations described above; instead, it is likely that the cooling rate of BDs is slowed down by atmospheric clouds, an effect that is not taken into account in current evolution models \citep{2015arXiv150306212D}.

The resulting physical parameters are presented in Table~\ref{tab:physpar}. This allowed us to compile a total of 25 objects with an estimated mass in the planetary regime ($<$\,13\,\MJup); they are individually discussed in the Appendix. These objects are all likely located within 10--60\,pc and will constitute a sample of choice for a detailed study of the connection between the physical properties of BDs and giant, gaseous exoplanets, e.g. using the James Webb Space Telescope \citep{2006SSRv..123..485G}.

In Figure~\ref{fig:mr_all}, we compare the masses and radii estimated for the objects in our sample with those of other known exoplanets and young BDs, as well as with BD radii measured by \cite{2014AJ....147...94D}. We show that our sample overlaps with the regime of giant, gaseous exoplanets. Our sample displays inflated radii and lower masses than field dwarfs, for given spectral types, which is expected for young, low-gravity low-mass stars and BDs.

\subsection{Space Density at the Deuterium-Burning Limit}\label{sec:imf}

Late-type members of YMGs provide the opportunity of measuring the low-mass end of the IMF which is still poorly constrained. The \emph{BASS} survey is still not complete enough to construct individual IMFs for the YMGs under study, but we can already put constraints on the population of objects near the planetary-mass boundary where our survey is particularly sensitive.

\begin{figure*}
	\centering
	\subfigure[Estimated Mass]{\includegraphics[width=0.45\textwidth]{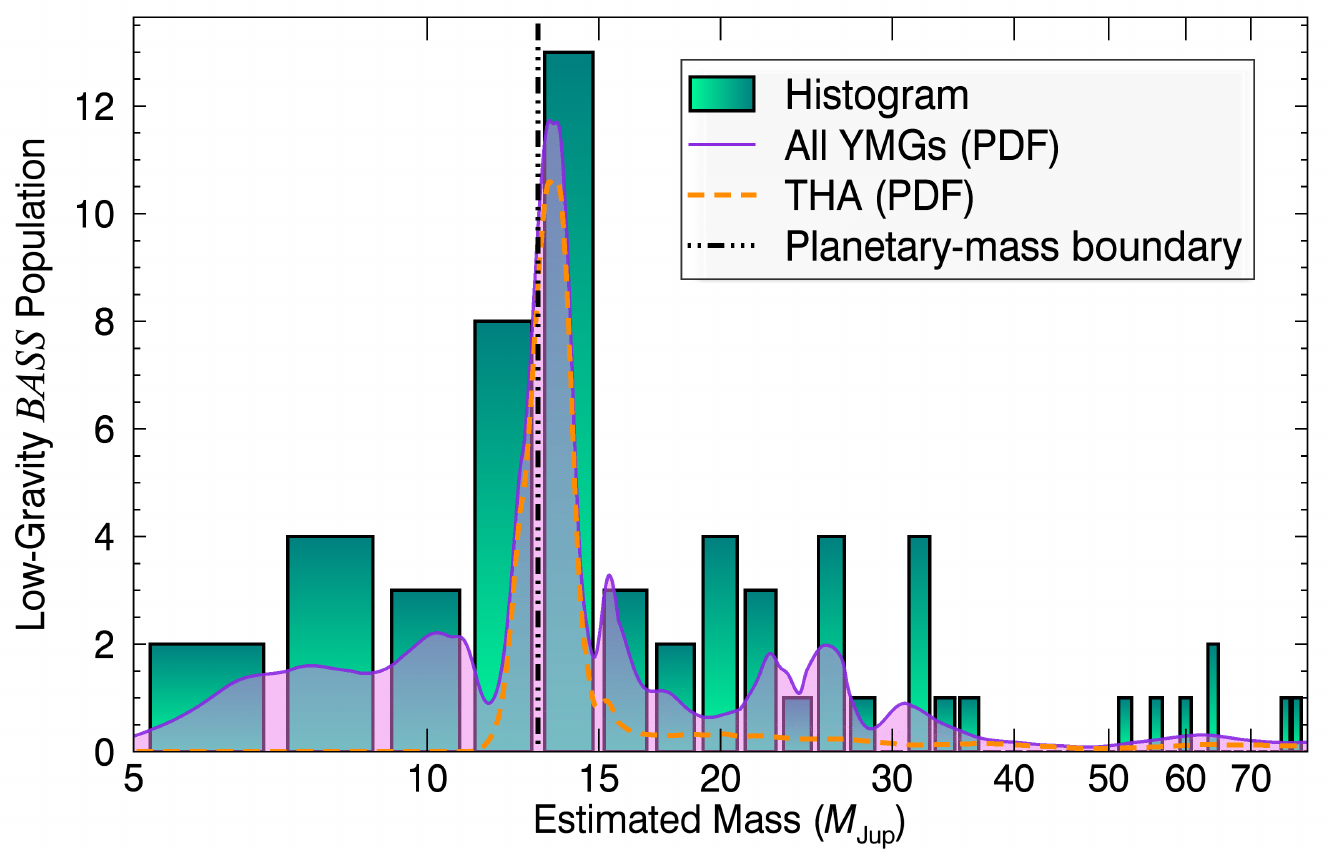}}
	\subfigure[Absolute $W1$ Magnitude]{\includegraphics[width=0.45\textwidth]{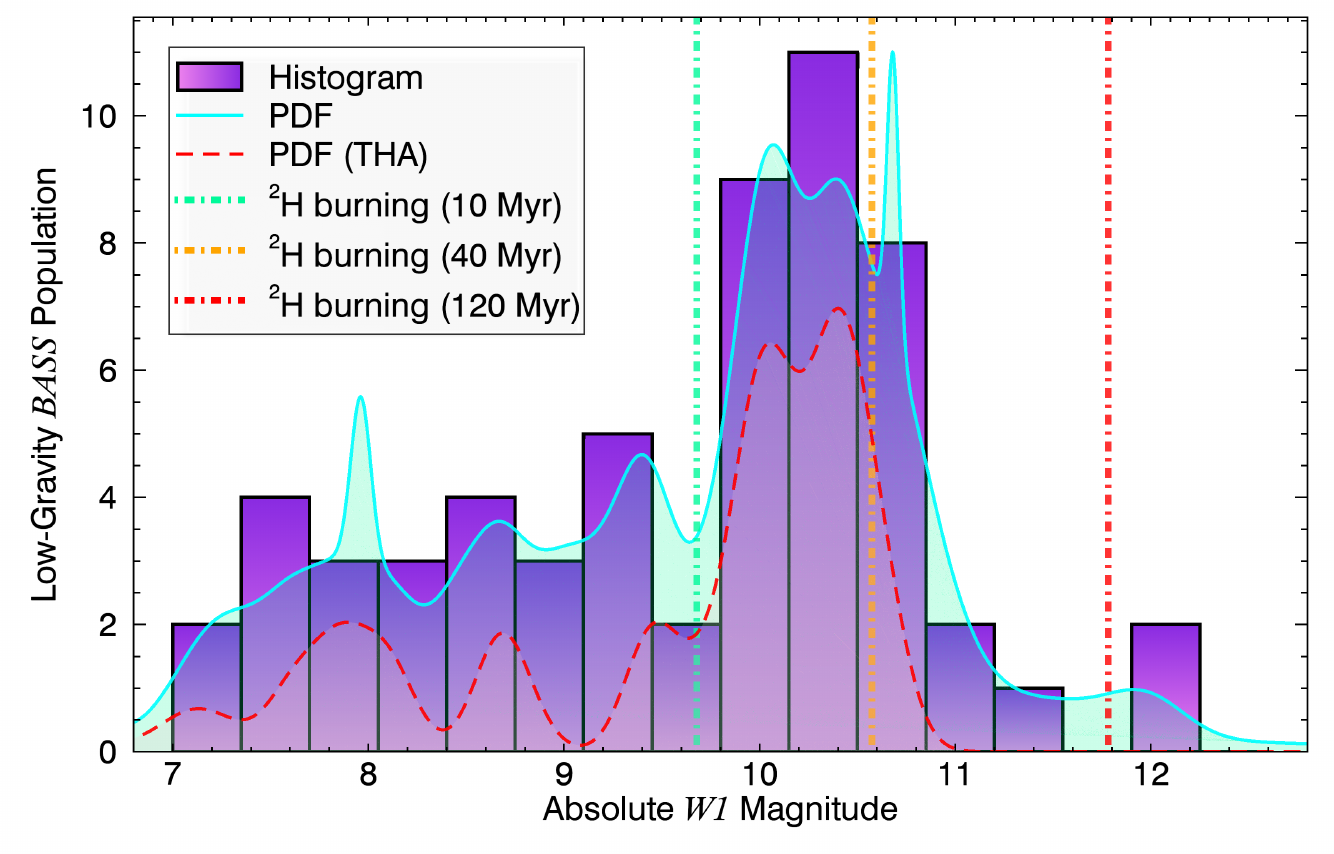}}
	\caption{Panel~a: Histogram of estimated masses (green bars) for low-gravity dwarfs in the \emph{BASS} sample, obtained from a comparison of NIR photometry and trigonometric or kinematic distances with BT-Settl--CIFIST2011 synthetic models. The continuous PDFs of different subsets of the candidates are indicated with different lines (see legend). They were obtained by combining the individual mass estimation PDFs directly, and they thus provide a histogram-like continuous distribution that include measurement errors and are independent of the binning. Panel~b: Histogram of absolute \emph{WISE} $M_{W1}$ magnitude of low-gravity dwarfs in the \emph{BASS} sample (purple bars), obtained from trigonometric or kinematic distances. The thick aqua distribution is a continuous distribution that does not include binning and takes account of measurement uncertainties, and was built in the same way as that of Panel~a. We find a larger number of objects as $M_{W1}$ gets fainter, up to $M_{W1}$\,$\sim$\,11 where we stop being sensitive. The most limiting aspect of our survey is the inclusion in the \emph{2MASS} catalog, with a limiting magnitude around $J \sim$\,16--17. The absence of a strong over-density is not in contradiction with Panel~a, because our sample is composed of objects at different ages ($\sim$\,12--120\,Myr), hence a given mass can correspond to a different temperature and absolute magnitude. The green, yellow and red vertical dashed lines correspond to the absolute $W1$ magnitudes of a 10, 40 and 120 Myr object, respectively.}
	\label{fig:histmass}
\end{figure*}

We display in Figure~\ref{fig:histmass} a histogram of the estimated masses of all objects in our sample. We also display in this Figure a probability density function (PDF) that represents a continuous analog of the histogram which is independent on the binning and that includes individual measurement errors. This PDF is obtained by normalizing the integral of each individual mass estimation PDF to unity and summing them over the full sample. In the case of absolute $W1$ magnitudes, the PDFs that correspond to individual measurements were taken as normalized gaussian distributions with a characteristic width that corresponds to the measurement error.

There are 15 objects in our sample of THA candidates that have estimated masses in the 12.5--14\,\MJup\ range, which corresponds to the planetary-mass limit. This peak-shaped distribution of estimated masses for the THA candidate members uncovered here is the combined effect of a selection bias (we observed the latest-type objects first) and the distance distribution of THA members ($\sim$\,30--70\,pc; Paper~II), as 12/15 of these objects are likely located within 50\,pc. Furthermore, we have identified a larger number of THA candidates compared to other YMGs, because its members are more easily identified in an all-sky search--the slightly larger distance of THA ensures that its members have a narrower distribution in space position and proper motion. The relatively large number of 12.5--14\,\MJup\ objects compared to objects in the 5--10\,\MJup\ or 15--75\,\MJup\ ranges is thus a selection effect.

Since our sample is biased on recovering objects more efficiently in the 12.5--14\,\MJup\ range, it remains useful to assess the space density of such objects. We will concentrate on the THA candidate members for this as they provide a larger sample. Assuming that we have uncovered all of the 12.5--14\,\MJup\ candidate members of THA in \emph{BASS} within 50\,pc (accounting for 65.6\% of the expected population according to our SKM model for THA) and correcting for the expected completeness of BASS for this association (90\%; \citealt{2015ApJ...798...73G}), we can expect that there are a total number of $20.3_{-5.1}^{+6.8}$ objects in THA that lie within this range of masses. The error was estimated assuming that the objects were drawn from a Poisson distribution, and they thus account for small number statistics.

Assuming that the population of 1.00--1.26\,\Msol\ stars is complete in THA ($N=14_{-3.3}^{+4.3}$ using Poisson statistics, see Figure~8 of \citealt{2014AJ....147..146K}) and adjusting a fiducial log-normal IMF peaking at 0.25\,\Msol with a width $\sigma = 0.5$\,dex \citep{2012EAS....57...45J}, we can expect a total of $356_{-47}^{+61}$ main-sequence stars in THA ($>$\,75\,\MJup) and only $0.56_{-0.13}^{+0.17}$~objects in the 12.5--14\,\MJup\ range (the ratio of 12.5--14\,\MJup\ to 1.00--1.26\,\Msol\ objects derived from that IMF is $0.04$). We thus seem to be uncovering at least $36.4_{-12.5}^{+16.6}$ times too many objects in this mass range, compared to the predictions of a typical log-normal IMF anchored on the 1.00--1.26\,\Msol\ population of THA.

%Figure : Mass Histogram / IMF

It is possible that this is a consequence of a fault in the evolution models rather than a true over-population. For example, one could argue that the models fail to reproduce the effects of clouds which have a larger impact on the spectra of less massive, cooler objects. This could lead us to misinterpret the masses of our 15 low-gravity THA candidates, assigning them 12.5--14\,\MJup\ while their true masses span a larger range. If this effect alone is to explain the over-population, the true range of masses for our 15 objects would have to be extended by 190\% in log space, which would mean that their true masses would span 4.5--39\,\MJup. This effect is thus unlikely to be the lone explanation of this over-population. It is also possible that the current age estimate of THA is wrong--e.g., $\beta$PMG, Upper~Scorpius, AB~Doradus and the Pleiades have recently been found to be slightly older than previously thought \citep{2005ApJ...628L..69L,2012ApJ...746..154P,2014ApJ...792...37M,2014MNRAS.438L..11B,2014MNRAS.445.2169M}. If it turns out that this is also the case for THA, our estimated masses would need to be shifted to larger values. As an example, doubling the age of THA would shift the estimated mass of a member from $\sim$\,13\,\MJup\ to $\sim$\,20\,\MJup. This effect alone would thus be insufficient to explain the large number of 12.5--14\,MJup\ THA candidates that we found. A similar shift of our estimated masses could be caused by systematics in evolution models (see our discussion in Section~\ref{sec:evmod}), although it is difficult to estimate the magnitude of this effect at this time. \cite{2015arXiv150306212D} has shown that masses from evolution model are likely under-estimated for dusty BDs at the L/T transition; it could be expected that the same effect is important in young L dwarfs. This would further accentuate the discrepancy between our observations and the predictions from a typical IMF.

\cite{2013ApJ...774...55B} noted that the age--absolute luminosity model sequences of $\sim$~13\,\MJup\ and $\sim$~25\,\MJup\ objects at different young ages overlap; such a pile-up in the isochrones could cause a degeneracy in our estimated masses and cause our method to mis-interpret true $\sim$~25\,\MJup\ objects as planetary-mass objects. However, there are several observations that make this explanation unlikely: (1) The likelihood method with which we estimate masses not only generates a measurement and error bars, but it also provides a continuous PDF for each individual mass estimate. If this effect is important, we would thus be able to observe double-peaked individual measurement PDFs, as well as a peak at $\sim$~25\,\MJup\ in the PDF displayed in Figure~\ref{fig:histmass}. Note that even if present, this effect would not introduce a second peak in the histogram, since it was constructed from the most probable values of the estimated masses only. (2) While the young age--absolute luminosity isochrones overlap at different masses, this effect is much more subtle in the individual $J$, $H$, $K_S$, $W1$ and $W2$ age--absolute magnitude isochrones. Furthermore, the slight overlap happens at slightly different ages and masses in the different filters, and allows to lift the degeneracy between $\sim$~13\,\MJup\ and $\sim$~25\,\MJup\ objects. This likely explains why we do not observe dual-valued mass estimate PDFs. (3) Performing a Monte Carlo analysis in which 20 and 40\,Myr isochrones are used to estimate the masses of a population of 20\,000 synthetic objects with true masses uniformly distributed between 4 and 80\,\MJup\ produces no over-density of estimated masses in the 12--14.5\,\MJup\ range. The absolute $J$, $H$, $K_S$, $W1$ and $W2$ magnitudes of these synthetic objects are obtained from the model isochrones themselves, hence this Monte Carlo analysis cannot be used to investigate systematics in the model cooling tracks. Instead, it only addresses the potential problem of overlapping isochrones that could produce degenerate mass estimates.

As a consequence of these observations, it does not appear that overlapping isochrones are the cause of the large population of 12--14.5\,\MJup\ THA candidates in our sample. We note that it is however possible that a fraction of these THA candidate members are contaminants in our analysis (i.e., young interlopers from other moving groups or associations, considered in BANYAN~II or not) despite their high Bayesian probability and the low expected contamination rate in this particular YMG. It will be necessary to measure the RVs and parallaxes for all 12 objects discussed here to assess this, but at this stage it seems that this effect would be the most likely explanation for this over-density. For example, only 5/12 of these objects would need to be interlopers in order for the over-density to become a 1$\sigma$ result.

If we assume that the over-density is real, it would mean that there is at least one isolated dwarf in the 12.5--14\,\MJup\ range for every $17.5_{-5.0}^{+6.6}$ main-sequence star in THA. Comparing with the space density of main-sequence stars in the solar neighborhood ($9.3 \times 10^{-2}$ stars\,pc$^{-3}$; \citealt{2005ASSL..327...41C}) and assuming that the ratio we observed in THA is valid in the field, this would amount to a field density of $5.3_{-2.9}^{+3.8} \times 10^{-3}$ dwarfs\,pc$^{-3}$ in the 12.5--14\,\MJup\ range in THA. At ages older than 2.5\,Gyr, they will all have temperatures below 450\,K that correspond to spectral types later than Y0, and will thus be hard to locate due to their extreme faintness \citep{2011ApJ...743...50C,2011ApJS..197...19K,2014ApJ...786L..18L,2014A&A...570L...8B}. This is significantly larger than the lower limit measured by \cite{2011ApJS..197...19K} that corresponds to at least one $\geq$\,Y0 dwarf for every 78 main-sequence star (or $1.2 \times 10^{-3}$ dwarfs\,pc$^{-3}$), especially when considering that the population of field $\geq$\,Y0 dwarfs is also probably composed of objects that span a large range of ages and thus masses. \cite{2011ApJS..197...19K} noted that their measurement is only a gross underestimation on the space density of Y-type dwarfs due to several biases. We note however that the IMF of YMGs might be different than that of the field, which could be yet another cause for this difference.

A less likely scenario is that our results could be an indication that we are approaching an up-turn in the IMF of isolated objects in THA with masses below the deuterium-burning limit: such an up-turn has already been hinted at by micro-lensing surveys in the galactic plane that measure $1.8^{+1.7}_{-0.8}$ Jupiter-mass object for every main-sequence star (corresponding to space density of $1.7^{+1.6}_{-0.7} \times 10^{-1}$ objects\,pc$^{-3}$; \citealt{2011Natur.473..349S}). Measurements of RV and distance for the complete set of YMG candidates in \emph{BASS} will be crucial to assess whether the observed over-density holds, and discovering YMG candidate members at even lower masses will provide a strong constraint on whether there is an up-turn in the IMF of YMGs.

\section{Summary and Conclusions}\label{sec:conclusion}

We presented a NIR spectroscopic follow-up of 241 candidate members of YMGs identified through the \emph{BASS}, \emph{LP-BASS} and \emph{PRE-BASS} samples. This allowed us to identify 108 new low-gravity M5--L5 candidate members of YMGs with estimated masses spanning the range of 7--189\,\MJup. Thirty-seven of these objects were previously known in the literature, but no signs of low gravity had been reported for them before this work. We complemented this unique sample with 22 low-gravity dwarfs from the literature to (1) build color--spectral type and absolute magnitude--spectral type sequences for field and young dwarfs; (2) show that some gravity-sensitive indices correlate with age in the 10--200\,Myr regime, albeit with a large scatter, such that low-resolution NIR spectroscopy does not allow a strong constraint on the age of an individual object; (3) we discuss some limitations of the current BT-Settl models, mainly their improper treatment of dust clouds in L-type dwarfs of all ages; and (4) show that we find an unexpectedly large number of isolated objects with estimated planetary masses in the Tucana-Horologium association, which might be caused by young interlopers from other moving groups. This study represents one of the first steps towards bridging the gap in our knowledge of the the space density of the lowest-mass BDs ($\sim$\,13\,\MJup; \citealt{2011ApJS..197...19K}) and potential isolated giant planets that were ejected from their stellar system ($\sim$\,1\,\MJup; \citealt{2011Natur.473..349S}). Additional figures, data and information on this work can be found on the website \url{www.astro.umontreal.ca/\textasciitilde gagne} and in the Montreal Spectral Library, which is located at \url{www.astro.umontreal.ca/\textasciitilde gagne/MSL.php}.

\acknowledgments

The authors would like to thank the anonymous referee who suggested to improve BANYAN~II with the inclusion of parallax motion and significantly helped to improve the quality of this paper, as well as make it more concise and clear. We would like to thank Robert Simcoe, Philippe Delorme, Michael~C. Cushing, Rebecca Oppenheimer, Am\'elie Simon, Gilles Fontaine, Sergio~B. Dieterich, Benjamin~M. Zuckerman, Andr\'e-Nicolas Chen\'e, Sarah~Jane Schmidt, Simon Coud\'e, Daniella C. Bardalez Gagliuffi and Jonathan~B. Foster for useful comments and discussions. We thank Katelyn~N. Allers, Michael~C. Liu, Federico Marocco and Brendan~P. Bowler for sharing data. We also thank all observatory staff and observers who helped us in this quest - Bernard Malenfant, Ghislain Turcotte, Pierre-Luc L\'evesque, Alberto Past\'en, Rachel Mason, Stuart Ryder, Rub\'en D\'iaz, St\'ephanie C\^ot\'e, John~P. Blakeslee, Mischa Schirmer, Andrew McNichols, Dave Griep, Brian Cabreira, Tony Matulonis, German Gimeno, Steve Margheim, Percy~L. Gomez, Ren\'e Rutten, Bernadette Rodgers, Tim~J. Davidge, Jaehyon~Rhee Jay, Inger J\o rgensen, Thomas L. Hayward, Andrew Cardwell, Blair~C. Conn, Eleazar~Rodrigo Carrasco~Damele, David~A. Krogsrud, Eduardo Marin, Erich Wenderoth, Fredrik~T. Rantakyro, Joanna~E. Thomas-Osip, Pablo Patricio Candia, Pascale Hibon, Cl\'audia Winge, Benoit Neichel, Peter Pessev, Matthew~B. Bayliss and Anne Sweet. This work was supported in part through grants from the Fond de Recherche Qu\'eb\'ecois - Nature et Technologie and the Natural Science and Engineering Research Council of Canada. This research has benefited from the SpeX Prism Spectral Libraries, maintained by Adam Burgasser at \url{http://pono.ucsd.edu/\textasciitilde adam/browndwarfs/spexprism}, as well as the M, L, T and Y dwarf compendium housed at \url{http://DwarfArchives.org} and maintained by Chris Gelino, Davy Kirkpatrick, and Adam Burgasser, whose server was funded by a NASA Small Research Grant, administered by the American Astronomical Society. This research made use of: the SIMBAD database and VizieR catalog access tool, operated at the Centre de Donn\'ees astronomiques de Strasbourg, France \citep{2000A&AS..143...23O}; data products from the Two Micron All Sky Survey (\emph{2MASS}; \citealp{2006AJ....131.1163S,2003yCat.2246....0C}), which is a joint project of the University of Massachusetts and the Infrared Processing and Analysis Center (IPAC)/California Institute of Technology (Caltech), funded by the National Aeronautics and Space Administration (NASA) and the National Science Foundation \citep{2006AJ....131.1163S}; the Extrasolar Planets Encyclopaedia (\url{exoplanet.eu}), which was developed and is maintained by the exoplanet TEAM; data products from the \emph{Wide-field Infrared Survey Explorer} (\emph{WISE}; \citealp{2010AJ....140.1868W}), which is a joint project of the University of California, Los Angeles, and the Jet Propulsion Laboratory (JPL)/Caltech, funded by NASA; the NASA/IPAC Infrared Science Archive (IRSA), which is operated by JPL, Caltech, under contract with NASA; the Infrared Telescope Facility (IRTF), which is operated by the University of Hawaii under Cooperative Agreement NNX-08AE38A with NASA, Science Mission Directorate, Planetary Astronomy Program; the Database of Ultracool Parallaxes maintained by Trent Dupuy \citep{2012ApJS..201...19D}; the Hale 5 m telescope at Palomar Observatory, which received funding from the Rockefeller Foundation; and of tools provided by Astrometry.net. This paper includes data gathered with the 6.5 meter Magellan Telescopes located at Las Campanas Observatory, Chile (CNTAC program CN2013A-135). Based on observations obtained at the Gemini Observatory through programs number GN-2013A-Q-118, GS-2013B-Q-79, GS-2014A-Q-55, GS-2014B-Q-72, GS-2014B-Q-47 and GS-2015A-Q-60. The Gemini Observatory  is operated by the Association of Universities for Research in Astronomy, Inc., under a cooperative agreement with the National Science Foundation (NSF) on behalf of the Gemini partnership: the NSF (United States), the National Research Council (Canada), CONICYT (Chile), the Australian Research Council (Australia), Minist\'{e}rio da Ci\^{e}ncia, Tecnologia e Inova\c{c}\~{a}o (Brazil) and Ministerio de Ciencia, Tecnolog\'{i}a e Innovaci\'{o}n Productiva (Argentina). All data were acquired through the Canadian Astronomy Data Center and part of it was processed using the Gemini IRAF package. This material is based upon work supported by AURA through the National Science Foundation under AURA Cooperative Agreement AST 0132798 as amended. This publication uses observations obtained at IRTF through programs number 2007B023, 2007B070, 2008A050, 2008B054, 2009A055, 2010A045, 2011B071, 2012A097, 2012B015, 2013A040, 2013A055, 2013B025, 2014B026 and 2015A026. The authors recognize and acknowledge the very significant cultural role and reverence that the summit of Mauna Kea has always had within the indigenous Hawaiian community. We are most fortunate to have the opportunity to conduct observations from this mountain.

\indent \emph{Facilities:} IRTF (SpeX), Magellan:Baade (FIRE), Gemini:South (Flamingos-2), Gemini:North (GNIRS), Hale (TripleSpec).

\clearpage
\onecolumngrid

%Table : Polynomial Coefficients
\tabletypesize{\footnotesize}
\renewcommand{\tabcolsep}{6pt}
\begin{deluxetable}{rcllllcllll}
\tablecolumns{17}
\tablecaption{Polynomial coefficients for spectral type-magnitude and spectral type-color diagrams.\label{tab:coeff}}
\tablehead{\colhead{Sequence} & \colhead{} & \multicolumn{4}{c}{Field Sequence} & \colhead{} & \multicolumn{4}{c}{Young Sequence}\\
\cline{3-6}
\cline{8-11}
\colhead{Name} & \colhead{} & \colhead{$c_0$} & \colhead{$c_1$} & \colhead{$c_2$} & \colhead{$\sigma$} & \colhead{} & \colhead{$c_0$} & \colhead{$c_1$} & \colhead{$c_2$} & \colhead{$\sigma$}
}
\startdata

M$_{\it J}$ & & $8.53$ & $3.08e-1$ & $2.56e-3$ & $0.66$ & & $4.97$ & $7.96e-1$ & $-1.14e-2$ & $1.16$\\
& & $\pm 3.07e-1$ & $\pm 5.66e-2$ & $\pm 2.42e-3$ & & & $\pm 5.65e-1$ & $\pm 1.02e-1$ & $\pm 4.35e-3$ & \\[3pt]
\hline\\[-5pt]
M$_{\it H}$ & & $8.11$ & $2.81e-1$ & $1.65e-3$ & $0.63$ & & $4.76$ & $7.37e-1$ & $-1.20e-2$ & $0.72$\\
& & $\pm 2.86e-1$ & $\pm 5.23e-2$ & $\pm 2.24e-3$ & & & $\pm 5.04e-1$ & $\pm 8.81e-2$ & $\pm 3.60e-3$ & \\[3pt]
\hline\\[-5pt]
M$_{\it K_S}$ & & $8.12$ & $2.05e-1$ & $3.86e-3$ & $0.60$ & & $4.69$ & $6.88e-1$ & $-1.22e-2$ & $0.56$\\
& & $\pm 2.74e-1$ & $\pm 5.02e-2$ & $\pm 2.19e-3$ & & & $\pm 4.74e-1$ & $\pm 8.16e-2$ & $\pm 3.28e-3$ & \\
\enddata
\tablecomments{All abscissa are spectral types $S_\mathrm{type}$, expressed in decimal value, where zero is M0, ten is L0, etc. A given sequence respects the equation $y = \sum_{i=0}^N c_i S_\mathrm{type}^i$. The scatter of the data with respect to a best-fitting sequence is given by $\sigma$. All sequences are valid in the M6--L8 range.}
\end{deluxetable}
\renewcommand{\tabcolsep}{6pt}

%Table : Physical Parameters
\LongTables
\tabletypesize{\footnotesize}
\renewcommand{\tabcolsep}{6pt}
\begin{deluxetable}{llclcclcc}
\tablecolumns{9}
\tablecaption{Physical Parameters.\label{tab:physpar}}
\tablehead{\colhead{\emph{2MASS}} & \colhead{Spectral} & \colhead{YMG} & \colhead{} & \multicolumn{2}{c}{Estimated from isochrones\tablenotemark{a}} & \colhead{} & \multicolumn{2}{c}{Estimated from SED fitting} \\
\cline{5-6}
\cline{8-9}
\colhead{Designation} & \colhead{Type} & \colhead{Age (Myr)} & \colhead{} & \colhead{Mass (\MJup)} & \colhead{Radius (\RJup)} & \colhead{} & \colhead{\Teff\ (K)\tablenotemark{b}} & \colhead{$\log g$\tablenotemark{c}}}
\startdata

\multicolumn{9}{c}{Field Objects}\\[2.5pt]\hline\\[-6pt]
02535980+3206373 & M6 & $\cdots$ & & $\cdots$ & $\cdots$ & & $\cdots$ & $5.5 \pm 0.5$\\
07522390+1612157 & M6 & $\cdots$ & & $\cdots$ & $\cdots$ & & $\cdots$ & $5.5 \pm 0.5$\\
00335534-0908247 & M7 & $\cdots$ & & $\cdots$ & $\cdots$ & & $\cdots$ & $4.5 \pm 1.0$\\
18393308+2952164 & M7 & $\cdots$ & & $\cdots$ & $\cdots$ & & $2800 \pm 200$ & $5.0 \pm 1.0$\\
22021125-1109461 & M7 & $\cdots$ & & $\cdots$ & $\cdots$ & & $2800 \pm 200$ & $5.0 \pm 1.0$\\
16553529-0823401 & M7 & $\cdots$ & & $\cdots$ & $\cdots$ & & $2700 \pm 100$ & $5.0 \pm 0.5$\\
\enddata
\tablenotetext{a}{Estimated masses and radii were derived from a comparison of the trigonometric or statistical distances, the age of the most probable YMG membership and the \emph{2MASS} and \emph{WISE} photometry with AMES-Cond evolutionary models (see the text for a detailed explanation).}
\tablenotetext{b}{Values obtained from the modified model fitting that uses \emph{WISE} $W1$, $W2$ photometry as well as NIR spectra in the $J$, $H$ and $K$ bands.}
\tablenotetext{c}{Values obtained from a weighted mean of $H$-band and $K$-band $\log g$ (see text for more information).}
\end{deluxetable}
\renewcommand{\tabcolsep}{6pt}

%Table : Planemos
% Table : Planemos
\LongTables
\tabletypesize{\footnotesize}
\renewcommand{\tabcolsep}{3pt}
\begin{deluxetable}{lclcccc}
\tablecolumns{7}
\tablecaption{Potential isolated planetary-mass objects.\label{tab:jwst}}
\tablehead{\colhead{\emph{2MASS}} & \colhead{} & \colhead{Spectral} & \colhead{Source} & \colhead{Moving} & \colhead{Estimated} & \colhead{Distance\tablenotemark{a}}\\
\colhead{Designation} & \colhead{Ref.} & \colhead{Type} & \colhead{Sample} & \colhead{Group} & \colhead{Mass (\MJup)} & \colhead{(pc)}}
\startdata

00303013--1450333 & 1 & L4--L6 $\beta$ & \emph{LP-BASS} & ARG & $10.8^{+0.4}_{-0.6}$ & $26.72 \pm 3.21$\tablenotemark{b}\\
00344300--4102266 & 2 & L1: $\beta$ & \emph{BASS} & THA & $12.8 \pm 0.4$ & $41.2 \pm 2.4$\\
01531463--6744181 & 3 & L3 $\beta$ & \emph{BASS} & THA & $12.9^{+0.3}_{-0.5}$ & $47.0 \pm 3.2$\\
03420931--2904317 & 2 & L0: $\beta$ & \emph{BASS} & THA & $12.7 \pm 0.4$ & $48.2^{+3.6}_{-3.2}$\\
03421621--6817321 & 4 & L4 $\gamma$ & \emph{BASS} & THA & $12.4 \pm 0.5$ & $48.6 \pm 3.6$\\
04185879--4507413 & 2 & L3 $\gamma$ & \emph{BASS} & THA & $12.9^{+0.3}_{-0.4}$ & $49.8^{+4.0}_{-3.6}$\\
05012406--0010452 & 3 & L4 $\gamma$ & \emph{BASS} & COL,CAR & $10.2^{+0.8}_{-1.0}$ & $14.7 \pm 2.8$\tablenotemark{b}\\
05120636--2949540 & 5 & L5 $\beta$ & \emph{BASS} & BPMG & $6.7^{+1.0}_{-0.9}$ & $10.9^{+4.4}_{-4.0}$\\
\enddata
\tablenotetext{a}{Kinematic distances estimated from moving group membership unless noted.}
\tablenotetext{b}{Trigonometric distance.}
\tablecomments{References to this Table~: \\ (1)~\citealt{2000AJ....120..447K}; (2)~This paper; (3)~\citealt{2008AJ....136.1290R}; (4)~\citealt{2007AJ....133..439C}; (5)~\citealt{2003AJ....126.2421C}; (6)~\citealt{2014ApJ...785L..14G}; (7)~\citealt{2006AJ....131.2722C}; (8)~\citealt{2014ApJ...792L..17G}.}

\end{deluxetable}
\renewcommand{\tabcolsep}{6pt}
\twocolumngrid

\appendix
\section{DISCUSSIONS ON INDIVIDUAL OBJECTS}\label{App:discussion}
\twocolumngrid

Several objects presented here deserve a detailed discussion, either because they display peculiar features, or were reported in the literature as candidate members of other YMGs. Additionally, optical spectra were available in the literature for some objects discussed here, and can 
serve as an independent assessment of low surface gravity.

\subsection{Potential Planetary-Mass Low-Gravity Candidate Members of YMGs}\label{sec:potplan}

We list in Table~\ref{tab:jwst} twenty potential isolated planetary-mass objects in our sample, ten of which were discovered as part of this work. A few of these objects deserving further discussion are listed below.

\textbf{2MASS~J05012406--0010452} was discovered by \cite{2008AJ....136.1290R} as an L4 dwarf in the optical, and was categorized as a low-gravity L4\,$\gamma$ by \cite{2009AJ....137.3345C}, using its optical spectrum. \cite{2013ApJ...772...79A} categorized it as a very-low gravity L3 dwarf in the NIR, whereas we categorize it as an L4\,$\gamma$ dwarf. \cite{2012ApJ...752...56F} measured a trigonometric distance of $13.1 \pm 0.8$\,pc. We recovered this object in \emph{BASS} as an ambiguous candidate member of Columba or Carina with respective Bayesian probabilities of 49\% and 17\%, taking the trigonometric distance measurement of \cite{2012ApJ...752...56F} into account. If this object is a member of either COL or CAR (both YMGs are coeval at 20--40\,Myr), it has an estimated mass of $10.2^{+0.8}_{-1.0}$\,\MJup. \cite{2014A&A...568A...6Z} independently measured a trigonometric distance of $19.6 \pm 1.4$\,pc, which is discrepant with that of \cite{2012ApJ...752...56F} at the 5$\sigma$ level. The reason for this large discrepancy is unclear; the measurement of \cite{2012ApJ...752...56F} used a smaller number of epochs (11 versus 21); however, they were spread across a larger temporal coverage (3\,yr versus 2\,yr). If we adopt the distance measurement of \cite{2014A&A...568A...6Z}, the CAR membership probability becomes negligible and that of COL becomes considerably smaller (7.2\%), although we also calculate a low field contamination probability (1.3\%). It will be necessary to better constrain the distance of this object to assess whether it is a viable candidate member of COL or CAR. Obtaining an RV measurement would also be useful for this.

\textbf{2MASS~J05120636--2949540} has been identified as an L4.5 dwarf in the optical by \cite{2003AJ....126.2421C,2008ApJ...689.1295K}, and \cite{2014ApJ...794..143B} obtained a NIR spectrum to categorize it as an L4.5\,$ \pm 2$ dwarf. In Paper~II, we determined that this object is a low-probability candidate member of $\beta$PMG. We used the NIR spectrum of \cite{2014ApJ...794..143B} to revisit its spectral classification: we find that this object is a very good match to our L5\,$\beta$ template; however, the method of \cite{2013ApJ...772...79A} assigns it an intermediate gravity. We note that the VO$_Z$ index is significantly larger than that of field L5 dwarfs, but \cite{2013ApJ...772...79A} only use this index within the L0--L4 spectral types, as later-type low-gravity dwarfs in their sample displayed similar VO absorption than that of field dwarfs of the same spectral types. However, only one low-gravity L5 dwarf was available at the time, hence it is possible that the VO$_Z$ index remains useful to discriminate low-gravity L5 dwarfs. For this reason, we adopt the L5\,$\beta$ spectral type. Due to its low-gravity features, this object is preserved as a candidate member of $\beta$PMG. This object has one of the lowest estimated masses among the YMG candidates presented here, with $6.7^{+1.0}_{-0.9}$\,\MJup. Its statistical distance associated with membership to $\beta$PMG is $10.9^{+4.4}_{-4.0}$\,pc, which makes it a valuable benchmark to study the atmosphere of planetary-mass objects.

\textbf{2MASS~J12074836--3900043} (2MASS~J1207--3900) was discovered as a candidate member of TWA in \emph{BASS}. Its discovery and NIR spectroscopic follow-up have been presented in \cite{2014ApJ...785L..14G}. They reported an optical spectral type L0\,$\gamma$ and a NIR spectral type L1\,$\gamma$. Here we used the spectra of several low-gravity candidate members of Upper Scorpius obtained by \cite{2008MNRAS.383.1385L} to define tentative templates for the spectral type L0\,$\delta$, which likely correspond to objects younger than $\sim$\,15--20\,Myr and have an even more triangular $H$-band continuum than the L0\,$\gamma$ type. Given that both the optical and NIR spectra of 2MASS~J1207--3900 are peculiar even in comparison to the best template matches (L0\,$\gamma$ and L1\,$\gamma$ respectively) and that its $H$ band continuum is more triangular than any $\beta$ or $\gamma$ template, we revised its spectral classification by comparing it to Upper Scorpius candidate members. We find that the best match is the L0\,$\delta$ template; however, 2MASS~J1207--3900 displays features that are attributable to a later spectral type (redder slopes at 1.2--1.35\,$\mu$m and 1.5--1.6\,$\mu$m). We thus suggest a tentative spectral type of L1\,$\delta$ for this object, but identifying other similar objects will be necessary to confirm this. If it is a member of TWA (5--15\,Myr), this object has an estimated mass of $12.1^{+1.4}_{-2.0}$\,\MJup\ and a statistical distance of $58.2^{+6.8}_{-6.4}$\,pc.

\textbf{2MASS~J12271545--0636458} was identified as an M9 dwarf by \cite{2003AJ....126.2421C} using optical spectroscopy. We identified it as a candidate member of TWA in \emph{PRE-BASS}, and NIR spectroscopy allowed us to categorize it as a low-gravity M8.5\,$\beta$ dwarf. It was initially rejected from the \emph{BASS} sample because of its low Bayesian probability, which is in part due to the fact that its kinematic distance of $32.5 \pm 3.2$\,pc if it is a member of TWA does not match its spectrophotometric distance ($63.2 \pm 11.4$\,pc). The latter estimate would place 2MASS~J12271545--0636458 at the far-end of the TWA members ($\sim$\,40--62\,pc; Paper~II; \citealp{2013ApJ...762..118W, 2014A&A...563A.121D}). This is reminiscent of 2MASS~J12474428--3816464, TWA~29 and TWA~31, which are young and seem to be located between TWA and SCC in terms of distance (\citealp{2003ApJ...599..342S, 2012ApJ...754...39S, 2014ApJ...785L..14G}; Paper~V). Measurements of distance and RV will be useful to assess whether this is a true member of TWA despite its small Bayesian probability. If it is a true member of TWA (5--15\,Myr) located at its statistical distance, this object has an estimated mass of $11.6^{+1.4}_{-1.9}$\,\MJup.

\textbf{2MASS~J12563961--2718455} was identified in \emph{PRE-BASS} as a low-probability candidate member of TWA. NIR spectroscopy revealed that this object is a low-gravity L3$ \pm 1$\,$\beta$ dwarf. The probability that this object belongs to TWA is lower than 20\%, but the field contamination probability is also very low at $<$\,0.1\%. This usually points out to either an incomplete SKM for the YMG or to contamination from a source not taken into account in BANYAN~II. The most likely a priori explanation would be that this object is a contaminant from SCC (located at $\sim$\,100--150\,pc; \citealp{2003A&A...404..913S}); however, the spectrophotometric distance of 2MASS~J12563961--2718455 ($43.1 \pm 3$\,pc) is not consistent with this hypothesis, even when its low gravity is taken into account. Using its \emph{2MASS} and \emph{WISE} photometry and comparing it with other known low-gravity L4 dwarfs, we can rule out a distance larger than 48.5\,pc at a 95\% confidence level, assuming this object is not an unresolved multiple system. We can hence conclude that as long as this object is not extremely peculiar for a low-gravity L4 dwarf or a multiple system composed of four equal-luminosity components, it cannot be a member of SCC. The statistical distance from BANYAN~II which is associated to the TWA hypothesis ($46.2^{+4.8}_{-4.4}$\,pc) is similar to those of bona fide members of TWA ($\sim$\,40--62\,pc; Paper~II; \citealp{2013ApJ...762..118W, 2014A&A...563A.121D}), hence this case is different from those of 2MASS~J12271545-0636458, 2MASS~J12474428--3816464, TWA~29 and TWA~31, which are young and seem to be located between TWA and SCC in terms of distance (\citealp{2003ApJ...599..342S, 2012ApJ...754...39S, 2014ApJ...785L..14G}; Paper~V). Obtaining a distance measurement for this object will be helpful to assess whether it is a member of TWA. Assuming an age of 5--15\,Myr and comparing its statistical distance from BANYAN~II with BT-Settl \emph{2MASS} and \emph{WISE} isochrones, the estimated mass of this object is $7.7^{+1.4}_{-1.5}$\,\MJup, amongst the lowest of all candidate YMG members reported here. Its statistical distance associated with membership to TWA is $44.6 \pm 5.2$\,pc.

\textbf{2MASS~J21324036+1029494} was discovered as an L4.5\,$\pm 1$ dwarf by \cite{2006AJ....131.2722C} using low-S/N NIR spectroscopy. We identified it as a candidate member of ARG from \emph{PRE-BASS}. The NIR spectrum obtained by \cite{2006AJ....131.2722C} is available in the SpeX PRISM Spectral Libraries, we thus retrieved it to assess whether it is a low-gravity dwarf. We categorize this object as an L4:\,$\beta$ dwarf. Its $H$-cont index \citep{2013ApJ...772...79A} is consistent with low-gravity objects; however, the quality of the data is not sufficient to assess whether its FeH$_Z$ and KI$_J$ indices are consistent with this. Obtaining a better-quality and higher-resolution NIR spectrum will be useful to confirm the spectral type of this object. If it is a member of ARG (30--50\,Myr), this object has an estimated mass of $11.4 \pm 0.4$\,\MJup\ and a statistical distance of $34.2 \pm 4.8$\,pc.

\subsection{Low-Gravity Candidate members of YMGs}\label{sec:lowgm}

\textbf{2MASS~J00413538--5621127} (DENIS-P~J00041353--562112) has been identified as a candidate nearby, red dwarf by \cite{2001A&A...380..590P}, and spectroscopically confirmed by \cite{2007AJ....133.2258S} as an active M8 dwarf. Using high resolution optical spectroscopy, \cite{2009ApJ...702L.119R} revised its spectral type to M7.5 and showed evidence that it displays Li and signatures of active accretion, which indicates that it is a young, $\sim$\,10\,Myr BD. Based on its position, proper motion and RV, they suggest that it could be a member of THA, or an ejected member of $\beta$PMG, which would make it the first accreting BD discovered in either of these associations. \cite{2010ApJ...722..311L} reported that it is a binary with estimated spectral types of M6.5\,$\pm 1$ and M8 from photometry. In Paper~II, we corroborated that it is a high-probability candidate member of THA, with estimated masses of 14--41\,\MJup\ and 18--41\,\MJup\ for the individual components. This object was retrieved in \emph{BASS} as a high-probability candidate of THA. We obtained NIR spectroscopy for the unresolved system, and categorize it as a very low-gravity M7.5\,$\gamma$ BD system, which is consistent with its young age.

\textbf{2MASS~J02590146--4232204} was identified by \cite{2013ApJ...774..101R} as a candidate member of COL with infrared excess indicative of the presence of a circumstellar disk host. We independently identified this object as a candidate member of COL in \emph{PRE-BASS}; however, it was subsequently rejected from \emph{BASS} because of its low membership probability and the fact that its \emph{WISE} colors did not survive the extragalactic filter defined by \cite{2011ApJS..197...19K}, which is likely a consequence of its infrared excess. NIR spectroscopy allowed us to categorize it as an M5\,$\gamma$ dwarf. This is consistent with the results of \cite{2013ApJ...774..101R}, who reported that this object displays weak \ion{Na}{1} absorption that is indicative of a low surface gravity. Including the RV measurement of $15.3 \pm 1.5$\,\kms\ from \cite{2013ApJ...774..101R}, we find that this object is a low-probability candidate member of COL: this conclusion differs from that of \cite{2013ApJ...774..101R}, which found that 2MASS~J02590146--4232204 is a candidate member of THA. Obtaining a trigonometric distance will be useful to assess whether this object is a member of COL or THA.

\textbf{2MASS~J03264225--2102057} has been identified as an L4 dwarf with Li absorption by \cite{2007AJ....133..439C}. Using the DUSTY evolution models \citep{2000ApJ...542..464C}, the presence of Li and the spectrophotometric absolute magnitude of this object, they determined that it should be younger than 500\,Myr and less massive than 50\,\MJup. We identified this object as a highly probable L5\,$\beta$/$\gamma$ candidate member of ABDMG in \emph{PRE-BASS}. \cite{2002AJ....124.1170D} measured a trigonometric distance of $32.3 \pm 1.6$\,pc that is consistent with membership to ABDMG. The presence of low-gravity feature in its optical and NIR and optical spectra puts a slightly stronger constraint on the age of 2MASS~J03264225--2102057, since it is expected that gravity-sensitive spectral indices remain useful only up to $\sim$\,200\,Myr \citep{2009AJ....137.3345C,2013ApJ...772...79A}. We therefore categorize this object as a low-gravity L5\,$\beta$/$\gamma$ dwarf. An RV measurement is needed before it can be assessed whether this object is a bona fide member of ABDMG.

\textbf{2MASS~J04493288+1607226} was identified in \emph{PRE-BASS} as a candidate member of $\beta$PMG, but was rejected from the \emph{BASS} sample because of its proximity with TAU. NIR spectroscopy revealed that it is a low-gravity M9\,$\gamma$ dwarf. We estimate a distance of $54.9 \pm 10.0$\,pc for this object by comparison with other low-gravity dwarfs. A distance larger than 82\,pc can be excluded at a 99\% confidence level, which is incompatible with membership to TAU ($140 \pm 20$\,pc; \citealt{2007ApJ...671.1813T}) unless it is an unresolved multiple with at least 3 individual equal-luminosity components. This scenario is unlikely, especially considering that the NIR spectrum of 2MASS~J04493288+1607226 is not reddened. We thus preserve this object as a candidate member of $\beta$PMG.

\begin{figure*}
	\centering
	\subfigure[Galactic position]{\includegraphics[width=0.48\textwidth]{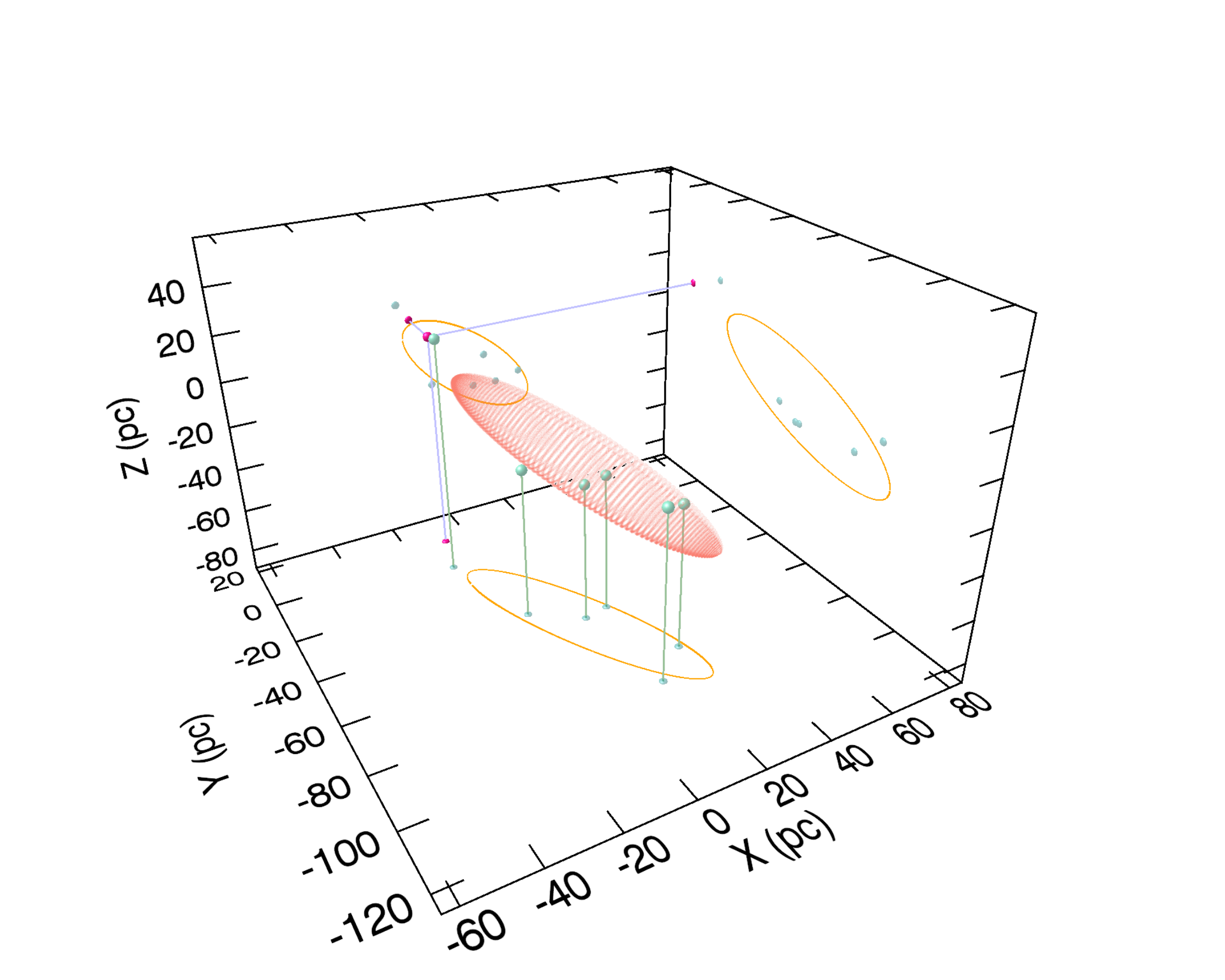}}
	\subfigure[Space velocity]{\includegraphics[width=0.48\textwidth]{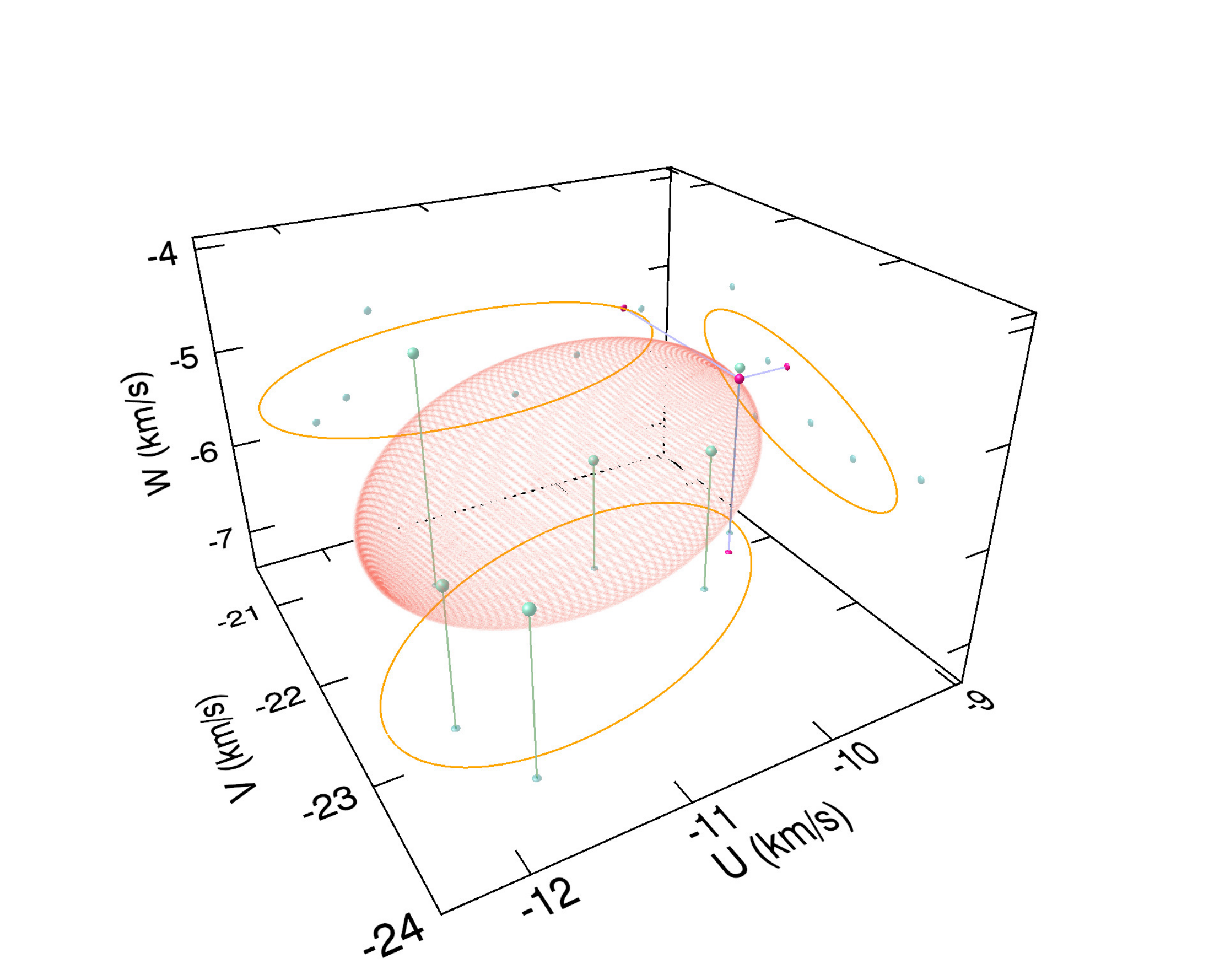}}
	\caption{Predicted galactic position $XYZ$ and space velocity $UVW$ of the CAR candidate member 2MASS~J11083081+6830169 (red point and its projections) using its statistical distance from BANYAN~II, compared with bona fide members of CAR (green points and their vertical projections on the $XY$ and $UV$ planes) and the SKM models of CAR (as defined in Paper~II; orange ellipsoid and its projections).}
	\label{fig:J1108XYZUVW}
\end{figure*}

\textbf{2MASS~J11083081+6830169} has been discovered by \cite{2000AJ....120.1085G} as an L1 dwarf in the optical with H$\alpha$ emission. We recovered this object in \emph{BASS} as a candidate member of ABDMG, and \cite{2002ApJ...575..484G} identified it as a candidate member of TWA. The RV of $-9.8 \pm 0.1$\,\kms\ measured by \cite{2010ApJ...723..684B} does not match the predicted RV of $-18.9 \pm 1.5$\,\kms\ for membership to ABDMG. It closely matches that of the CAR hypothesis ($-9.7 \pm 0.8$), but it still obtains a very low Bayesian probability of being a member of CAR. Its statistical distance ($15.3 \pm 0.8$\,pc) places it right into the locus of known young L dwarfs in both an $M_{W1}$ versus $J - K_S$ and $M_{W1}$ versus $H - W2$ CMDs. This distance places it at only 0.27\,\kms\ of the CAR bona fide member HIP~33737 in $UVW$ space, and at 17.2\,pc of the CAR bona fide member GJ~2079 in $XYZ$ space. We show in Figure~\ref{fig:J1108XYZUVW} its $XYZUVW$ position at its most probable distance: it seems that this object has a most probable position that is consistent with bona fide members of CAR, but our SKM fails to represent this. It can be expected that our SKM of CAR is not accurate because it was derived from a small number of bona fide members. Furthermore, both the NIR spectrum that we obtained and the optical spectrum from the RIZzo spectral library display clear signs of low-gravity and allowed us to categorize it as an L1\,$\gamma$ dwarf, which is consistent with membership to a YMG. A measurement of this object's trigonometric distance will be useful to assess whether or not it is a member of CAR, but we note that it is likely a member despite its low Bayesian membership probability.

%Figure : J1108 and CAR : XYZUVW

\textbf{2MASS~J12265135--3316124} (TWA~32) has been identified by \cite{2011ApJ...727....6S} as an UV-bright M6.5 low-mass star. They measured strong H$\alpha$ emission and Li absorption, as well as an RV of $7.15 \pm 0.26$\,\kms. They used this information as well as a photometric distance ($53 \pm 5$\,pc) to identify it as a new member of TWA, and they noted that it is a $656.1 \pm 0.4$\,mas visual binary with near-equal luminosity. \cite{2011ApJ...727...62R} independently discovered this object and measured strong H$\alpha$ and \ion{He}{1} emission at 5876\,\AA\ and 6678\,\AA, as well as strong Li absorption. They argued that the H$\alpha$ full width at 10\% of 270\,\kms\ is consistent with this object being a classical T~Tauri star. They measured an RV of $14.8 \pm 3$\,\kms\ and note that its $UVW$ space velocity is consistent with TWA and SCC. We recovered this object in \emph{PRE-BASS} as a candidate member of TWA, and obtained NIR spectroscopy that allowed us to assign it a spectral type of M5.5\,$\gamma$. At this spectral type, only the weaker \ion{Na}{1} absorption is a useful low-gravity indicator. We adopted the RV measurement of \cite{2011ApJ...727....6S} which is more precise, and, like them, found that this object is a strong candidate member of TWA. The BANYAN~II statistical distance corresponding to the TWA hypothesis is $61.8^{+6.4}_{-6.0}$\,pc, which is consistent with the photometric estimate of \cite{2011ApJ...727....6S} that takes its binary nature into account. The SKMs of BANYAN~II do not take SCC into account, which includes the Lower Centaurus Crux (LCC) and the Upper Centaurus Lupus (UCL) regions, hence our result does not preclude membership to SCC. The space velocity $UVW$ for this object is ($-8.6 \pm 1.4$, $-15.7 \pm 1.1$, $-3.4 \pm 1.1$)\,\kms\ \citep{2011ApJ...727....6S}, at 4.6\,\kms\ from the kinematic center of TWA (Paper~II), 4.6\,\kms\ from that of UCL and 4.2\,\kms\ from that of LCC \citep{2003A&A...404..913S}. Its kinematics are thus consistent with SCC and TWA; however, its photometric distance is not consistent with the distance of this complex ($\sim$\,100--150\,pc; \citealp{2003A&A...404..913S}), whereas it is consistent with that of TWA members ($\sim$\,40--62\,pc; Paper~II; \citealp{2013ApJ...762..118W}; \citealp{2014A&A...563A.121D}). We conclude that TWA~32 is a likely member of TWA, unless it is a multiple system composed of at least three equal-luminosity components. A trigonometric distance measurement will be useful to assess this.

\textbf{2MASS~J20391314--1126531} was discovered as an M9 dwarf by \cite{2003AJ....126.2421C} using optical spectroscopy. \cite{2010MNRAS.409..552G} reported that it is a candidate member of the Pleiades stream. \cite{2005A&A...430..165F} demonstrated that the Pleiades stream is not a moving group but rather a dynamical stream of stars without a common origin. We identified 2MASS~J20391314--1126531 as a candidate member of ABDMG as part of \emph{PRE-BASS} and obtained NIR spectroscopy which revealed that this is a low-gravity M7\,$\beta$ dwarf. The RV of $-18.0 \pm 2$\,\kms\ that was measured by \cite{2010MNRAS.409..552G} is consistent with a membership to ABDMG, and the fact that it has a low gravity indicates that it might not be a contaminant from the Pleiades stream. A measurement of its distance will be needed to assess this.

\subsection{Candidate members of YMGs with no age constraint}\label{sec:noconstraintmg}

\textbf{2MASS~J03582255--4116060} has been discovered by \cite{2007AJ....133..439C} as an L5 BD in the optical. We identified it as a low-probability candidate member of $\beta$PMG as part of \emph{BASS}. $R \sim 75$ NIR spectroscopy allowed us to categorize it as a peculiar L6 dwarf. Its continuum is redder and its $H$ band is slightly more triangular than our field L6 template, however it is unclear at this time if these effects are due to a low gravity or not. Obtaining a higher-resolution spectrum would be useful to assess this. If we assume an age of 20--26\,Myr and the BANYAN~II statistical distance associated with the $\beta$PMG hypothesis ($18.1 \pm 3.2$\,pc) and compare its \emph{2MASS} and \emph{WISE} photometry with the BT-Settl isochrones, we find that this object has one of the lowest estimated mass of all candidate YMG members reported here, with $8.2 \pm 0.6$\,\MJup.

\textbf{2MASS~J08095903+4434216} was identified by \cite{2004AJ....127.3553K} and confirmed by \cite{2006AJ....131.2722C} as an L6 dwarf. In Paper~V, we identified it as a candidate member of ARG as part of \emph{BASS}. We used its NIR spectrum to revise its spectral type to L6\,pec\,(red) from a visual comparison with field and low-gravity templates. This object has a red continuum and red NIR colors for its spectral type, with $J - K_S = 2.02$ and $J - W2 = 3.63$, compared with median values of $J - K_S = 1.7 \pm 0.3$ and $J - W2 = 2.9 \pm 0.4$ for field L6 dwarfs (Figure~\ref{fig:CC2}). The low-resolution gravity classification scheme of \cite{2013ApJ...772...79A} categorizes it as an intermediate-gravity L5.4 dwarf due to its $H$--cont, KI$_J$ and FeH$_Z$ indices. However, it is visually a better match to the field L6 template than the field L5 template, albeit it displays a slightly redder continuum. Adopting a spectral type of L6, only the $H$--cont index remains useful and categorizes it as an intermediate-gravity dwarf, but this index alone does not reject the possibility that this object is a dusty dwarf in the field. \cite{2014AJ....147...34S} demonstrated that the H$_2(K)$ index defined by \cite{2013MNRAS.435.2650C} seems to be gravity-sensitive up to at least L8; we obtain a value of H$_2(K) = 1.056 \pm 0.008$ for 2MASS~J08095903+4434216, which is slightly lower than the typical values for field L6 dwarfs ($1.06 \pm 0.01$; see Figure~\ref{fig:agecalib2k} of this work and Figure~10 of \citealt{2014AJ....147...34S}). It is unclear at this time if this object is a low-gravity L6 dwarf; a higher resolution (R\,$\gtrsim$\,750) NIR spectrum will be useful to confirm if this object is a very low-mass, very late-type candidate member of ARG, or more massive and dusty field interloper. At the age of ARG (30--50\,Myr), this object would have one of the lowest estimated masses amongst all YMG candidates presented here, with $8.1 \pm 0.8$\,\MJup. Its statistical distance associated with membership to ARG is $15.3 \pm 2.0$\,pc.

\textbf{2MASS~J23512200+3010540} was discovered by \cite{2010ApJS..190..100K} as L5.5 dwarf in the optical, and as an unusually red L5.5 dwarf in the NIR. In Paper~II, we identified this object as a candidate member of ARG, and it was recovered as such in \emph{PRE-BASS}. We used its NIR spectrum to categorize it as a peculiar L5 dwarf. Only the $H$--cont index is indicative of a possible young gravity; it is thus a likely scenario that this object is a dusty field interloper. Obtaining a higher-resolution spectrum would be useful to assess this. If it is a member of ARG (30--50\,Myr), this object has an estimated mass of $10.0^{+0.6}_{-0.7}$\,\MJup\ and a statistical distance of $20.1 \pm 1.8$\,pc.

\subsection{Interlopers from the Field or Other Regions}\label{sec:fieldinterlop}

\textbf{2MASS~J00174858-0316334} was identified as a candidate member of ABDMG as part of \emph{PRE-BASS}. NIR spectroscopy revealed that this is a reddened low-gravity M7\,$\beta$ dwarf. We de-reddened its spectrum using the \emph{fm\_unred.pro} IDL routine based on the extinction law of \cite{1999PASP..111...63F} and visually compared it with our M7\,$\beta$ template to determine that its total extinction is $A(V) = 2.5$. We used the parametrization of \cite{1999PASP..111...63F} with a total-to-selective extinction of $R(V) = 3.1$. This reddening is unlikely caused by interstellar dust, since this object is far from the galactic plane ($b = -64.8$\textdegree) and has a spectro-photometric distance of only $65.1 \pm 11.5$\,pc (low gravity was considered in this estimate). It can be expected that this object is still embedded in its formation material, which indicates that it is not a member of ABDMG, but rather a member of another young star-forming region that might not be known. It would be interesting to investigate whether other very young objects can be found in its vicinity.

\begin{figure}
	\centering
	\includegraphics[width=0.48\textwidth]{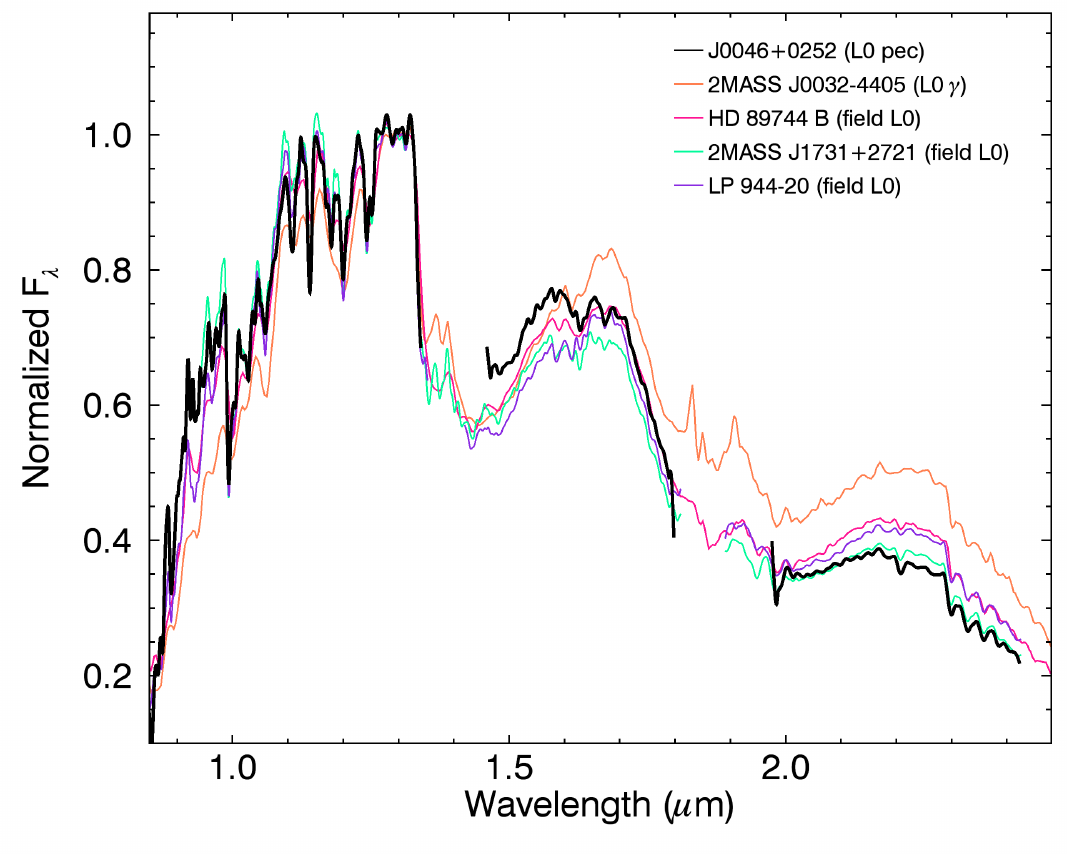}
	\caption{NIR spectrum of the peculiar L0 dwarf 2MASS~J00461551+0252004 that was recovered as a candidate member of ABDMG in \emph{PRE-BASS}. The $H$-band bump at 1.57\,$\mu$m could be a hint of an unresolved T-type companion.}
	\label{fig:J0046_pec}
\end{figure}

\textbf{2MASS~J00461551+0252004} was identified as a candidate member of ABDMG in \emph{PRE-BASS}. NIR spectroscopy revealed that it is a peculiar L0 dwarf with no indication of low gravity. The $H$-band bump at 1.57\,$\mu$m is significantly stronger than that of field dwarfs, which could hint at an unresolved T-type component (Figure~\ref{fig:J0046_pec}); however, the spectral indices constructed by \cite{2014ApJ...794..143B} do not categorize it as a likely L-type + T-type binary. The cause of its peculiar properties is thus unclear.

%Figure : J0046

\begin{figure}
	\centering
	\includegraphics[width=0.48\textwidth]{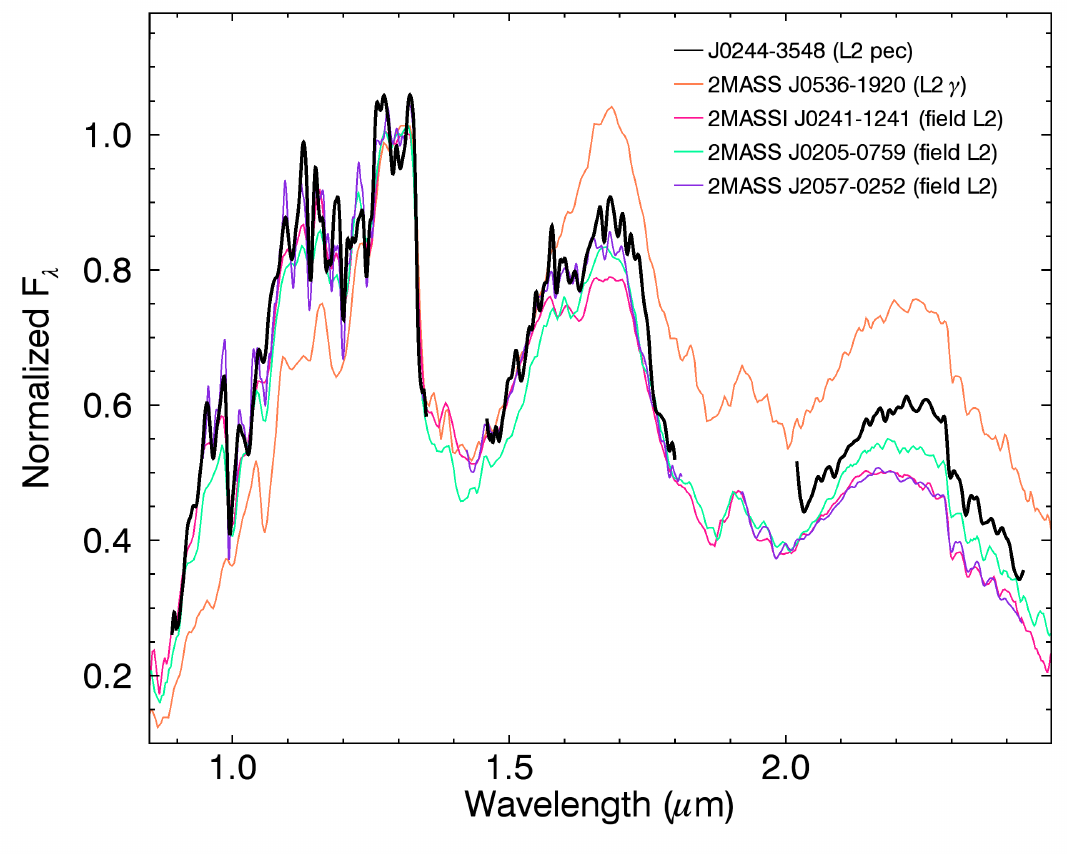}
	\caption{NIR spectrum of the peculiar L1 dwarf 2MASS~J02441019--3548036 that was recovered as a candidate member of THA in \emph{BASS}. Its NIR continuum is redder than usual and the shape of its $H$ band is peculiar. This could be explained by a dusty atmosphere; however, this object is likely older than the YMGs considered here.}
	\label{fig:J0244_pec}
\end{figure}

\textbf{2MASS~J02441019--3548036} was discovered as candidate member of THA in \emph{BASS}. NIR spectroscopy ($R\sim 750$) reveals that it is an L2 dwarf that lacks the weaker alkali lines or stronger VO absorption that are typical of low-gravity dwarfs. However, its continuum is unusually red for an L2 dwarf and the shape of its $H$ band is unusual (Figure~\ref{fig:J0244_pec}). This could be explained by an unusually dusty atmosphere or an unresolved later-type companion; however, the classification of \cite{2014ApJ...794..143B} based on various spectral indices does not categorize it as a candidate binary. We reject it as a candidate member of THA, as the weaker-than-usual alkali lines are not consistent with a young age even if this object is dusty or multiple.

%Figure : J0244

\textbf{2MASS~J02530084+1652532} (Teegarden's star) was discovered by \cite{2003ApJ...589L..51T} as a nearby ($2.43 \pm 0.54$\,pc) M6.5 dwarf; \cite{2006AJ....132.2360H} refined its distance measurement to $3.85 \pm 0.01$\,pc. \cite{2011A&A...529A..44W} identified this object as a potential low-gravity dwarf from atmosphere model fitting, and we identified it as a candidate member of ARG in Paper~II, but its distance is not consistent with this possibility. We obtained the NIR spectrum of \cite{2012ApJ...745...26B} from the SpeX PRISM spectral library and categorized it as an M7.5\,$\beta$ dwarf. The classification scheme of \cite{2013ApJ...772...79A} assigns it an intermediate gravity, which is consistent with our visual comparison. This is due to a low FeH$_Z$ index ($1.0699 \pm 0.0040$), a low KI$_J$ index ($1.0511 \pm 0.0065$) and a high $H$-cont index ($0.9974 \pm 0.0049$) compared with field M7.5 dwarfs. This object has a relatively blue \emph{2MASS} $J - K_S$ color ($0.809 \pm 0.053$) for an M7.5 dwarf (Figure~\ref{fig:JK}), which is not expected for a low-gravity dwarf. We used the NIR spectrum to measure its synthetic NIR colors and obtained $J - K_S = 0.869 \pm 0.077$ (assuming a 5\% uncertainty in the \emph{2MASS} photometric zero points), which places it closer to the locus of low-gravity and field M7 dwarfs, albeit still on the blue end. Its \emph{WISE} $W1 - W2$ color ($0.265 \pm 0.034$) is consistent with field and low-gravity M7.5 dwarfs (Figure \ref{fig:W1W2}). Obtaining a higher-resolution NIR spectrum will be useful to assess whether alkali lines are weaker than usual, which would confirm if this object has a low gravity or not. Another explanation could be that this object is an unresolved binary. If Teegarden's star is young, it could be a member of a YMG that is not considered here, and it would thus be interesting to measure its RV. It is worthwhile mentioning that this object would be the nearest low-gravity dwarf if this is confirmed, a record currently held by LP~944--20 (an L0\,$\beta$ at $6.41 \pm 0.04$\,pc that is a candidate member of the Castor stream; \citealp{2001ApJ...548..908L,2013ApJ...772...79A,2014AJ....147...94D,1998A&A...339..831B}).

\textbf{2MASS~J03140344+1603056} was identified by \cite{2007AJ....133.2258S} as an L0 dwarf with H$\alpha$ emission. \cite{2010A&A...512A..37S} measured its RV and used its kinematics to assign it as a candidate member of UMA. We initially identified this object as a low-probability candidate member of $\beta$PMG in \emph{PRE-BASS}, but was later rejected because of its large contamination probability, as well as its position on a $M_{W1}$ versus $H-W2$ diagram that is not consistent with young BDs at the most probable statistical distances obtained from BANYAN~II. A NIR follow-up allowed us to categorize it as a peculiar M9 dwarf with no apparent sign of low gravity from the classification scheme of \cite{2013ApJ...772...79A} or a visual comparison with spectroscopic standards. However, it is unclear at what exact age signs of low-gravity stop being apparent in moderate-resolution NIR spectra, and \cite{2013ApJ...772...79A} suggest that this might take place around $\sim$\,200\,Myr. We thus reject any possible membership with the younger moving groups considered here, but our data is insufficient to corroborate its possible membership to UMA.

\textbf{2MASS~J04070752+1546457} has been identified as an L3.5 dwarf by \cite{2008AJ....136.1290R} from optical spectroscopy. We identified this BD as an ambiguous candidate member of $\beta$PMG and COL as part of \emph{PRE-BASS}, but we subsequently rejected it because of its alignment with TAU. NIR spectroscopy allowed us to categorize it as field L3 BD. It displays marginal signs of low-gravity (weaker FeH and slightly weaker alkali line widths); however, all other features as well as a visual comparison with spectroscopic standards are consistent with a field L3 BD. It could be interesting to investigate whether this object has a peculiar metallicity or a slightly young age ($\sim$\,200 to a few hundred Myrs), but it is most probably not a member of any YMG considered here.

\textbf{2MASS~J05243009+0640349} has been identified as a potential member of $\beta$PMG in \emph{PRE-BASS}. It has been subsequently excluded from \emph{BASS} because of its low Bayesian probability, but its NIR spectrum allowed us to categorize it as a low-gravity M5.5\,$\beta$. The only useful sign of low gravity for this spectral type is the weaker \ion{Na}{1} absorption. This object has a low galactic latitude ($b = -15.92$\textdegree) and is located within the Orion~II super bubble \citep{1974ApJ...191L.121G} at only 10\farcm6 of the Ori~C~11 core (see Figure~12b of \citealp{1994ApJS...95..457W}; the B1950 coordinates of this object are 05h21m48.67s, +06\textdegree37$'$54\farcs5). These clouds are located at significantly larger distances ($\sim$\,400--500\,pc;  \citealp{2014ApJ...786...29S}) compared to the YMGs considered here. Using the absolute \emph{2MASS} and \emph{WISE} photometry of known young M5.5 dwarfs, we estimate a spectrophotometric distance of $42.0 \pm 7.7$\,pc for 2MASS~J05243009+0640349 and exclude a distance larger than 63\,pc at a 99\% confidence level, assuming it is not a multiple system. This discrepancy, supplemented with the fact that its spectrum does not seem reddened by interstellar dust, makes it unlikely that this object is a member of the Orion Molecular Complex (OMC) even though it is clearly young. Even when its youth is taken into account, this object has a very low probability of being a member of $\beta$PMG. It will be useful to obtain a distance and RV measurement to investigate whether this object is a member of another YMG that is not considered here.

\textbf{2MASS~J05271676+0007526} has been identified as a potential member of $\beta$PMG in \emph{PRE-BASS}. Its low Bayesian probability as well as color filters ($H - K_S > 0.269$ and $VR - J \geq 2.63$; Paper~II) excluded it from the \emph{BASS} sample. This object has a low galactic latitude ($b = -18.57$\textdegree) and is located in the vicinity of the OMC \citep{1974ApJ...191L.121G, 2014ApJ...786...29S}. Acquisition images obtained with SpeX revealed that this is a 2\farcs4 visual binary. We obtained resolved NIR spectroscopy and determined that both components are reddened early M dwarfs. We de-reddened both spectra using the \emph{fm\_unred.pro} IDL routine based on the extinction law of \cite{1999PASP..111...63F} and visually compared the results with NIR spectroscopic standards to determine the best matching spectral types and total extinction $A(V)$, using the parametrization of \cite{1999PASP..111...63F} with a total-to-selective extinction of $R(V) = 3.1$. We find a best match of $A(V) = 0.93$ with spectral types M0 + M3. We note that the $H$-band continuum of both objects has a rounded triangular shape, which is only seen at those spectral types for very young ($\lesssim$\,5\,Myr) objects. This system is thus likely very young and still embedded in its formation material. Using the \emph{2MASS} $J$ magnitude of the unresolved system with the young absolute magnitude-spectral type sequences of \cite{2013ApJ...762...88M}, we estimate a distance of $\sim$\,450\,pc. We conclude that this system is a probable very young low-mass star member of OMC.

\textbf{2MASS~J08503593+1057156} (2MASS~J0850+1057) was first identified from the \emph{2MASS} survey as an L6 BD by \cite{1999ApJ...519..802K}. Subsequently, \cite{2001AJ....121..489R} and \cite{2008A&A...481..757B} identified and confirmed that this object is a 0\farcs16 binary system, and \cite{2011AJ....141...70B} used a template fitting method constrained by the flux ratio of its individual components to assign them spectral types of L7 and L6. They noted the surprising fact that the brighter primary component is the one that gets assigned a later spectral type. They argue that this could be explained either by youth or the latest-type component being an unresolved binary. \cite{2011AJ....141...71F} subsequently identified the NLTT~20346 M5+M6 binary system as a very wide ($\sim$\,7700\,AU) co-moving companion to 2MASS~J0850+1057. They assign an age estimate of 250--450\,Myr for NLTT~20346 based on X-ray luminosity, but they note that this estimate is discrepant with that based on H$\alpha$ emission ($6.3 \pm 1.0$\,Gyr and $6.5\pm 1.0$\,Gyr for its respective components). They measure a systemic RV of $26 \pm 9$\,\kms\ for NLTT~20346, and a trigonometric distance of $29 \pm 7$\,pc for 2MASS~J0850+1057. They note that this latter measurement is not precise enough to discriminate between two previous inconsistent measurements in the literature ($38 \pm 6$\,pc from \citealp{2004AJ....127.2948V} and $25.6 \pm 2.3$ pc from \citealp{2002AJ....124.1170D}). Using their proper motion measurement of $\mu_\alpha = -144 \pm 6$\,\masyr and $\mu_\delta = -38 \pm 6$\,\masyr, they argued that a faint background contaminant was blended at the epochs used for previous distance measurements, which could explain the discrepancy.

\cite{2012ApJS..201...19D} independently measured a proper motion of $-144.2 \pm 0.6$\,\masyr, $\mu_\delta = -12.6 \pm 0.6$\,\masyr\ and a distance of $33.2 \pm 0.9$\,pc for 2MASS~J0850+1057. They also refined the photometry of its resolved components, and used these new measurements to draw different conclusions than those outlined above. First, they used a similar analysis to that of \cite{2011AJ....141...70B} with their updated photometry to argue that the spectral types of the components are rather L6.5\,$\pm 1$ and L8.5\,$\pm 1$, with the fainter component now associated with the later spectral type. This conclusion does away with the need to invoke youth or any additional component, which was previously based on flux reversal \citep{2011AJ....141...70B}. They thus argued that 2MASS~J0850+1057 is a BD system displaying no notable peculiarity. Furthermore, they use their new proper motion measurement at 6.7$\sigma$ from that of NLTT~20346 with the criterion of \cite{2007AJ....133..889L} to argue that the two systems are likely random alignments, and thus not gravitationally linked.

\begin{figure}
	\centering
	\includegraphics[width=0.48\textwidth]{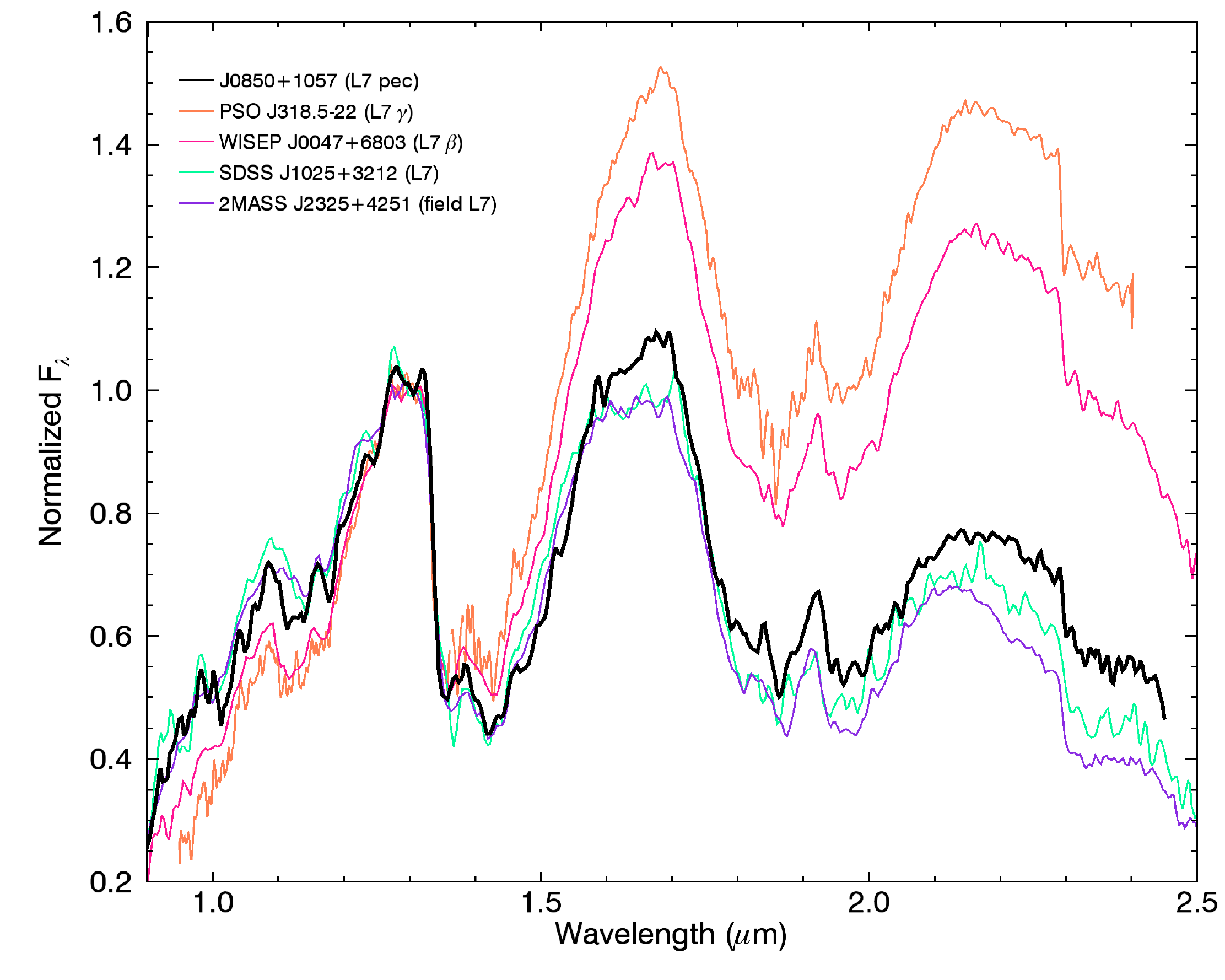}
	\caption{NIR spectrum of the peculiar L7 dwarf 2MASS~J0850+1057 that was recovered as a candidate member of ARG in \emph{BASS}. This system is a binary with resolved spectral types of L6.5\,$\pm 1$ and L8.5\,$\pm 1$ estimated from photometry. The NIR continuum of 2MASS~J0850+1057 is redder than field L7 dwarfs which is likely an effect of its binary nature, and its kinematics are not consistent with those of ARG.}
	\label{fig:J0850_pec}
\end{figure}

We measure a proper motion of $-141.1 \pm 7.7$\,\masyr\ and $\mu_\delta = -13.1 \pm 9.5$\,\masyr, based on \emph{2MASS} and \emph{ALLWISE}. The $\mu_\alpha$ component is consistent with both measurements, whereas the $\mu_\delta$ component is at 2.2$\sigma$ and 0.05$\sigma$ respectively from the measurements of \cite{2011AJ....141...71F} and \cite{2012ApJS..201...19D}. Our measurement thus favors the later one, but our precision is $\sim\,16$ times lower. We find that even if both components seem to display no peculiarity in their relative fluxes, the unresolved system seems to be unusually red for its absolute magnitude (Figure~\ref{fig:J0850_pec}). This could be explained either by additional unresolved later-type components, or the presence of thicker/higher clouds in their atmosphere compared to field BDs. We used the $R \sim 120$ NIR spectrum for the unresolved BD system to assign a spectral type of L7. For such a late spectral type, the only features known to be gravity-sensitive in a low-resolution NIR spectrum are the $H$-cont index of \cite{2013ApJ...772...79A} and the H$_2(K)$ index of \cite{2013MNRAS.435.2650C}. We find values of $H$-cont\,$= 0.872 \pm 0.025$ and H$_2(K) = 1.055 \pm 0.014$, both being only marginally consistent with a low surface gravity. The parallax and proper motion measurements of \cite{2012ApJS..201...19D} preclude a possible membership to ARG; however, obtaining a higher-resolution NIR spectrum for this system would be interesting to assess whether it displays signs of low surface gravity. If this system is younger than $\sim$\,200\,Myr, the individual mass of each component would be well below the deuterium burning limit, which would make it a remarkable benchmark system to understand the properties of planetary-mass objects.

%Figure : J0850

\begin{figure}
	\centering
	\includegraphics[width=0.48\textwidth]{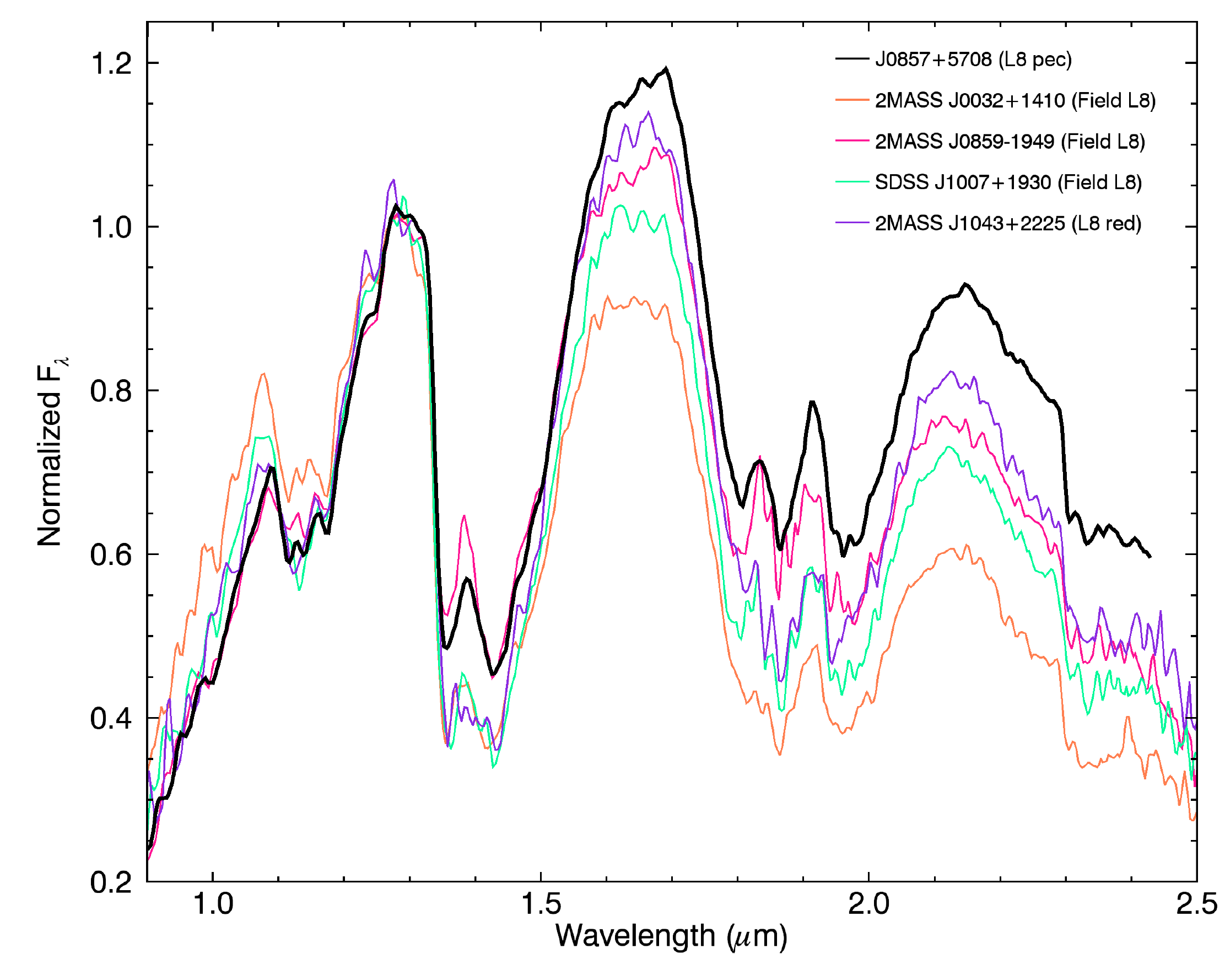}
	\caption{NIR spectrum of the peculiar L8 dwarf 2MASS~J08575849+5708514 that was recovered as a candidate member of ARG in \emph{BASS}. Its red NIR continuum could be an indication of a low surface gravity, but an RV measurement that was obtained from low-S/N data is not consistent with membership to ARG.}
	\label{fig:J0857_pec}
\end{figure}

\textbf{2MASS~J08575849+5708514} has been discovered by \cite{2002ApJ...564..466G} as an L8\,$\pm 1$ BD. \cite{2009ApJ...702..154S} used atmosphere model fitting to determine that this object is unusually cloudy and seems to have a low surface gravity ($\log g = 4.5$). Using our visual comparison with spectral templates, we categorize this object as a peculiar L8 dwarf (Figure~\ref{fig:J0857_pec}). No indices from \cite{2013ApJ...772...79A} are gravity-sensitive for such a late spectral type, but the H$_2(K)$ index defined by \cite{2013MNRAS.435.2650C} seem to remain useful \citep{2014AJ....147...34S}. We find a weaker H$_2(K)$ value ($1.102 \pm 0.006$) compared with typical field L8 dwarfs ($1.12 \pm 0.02$), which could be an indication of a lower surface gravity. We identified this object as a highly probable candidate member of ARG in \emph{BASS}, with an estimated mass of $8.5 \pm 0.8$\,\MJup\ and an estimated distance of $8.9 \pm 0.8$\,pc. However, \cite{2010AJ....139.1808S} measured an RV of $-123.5 \pm 20.0$ \kms, which is not consistent with membership to ARG or even with the kinematics of any young BD in the solar neighborhood. This large RV thus seems contradictory with its unusually red colors and tentative indications of a lower surface gravity, but it was measured from a low-signal optical spectrum. It is likely that this object is not a young member of ARG but rather an interloping cloudy object from the field. However, obtaining an RV measurement from higher-S/N data will be useful to assess this.

%Figure : J0857

\textbf{2MASS~J11335700--7807240} was identified by \cite{2007ApJS..173..104L} in the optical as an M8 dwarf in a search for new members of the Chamaeleon I (CHA) star-forming region. They rejected it because it lacks low-gravity indications in its optical spectrum. We independently recovered this object in \emph{PRE-BASS} as a candidate member of CAR and obtained a NIR spectrum. We categorize this object as a peculiar M6$ \pm 1$ dwarf; its $H$ band is more triangular and the slope of its $K$ band is bluer. This could be indicative of a low surface gravity, but it lacks all the other usual signatures: only the 1.253\,$\mu$m \ion{K}{1} line is slightly weaker than that of field M6 dwarfs. All other \ion{K}{1} lines, the \ion{Na}{1} doublet and FeH absorption are all consistent with a field M6 dwarf : the classification scheme of \cite{2013ApJ...772...79A} thus categorizes this object as a field-gravity M6 dwarf. It is unclear what is the source of the peculiar features in this object's NIR spectrum. It is possible that the triangular-shaped $H$ band could be caused by dust in its photosphere \citep{2013ApJ...772...79A}, but this would be unusual at such an early spectral type, and neither its $J - K_S$ color ($1.01 \pm 0.04$) or its $J - W2$ color ($1.53 \pm 0.04$) are redder than those of field M6 dwarfs, which would be unexpected for a dusty object. We thus categorize this object as a peculiar M6 dwarf and reject it as a candidate member of CAR.

\textbf{2MASS~J11555389+0559577} was discovered by \cite{2004AJ....127.3553K} as an L7.5 dwarf using NIR spectroscopy. We recovered this object in \emph{PRE-BASS} as a candidate member of ARG. \cite{2012ApJ...752...56F} measured a trigonometric distance of $17.27 \pm 3.04$\,pc and \cite{2010AJ....139.1808S} used a low-quality optical spectrum from SDSS to categorize it as an L0 dwarf and measure an RV of $136.8 \pm 20.0$\,\kms. If we include only the trigonometric measurement, it remains a modest candidate of ARG; however, the RV measurement is not consistent with this, nor with the kinematics of nearby, young dwarfs \citep{2009AJ....137....1F}, much like the case of 2MASS~J08575849+5708514. We retrieved the NIR spectrum of this object from the SpeX PRISM Libraries and categorize it as a peculiar L6--L8 dwarf. It lacks the triangular-shaped $H$-band continuum that would be expected for a young object ($H$-cont $= 0.8230 \pm 0.0087$), and its NIR colors are consistent with those of field dwarfs. We measured the gravity-sensitive H$_2$($K$) index defined by \cite{2013MNRAS.435.2650C} and find a value of $1.0862 \pm 0.0085$, which is consistent with field L6--L7 dwarfs \citep{2014AJ....147...34S}. Higher-resolution NIR spectroscopy as well as an RV measurement derived from a high signal-to-noise spectrum would be needed to completely rule out low gravity, but it is very likely that this object is a regular BD; we thus reject it as a candidate member of ARG.

%Figure : J2048
\begin{figure}
	\centering
	\includegraphics[width=0.48\textwidth]{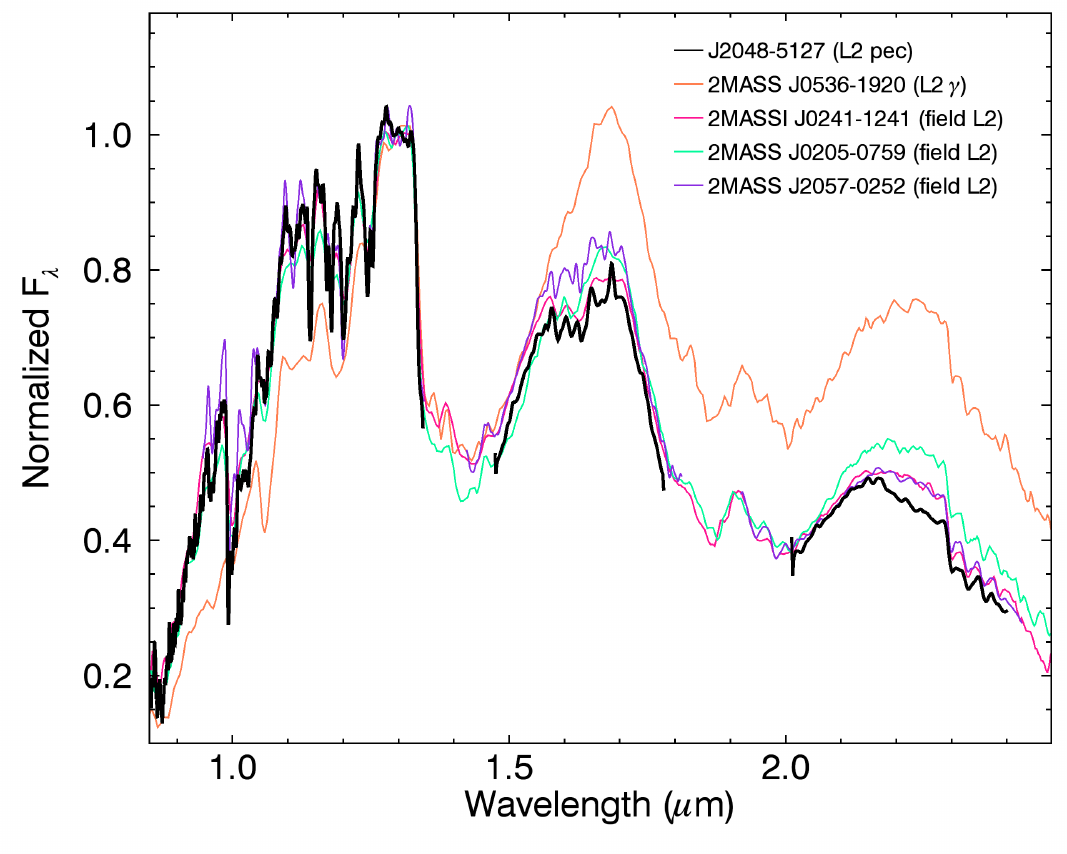}
	\caption{NIR spectrum of the peculiar L2 dwarf 2MASS~J20484222--5127435 that was recovered as a candidate member of THA in \emph{BASS}. It lacks indications of a low surface gravity and is thus not a likely member of THA. Its $K$ band is peculiar, as it resembles those of later-type L dwarfs.}
	\label{fig:J2048_pec}
\end{figure}

%Figure : J2206-6116
\begin{figure}
	\centering
	\includegraphics[width=0.48\textwidth]{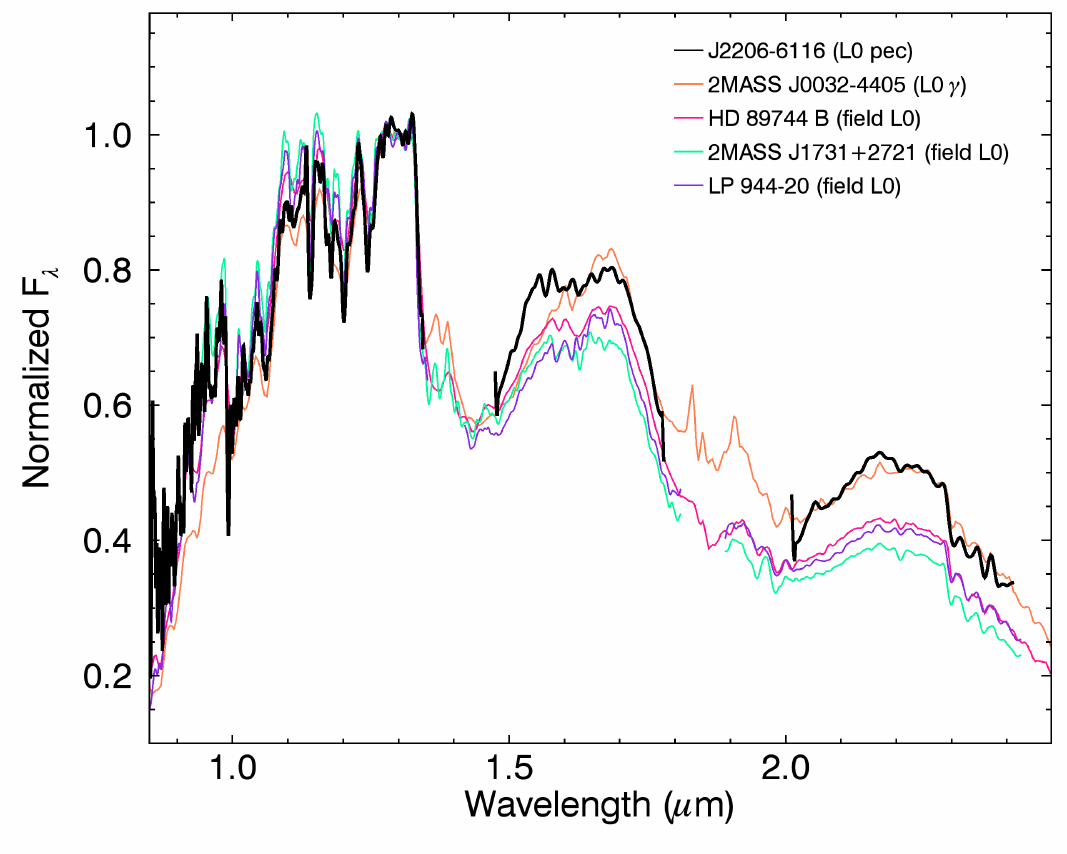}
	\caption{NIR spectrum of the peculiar L2 dwarf 2MASS~J22062157--6116284 that was recovered as a candidate member of THA in \emph{PRE-BASS}. It lacks indications of a low surface gravity and is thus not a likely member of THA. Furthermore, its $H$-band bump at $\sim$\,1.57\,$\mu$m is stronger than usual.}
	\label{fig:J2206_pec}
\end{figure}

%Figure : J2315-4747.
\begin{figure}
	\centering
	\includegraphics[width=0.48\textwidth]{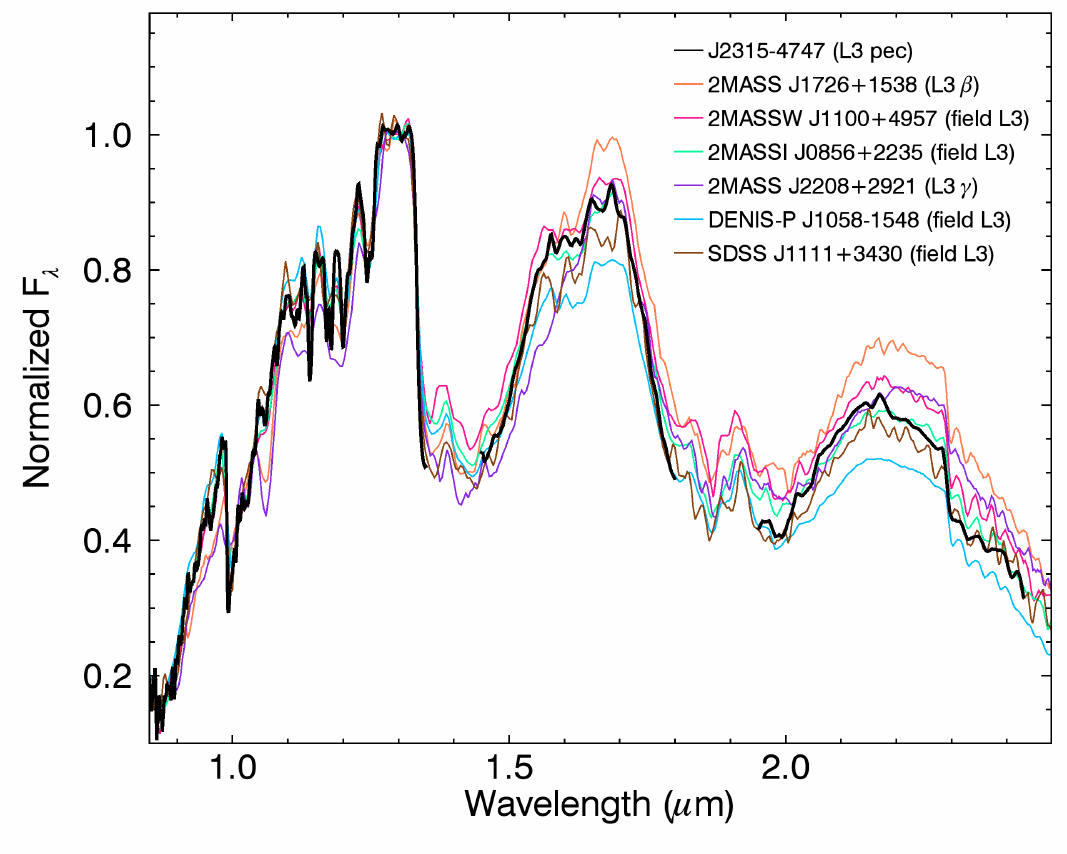}
	\caption{NIR spectrum of the peculiar L3 dwarf 2MASS~J23155665--4747315 that was recovered as a candidate member of THA in \emph{BASS}. It lacks indications of a low surface gravity and is thus not a likely member of THA. Its $J$ band is similar to a field L3 dwarf with stronger FeH absorption, and its $H$ and $K$ bands are consistent with a later spectral type.}
	\label{fig:J2315_pec}
\end{figure}

%Figure : J2339+3507.
\begin{figure}
	\centering
	\includegraphics[width=0.48\textwidth]{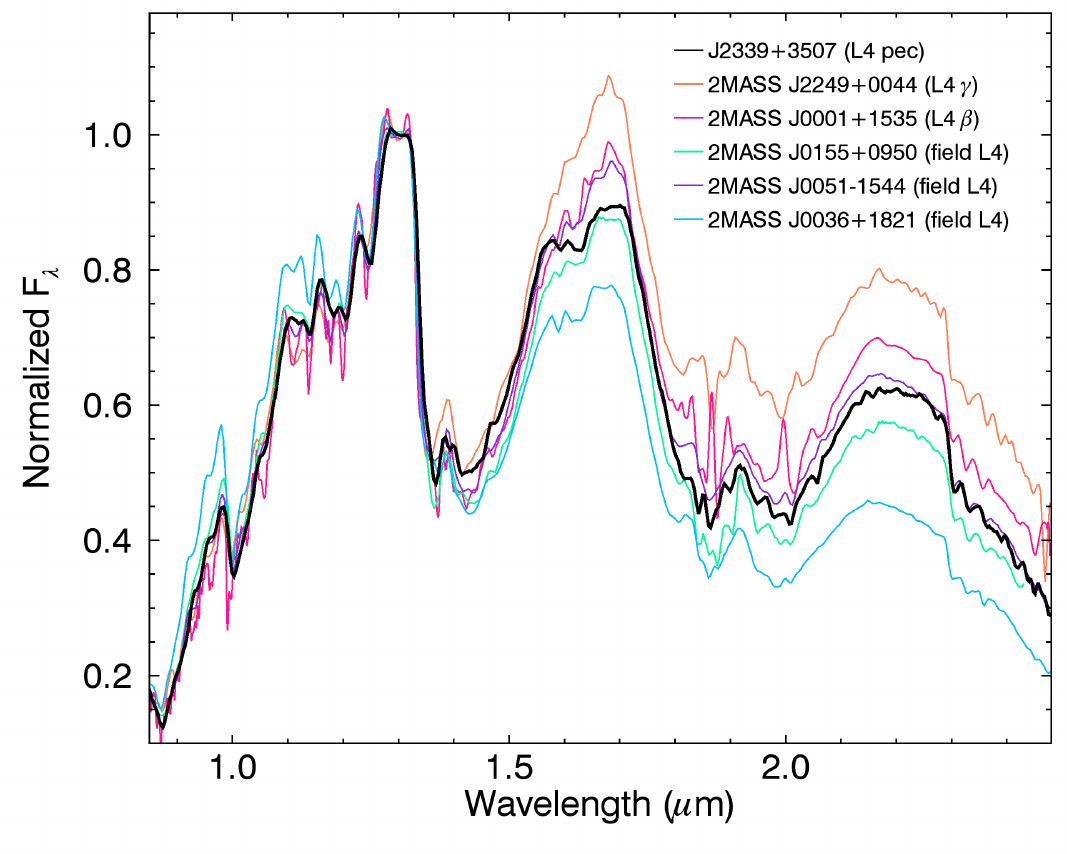}
	\caption{NIR spectrum of the peculiar L4 dwarf 2MASS~J23392527+3507165 that was recovered as a candidate member of $\beta$PMG in \emph{BASS}. It lacks indications of a low surface gravity and is thus not a likely member of $\beta$PMG. Furthermore, its $H$-band bump at $\sim$\,1.57\,$\mu$m is stronger than usual.}
	\label{fig:J2339_pec}
\end{figure}

\textbf{2MASS~J20484222--5127435} was identified as a candidate member of THA as part of \emph{BASS}. NIR spectroscopy revealed that its $J$ and $H$ bands are similar to a field L2 dwarf, but its $K$ band is significantly different, and similar to the $K$ band of field L5 dwarfs (Figure~\ref{fig:J2048_pec}). This object is thus unlikely young and we reject it as a candidate member of THA; however, it is unclear what is the cause of its peculiar $K$ band. It would be worthwhile investigating whether this is an early-L / mid-L binary from high-resolution imaging or an RV follow-up.

\textbf{2MASS~J22062157--6116284} has been identified as a candidate member of THA in \emph{PRE-BASS}. We obtained NIR spectroscopy and categorized it as a peculiar L0 $\pm 1$ dwarf because its $H$-band flux at $\sim$\,1.57\,$\mu$m is stronger than usual (Figure~\ref{fig:J2206_pec}). This could indicate the presence of an unresolved T-type component; however, the index-based scheme of \cite{2014ApJ...794..143B} indicates that this scenario is unlikely.

\textbf{2MASS~J23155665--4747315} has been identified as a candidate member of THA in \emph{BASS}. We obtained NIR spectroscopy and categorized it as a peculiar L3 dwarf; its $J$ band is similar to an L3 dwarf albeit with stronger FeH absorption, and its $H$ and $K$ bands are similar to our field L5 template (Figure~\ref{fig:J2315_pec}).

\textbf{2MASS~J23310161--0406193} (Koenigstuhl~3~BC) was discovered by \cite{2000AJ....120.1085G} as an M9 dwarf in the optical, and \cite{2003AJ....125.3302G} demonstrated that it is an M8 + L3, 0\farcs58 binary system. \cite{2007ApJ...667..520C} discovered that this system is a very wide 451\farcs8 comoving system to the F8 star HR~8931 (HD~221356). Koenigstuhl~3~BC was identified as a candidate member of ABDMG in \emph{LP-BASS}, but measurements of RV and distance for the co-moving star HR~8931 ($-12.86 \pm 0.09$\,\kms\ and $26.12 \pm 0.37$\,pc; \citealp{2002ApJS..141..503N,2007A&A...474..653V}) preclude a possible membership to all YMGs considered here. We categorize its unresolved spectrum as a peculiar M8 dwarf; our best-matching NIR template is M8\,$\gamma$, however it presents several differences with it and lacks several low-gravity indications such as weaker alkali lines. We thus conclude that the peculiar nature of this spectrum is likely related to its binary nature, which further rules out a possible membership to ABDMG.

\textbf{2MASS~J23392527+3507165} has been discovered as an L3.5 BD in the optical by \cite{2008AJ....136.1290R}, and \cite{2010ApJ...710.1142B} categorized it as an L4.5 BD in the NIR. We recovered this object as a candidate member of $\beta$PMG in \emph{BASS}. We used its NIR spectral type to categorize it as a peculiar L4 BD that has a stronger $H$-band peak at $\sim$\,1.57\,$\mu$m (Figure~\ref{fig:J2339_pec}). However, it is unlikely that this object is young as it lacks the usual low-gravity indications. We thus reject it as a candidate member of $\beta$PMG. The peculiar $H$-band feature described above can be an effect of an unresolved T-type companion; however, the index-based scheme of \cite{2014ApJ...794..143B} indicates that this scenario is unlikely.

%\input{BANYANVII_017_sub02.bbl}.

%\bibliographystyle{apj}
%\bibliography{/Users/gagne/ApJ_Library,/Users/gagne/ApJ_Library_Old}

\end{document}